\documentclass[trackchanges]{aastex701}

\usepackage{threeparttable}
\usepackage{booktabs}
\usepackage{multirow}
\usepackage{subcaption}

\usepackage{adjustbox}
\usepackage{longtable}
\usepackage{rotating}
\usepackage{float} 
\usepackage{textcomp} 
\usepackage{hyperref}
\usepackage{amsmath}
\usepackage[normalem]{ulem}
\usepackage{enumitem}

\newcommand{\coonezero}{CO(1\textrm{--}0)\ }
\newcommand{\cotwoone}{CO(2\textrm{--}1)\ }

\newcommand{\vzgal}{V\textit{z}{-}GAL\ }
\newcommand{\zgal}{\textit{z}{-}GAL\ }
\newcommand{\aco}{$\alpha_{\rm CO}$ }

\usepackage{xcolor}

\begin{document}

\title{V\textit{z}-GAL Dusty Star-Forming Galaxies: Revisiting the CO--H$_2$ Conversion Factor Tension}

\correspondingauthor{Prachi Prajapati}

\author[orcid=0000-0002-3094-1077]{Prachi Prajapati}
\altaffiliation{Doctoral Researcher at International Max Planck Research School (IMPRS) for Astronomy \& Astrophysics}
\affiliation{Institute for Astrophysics, University of Cologne, Z{\"u}lpicher Straße 77, 50937 Cologne, Germany}
\affiliation{Max Planck Institute for Radio Astronomy (MPIfR), Auf dem H{\"u}gel 69, 53121 Bonn, Germany}
\email[show]{prajapati@ph1.uni-koeln.de}

\author[orcid=0000-0003-4678-3939]{Axel Weiss}
\affiliation{Max Planck Institute for Radio Astronomy (MPIfR), Auf dem H{\"u}gel 69, 53121 Bonn, Germany}
\email{}

\author[orcid=0000-0001-9585-1462]{Dominik Riechers}
\affiliation{Institute for Astrophysics, University of Cologne, Z{\"u}lpicher Straße 77, 50937 Cologne, Germany}
\email{}

\author[orcid=0000-0002-5268-2221]{Tom J.~L.~C. Bakx}
\affiliation{Departement of Space, Earth \& Environment, Chalmers University of Technology, Chalmersplatsen 4, Gothenburg SE-412 96, Sweden}
\email{}

\author[orcid=0000-0002-3952-8588]{Leindert A. Boogaard}
\affiliation{Leiden Observatory, Leiden University, PO Box 9513, 2300 RA Leiden, The Netherlands}
\email{}

\author[orcid=0009-0007-2281-4944]{Diana Ismail}
\affiliation{Université de Strasbourg, CNRS, Observatoire astronomique de Strasbourg, UMR 7550, 67000 Strasbourg, France}
\email{}

\author[orcid=0000-0003-2027-8221]{Pierre Cox}
\affiliation{Sorbonne Université, Université Paris 6 and CNRS, Institut d’Astrophysique de Paris, 98bis boulevard Arago, 75014 Paris, France}
\email{}

\author[orcid=0000-0002-7892-396X]{Andrew J. Baker}
\affiliation{Department of Physics and Astronomy, Rutgers University---New Brunswick, NJ 08854-8019, USA}
\affiliation{Department of Physics and Astronomy, University of the Western Cape, Bellville 7535, Cape Town, South Africa}
\email{}

\author[orcid=0000-0002-7176-4046]{Roberto Neri}
\affiliation{Institut de Radioastronomie Millim\'etrique (IRAM), 300 rue de la Piscine, 38406 Saint-Martin-d'H{\`e}res, France}
\email{}

\author[orcid=0000-0003-1939-5885]{Matthew Lehnert}
\affiliation{Sorbonne Université, Université Paris 6 and CNRS, Institut d’Astrophysique de Paris, 98bis boulevard Arago, 75014 Paris, France}
\affiliation{Universit\'e Lyon 1, ENS de Lyon, Centre de Recherche Astrophysique de Lyon (UMR5574), 69230 Saint-Genis-Laval, France}
\email{}

\author[orcid=0000-0002-8117-9991]{Chentao Yang}
\affiliation{Department of Space, Earth \& Environment, Chalmers University of Technology, Gothenburg SE-412 96, Sweden}
\email{}

\author[orcid=0000-0002-0071-3217]{Emilio Romano-Díaz}
\affiliation{Argelander Institute für Astronomie, Auf dem H{\"u}gel 71, D-53121 Bonn, Germany}
\email{}

\author[orcid=0000-0002-4205-9567]{Hiddo S. B. Algera}
\affiliation{Institute of Astronomy and Astrophysics, Academia Sinica, 11F of Astronomy-Mathematics Building, No.1, Sec. 4, Roosevelt Rd, Taipei 106319, Taiwan, R.O.C}
\email{}

\author[orcid=0000-0002-0320-1532]{Stefano Berta}
\affiliation{Institut de Radioastronomie Millim\'etrique (IRAM), 300 rue de la Piscine, 38406 Saint-Martin-d'H{\`e}res, France}
\email{}

\author[orcid=0009-0009-9483-8763]{Edoardo Borsato}
\affiliation{Dipartimento di Fisica e Astronomia ``G. Galilei," Universit\`a di Padova, vicolo dell’Osservatorio 3, I-35122 Padova, Italy}
\email{}

\author[orcid=0000-0001-7387-0558]{Kirsty M. Butler}
\affiliation{Departement of Space, Earth \& Environment, Chalmers University of Technology, Chalmersplatsen 4, Gothenburg SE-412 96, Sweden}
\email{}

\author[orcid=0000-0002-3892-0190]{Asantha Cooray}
\affiliation{University of California Irvine, Department of Physics \& Astronomy, FRH 2174, Irvine, CA 92697, USA}
\email{}

\author[orcid=0000-0002-0675-0078]{Bethany Jones}
\affiliation{Institute for Astrophysics, University of Cologne, Z{\"u}lpicher Straße 77, 50937 Cologne, Germany}
\email{}

\author[orcid=0000-0003-4357-3450]{Am\'elie Saintonge}
\affiliation{Max Planck Institute for Radio Astronomy (MPIfR), Auf dem H{\"u}gel 69, 53121 Bonn, Germany}
\email{}

\author[orcid=0000-0001-5434-5942]{Paul van der Werf}
\affiliation{Leiden Observatory, Leiden University, PO Box 9513, 2300 RA Leiden, The Netherlands}
\email{}

\begin{abstract}

The CO luminosity-to-H$_2$ mass conversion factor ($\alpha_{\rm CO}$) remains a debated uncertainty in determining molecular gas masses of high-redshift dusty star-forming galaxies (DSFGs). Dynamical mass constraints have often favored $\alpha_{\rm CO}=0.8$~$\rm M_{\odot}~{(K~km~{s}^{-1}~{pc}^{2})}^{-1}$, whereas dust- and radiative-transfer-based methods imply higher values. We revisit this ``tension" using the largest homogeneous sample of 21 unlensed $z\sim1-4$ DSFGs, with securely measured \coonezero luminosities from the VLA \vzgal survey and resolved ($\sim{0.1}^{\prime\prime}$) ALMA 1~mm dust continuum imaging. For 12 galaxies with robust modeling constraints, we derive molecular gas masses using dust spectral energy distribution modeling and the TUNER LVG framework, adopting a solar-metallicity gas-to-dust mass ratio of 100. Although not fully independent due to shared assumptions on dust properties, these approaches yield mutually consistent gas masses corresponding to $\alpha_{\rm CO}\sim1.5-11.5$~$\rm M_{\odot}~{(K~km~{s}^{-1}~{pc}^{2})}^{-1}$, with a median near the Galactic $\alpha_{\rm CO}=4.3$~$\rm M_{\odot}~{(K~km~{s}^{-1}~{pc}^{2})}^{-1}$. Isotropic virial dynamical masses agree with these gas masses when realistic molecular gas sizes are adopted, while our proposed ``mixed" (rotating, pressure-supported, thick-disk) estimator systematically underestimates dynamical masses, producing low $\alpha_{\rm CO}$ limits. Using GN20 ($z=4.055$) as a case study, we show that resolved gas geometry and kinematics reconcile the discrepancy with LVG-derived $\alpha_{\rm CO}$. Our results suggest that current data do not require $\alpha_{\rm CO}=0.8~\rm M_{\odot}~(K~km~s^{-1}~pc^{2})^{-1}$, and intermediate to near-Galactic values remain dynamically viable given uncertainties in gas geometry, dust properties, and gas-to-dust ratios. Further progress in calibrating $\alpha_{\rm CO}$ in the early universe will require resolved molecular gas observations, physically motivated ISM modeling, and stringent constraints on dust properties.

\end{abstract}


\section{Introduction} \label{sec:intro} 

Cold molecular gas is a fundamental component in the formation and evolution of galaxies across cosmic time \citep{Saintonge+2017,riechers2019coldz,tacconi2020,walter+2020}. Commonly traced via $^{12}$C$^{16}$O --- hereafter, CO --- rotational transitions, it both fuels star formation and serves as a sensitive probe of galactic gas dynamics. CO(1--0), with $\rm \nu_{rest} = 115.271$~GHz, is the ground-state rotational transition and can be excited even in the most extended, cold gas reservoirs. \coonezero line luminosities, $L^{\prime}_{\rm CO(1-0)}$ (in units of $\rm K~km~{s}^{-1}~{pc}^{2}$) are, therefore, commonly used to derive total molecular gas masses ($M_{\rm H_2}$) via the CO--H$_2$ conversion factor \cite[$\alpha_{\rm CO}$ in units of $\rm M_{\odot}~{(K~km~{s}^{-1}~{pc}^{2})}^{-1}$; see][]{solomon97}. 

\begin{equation}
    M_{\rm H_2} \, {\rm [M_{\odot}]} = \alpha_{\rm CO} \,\, L^{\prime}_{\rm CO(1-0)}
    \label{eqn:acodef}
\end{equation}

Within the Milky Way and nearby main-sequence galaxies, \aco is known to vary with local physical conditions, including temperature, density, and metallicity \citep[e.g.,][]{leroy2011,kohno2024}. However, direct, non-dynamical estimates of the \aco conversion factor in the local universe are themselves subject to significant systematic uncertainties. Calibrations based on $\gamma$-ray emission, visual and infrared extinction, X-ray absorption, dust emission, and isotopologue measurements generally agree at $\sim$15--25\% level, but exhibit an overall spread approaching a factor of about 2 in some environments \citep[see e.g.,][]{lee2014}. Importantly, these differences are not purely statistical but appear to be systematically offset between methods, suggesting that residual physical and methodological uncertainties remain even in nearby systems where such calibrations are most applicable. On galaxy-integrated scales, additional uncertainty is expected due to the optical thickness of the low-\textit{J} CO transitions, which can produce intrinsic variations in CO emissivity across different gas phases and environments \citep{bolatto2013}. Such effects are already evident in resolved observations of local molecular clouds and complexes \citep[e.g.,][]{lee2018}. 

On the other hand, local ultra-luminous infrared galaxies or ULIRGs and high-\textit{z} dusty star-forming galaxies (DSFGs) rarely have direct measurements of $\alpha_{\rm CO}$. Instead, their \aco is typically inferred from unresolved, global CO observations and indirect constraints, relying on assumptions about gas geometry, excitation conditions, dynamics, gravitationally supported motions, and virialization that may break down in extreme star-forming environments \citep{bolatto2013,friascastillo2022}. A long-standing result from these approaches is that submillimeter-bright dusty systems across redshifts appear to convert gas into stars more rapidly than coeval main-sequence galaxies, implying short depletion times and highly efficient star formation such as that seen in local ULIRGs/mergers \citep[e.g.,][]{friascastillo23,berta+2023,hagimoto2023,prajapati2026}. In the same systems, masses inferred from dynamical constraints using integrated CO linewidths are often lower than the molecular gas masses obtained from CO luminosity when adopting a Milky Way-calibrated $\alpha_{\rm CO}$, implying that a systematically lower inferred CO--H$_2$ conversion factor is appropriate for ULIRGs and DSFGs \citep[e.g.,][]{downes_solomon_1998,aravena+2016}. This discrepancy is commonly referred to as the ``\aco tension".

A key open question is whether the apparent tension between galaxy-integrated molecular gas masses inferred from CO luminosities and dynamical constraints reflects genuine variations in the physical conversion factor across populations or instead arises from observational limitations. While studies of nearby molecular clouds and ULIRGs generally benefit from well-constrained dynamics, remaining discrepancies there may instead reflect assumptions about the physical state of the gas (e.g., virial equilibrium or cloud structure). In contrast, high-\textit{z} DSFGs are affected by substantially larger observational uncertainties. The absence of integrated \coonezero measurements in many studies introduces uncertainties associated with gas excitation corrections. Even when such data are available, complex kinematic structures --- including rotation, turbulence, inflows/outflows, and merger-driven motions --- are generally reduced to a single integrated linewidth. In addition, dynamical mass estimates depend on source sizes, geometry, and inclination, parameters that are often poorly constrained, thereby compounding the overall uncertainty. These quantities are typically inferred from resolved dust continuum morphology because directly imaging the cold molecular gas at comparable resolution remains observationally challenging. Altogether, these may systematically bias gas mass and dynamical estimates, making it difficult to determine whether the inferred global, low \aco values in high-\textit{z} DSFGs are intrinsic or observationally driven. 

Spatially resolved observations of inherently faint \coonezero offer a direct way to break these degeneracies by independently constraining gas extent and kinematics for high-\textit{z} DSFGs. However, such studies remain limited at high redshift due to sensitivity constraints. They are typically restricted to the brightest or strongly lensed galaxies and quasars \citep[e.g.,][]{riechers+2011lensed,swinbank11,sharon13}, where interpretations may be affected by additional uncertainties from lensing reconstruction models. Existing high-resolution \coonezero and \cotwoone studies of unlensed high-\textit{z} DSFGs \citep[e.g.,][]{friascastillo2022} remain limited to only a handful of objects (e.g., GN20 at $z\sim4$; \citealt{hodge2012}) and often lack the sensitivity required for robust dynamical modeling. Consequently, they have not yet determined whether the apparent low \aco values in high-\textit{z} DSFGs arise primarily from observational limitations or reflect genuine physical variations. Discrepancies between the molecular gas mass estimates and dynamical constraints remain even when masses are inferred from integrated [CII] or (typically optically thin) [CI] observations\footnote{Assuming all C$^{+}$ or C is directly related to star formation, i.e., no formation due to shock compression.}, lacking a systematic comparison of conversion factors across various tracers for large samples of DSFGs \citep[e.g.,][]{dunne+2022,eales2024}.

This motivates a re-examination of the apparent \aco tension in high-\textit{z} DSFGs, where observational limitations are likely to play a much larger role than in nearby systems. Quantifying the impact of unresolved kinematics and uncertain source sizes is essential for determining whether these observational biases largely drive the low conversion factors inferred for current high-\textit{z} statistical samples or whether a residual discrepancy remains after improved dynamical constraints are obtained. A previous study of homogeneously selected unlensed, high-\textit{z} DSFGs reports 11 robust \coonezero detections \citep{friascastillo23}. Further, a handful of additional \coonezero studies of unlensed galaxies exist, such as \cite{carilli10,ivison+2011,sharon+2016,kaasinen2019}. 

To address limitations arising from heterogeneous selection effects and limited statistics, the VLA \vzgal \coonezero large program\footnote{\href{https://vzgal.uni-koeln.de/}{vzgal.uni-koeln.de}} \citep{stanley+2023,prajapati2026} has studied \coonezero emission in 106 Herschel/SPIRE 500$\mu$m-selected high-\textit{z} DSFGs. These 106 sources were selected from a parent sample of 135 DSFGs detected in higher-\textit{J} CO lines and far-infrared continuum by the NOEMA \zgal survey \citep{cox+2023}. In this paper, we focus on the unlensed sources identified at $z=1-6$ within the V\textit{z}-GAL sample. In Section~\ref{sec:obs}, we describe the identification of the target sample and available data. In Section~\ref{sec:analysis}, we analyze the data to infer resolved dust sizes and inclination angles. We also derive their gas masses using two complementary methods, one using dust spectral energy distribution modeling and another involving a large velocity gradient model \citep[LVG; see][]{weiss2007,boogard2026}. Section~\ref{sec:mdyn} explores an isotropic virial and a rotational thick-disk estimator of dynamical masses for DSFGs, and compares them to the molecular gas masses.\footnote{The dynamical constraints explored in this work should be interpreted in the context of a non-negligible intrinsic uncertainty floor in the standard CO--H$_2$ conversion factors, even in the local universe, alongside observational and modeling uncertainties in the gas mass and dynamical measurements.} Section~\ref{sec:discussion} discusses the inferences of the comparison in the context of the \aco tension, followed by a summary in Section~\ref{sec:summary}. Throughout the work, we use a spatially flat Lambda Cold Dark Matter ($\Lambda$CDM) cosmology with $ {H}_\mathrm{0}$ = 67.4 $\mathrm{km} \ \mathrm{s}^{-1} \mathrm{Mpc}^{-1}$ and ${\Omega}_\mathrm{M}$ = 0.315 \citep{planckcosmo2020}. Unless otherwise stated, all uncertainties on median values reported in this paper are given as the median absolute deviation (MAD).

\section{Sample and Data} \label{sec:obs} 

\subsection{Sample Selection: \vzgal Unlensed DSFGs} \label{subsec:sampleselection}

The high-\textit{z} dusty galaxies in the \vzgal sample are originally selected based on $S_{500\mu{\rm m}} > 80$~mJy \citep[see Figure 1 of][]{prajapati2026}. They have robust positions and redshifts based on higher-\textit{J} CO line observations from the NOEMA \zgal program \citep{cox+2023}, which consists of 135 sources with $S_{500\mu{\rm m}} > 80$~mJy selected from $\sim700$~deg$^2$ of the largest fields observed with the Herschel/SPIRE instrument. Assuming optically thin dust emission, \citet{ismail+2023} modeled the integrated dust spectral energy distributions (SEDs) of the sample, while \citet{berta+2023} studied the gas properties. Recent ALMA 1 mm (rest-frame $\sim$300$\mu$m at the median $z\sim2.3$ of the sample) dust continuum observations at $\sim{0.1}^{\prime\prime}$ (Bakx et al., in prep.) show the resolved morphologies of these dusty sources, which are used here along with the HST/F110W imaging \citep{Borsato2024}, Subaru/HSC, or Sloan Digital Sky Survey (SDSS) imaging of foreground objects to separate lensed from unlensed sources.

\begin{figure}[h]
\centering
\includegraphics[width=0.45\textwidth]{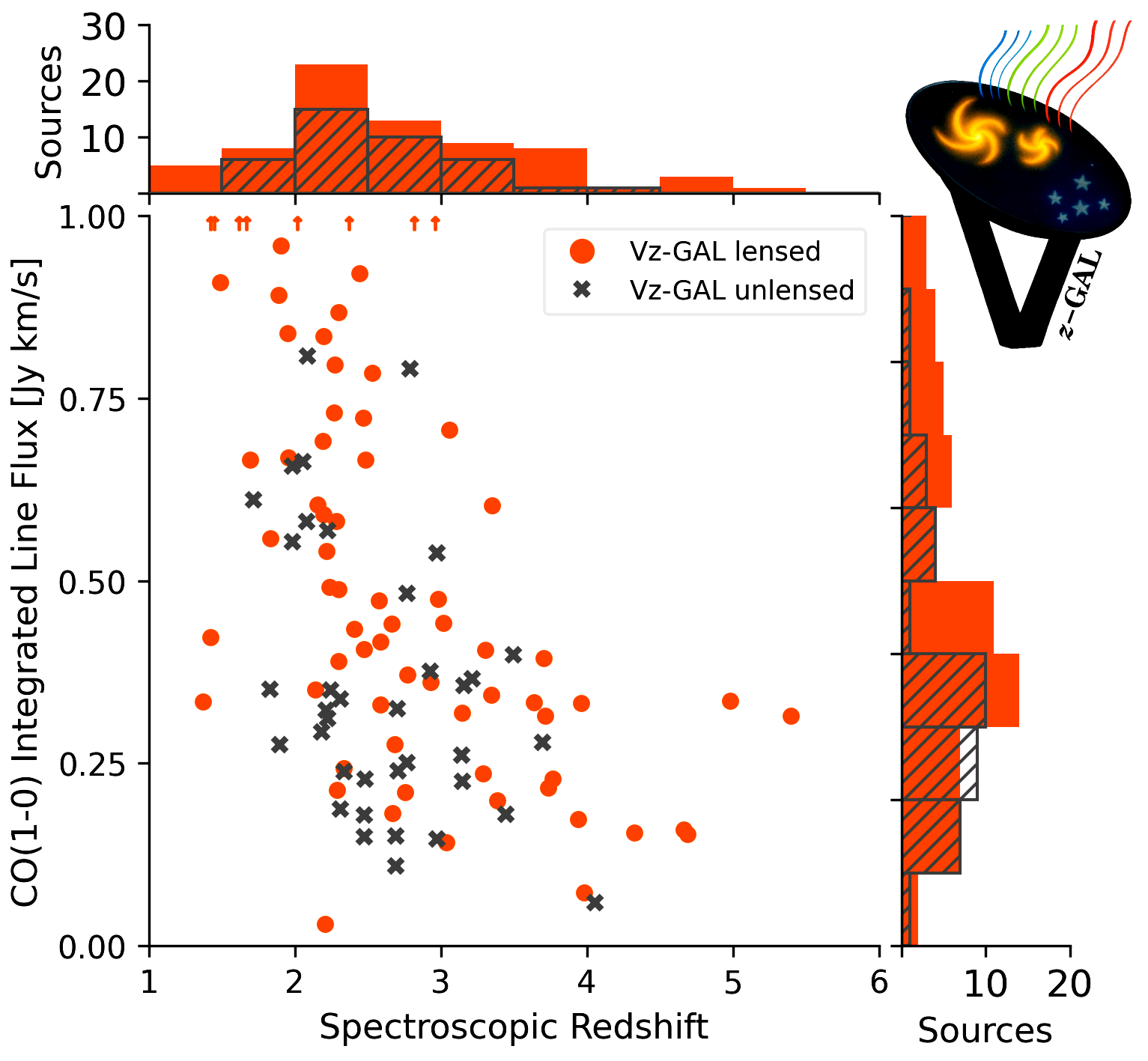}
\caption{Classification of the \vzgal high-\textit{z} dusty galaxies as lensed or unlensed sources, their \coonezero line fluxes and redshift distribution. We find that DSFGs at lower line fluxes show a higher fraction of unlensed galaxies, as expected.} 
\label{fig:selection}
\end{figure}

For this paper, we visually classify a \vzgal galaxy as gravitationally lensed if it satisfies one or both of the following criteria (see Figure~\ref{fig:selection}): (i) the ALMA dust continuum exhibits clear lensing signatures, such as extended morphologies with arcs, rings, or Einstein crosses, and/or (ii) a foreground object is present along the line of sight in the HST imaging that could plausibly act as a lens for the DSFG in the background. Galaxies that instead display compact dust emission peak(s) and lack any evident foreground lensing source are classified as unlensed DSFGs (see also Appendix $\S$~\ref{app:sources_fig}).

\begin{figure}[h]

    \centering
    \includegraphics[width=\textwidth]{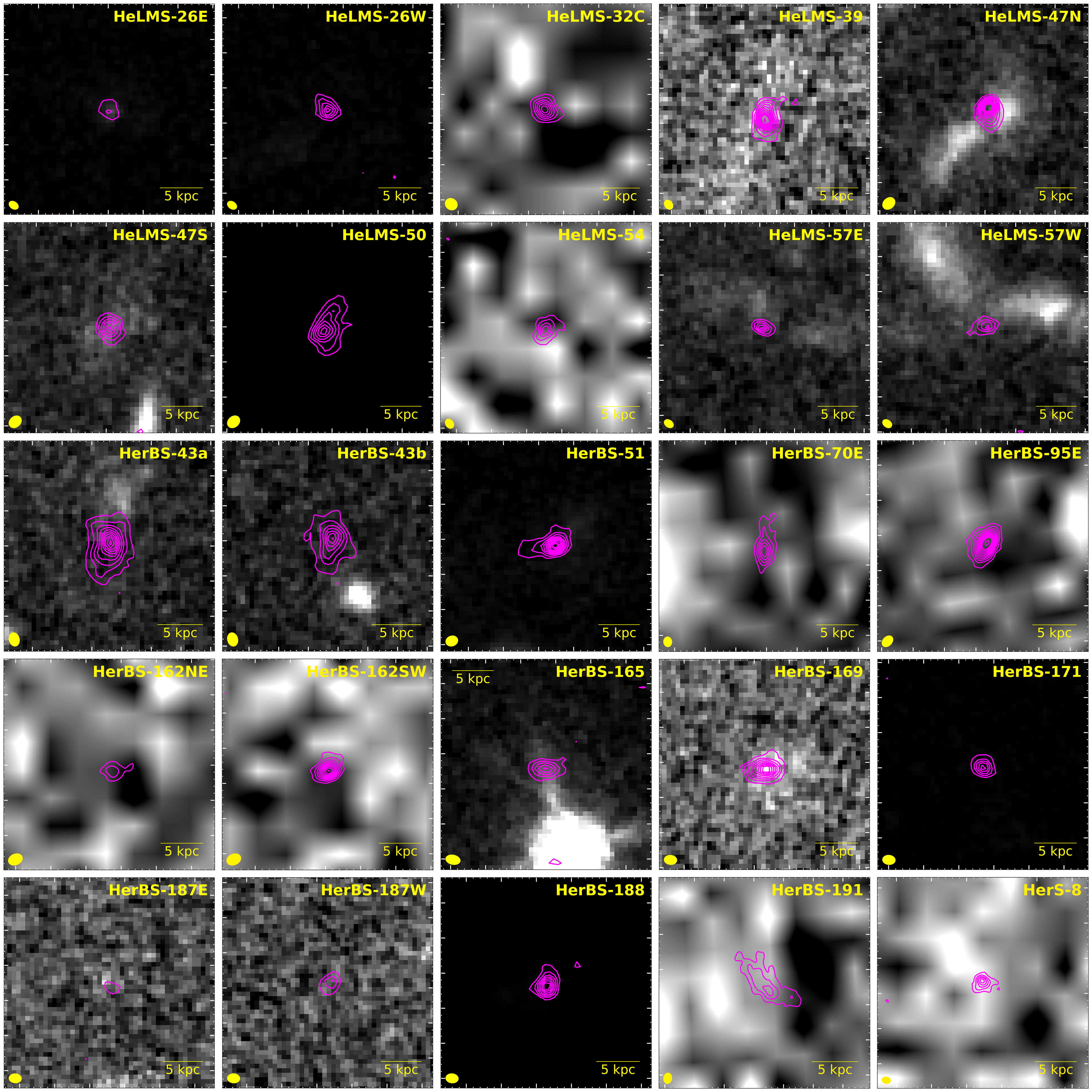}

\caption{ALMA 1~mm dust contours of the selected \vzgal unlensed sources (Table~\ref{tab:sources}) overlaid onto the optical images (HST/F110W, Subaru/HSC Y-band, and the SDSS) of the fields. Each image panel shown here is about $3^{\prime\prime}$ wide; i.e., it shows an area within roughly one synthesized beam of VLA \vzgal \coonezero observations. All the contour levels start at 4$\sigma$ and increase in steps of 4$\sigma$ up to 40$\sigma$, followed by a 10$\sigma$ increase up to 100$\sigma$. ALMA beam sizes are shown in the lower left corners of each panel. We also show a 5~kpc scale at the corresponding redshift for each source.}
\label{fig:source_grid}
\end{figure}

We find 19 Herschel fields with 27 isolated dusty galaxies (Table~\ref{tab:sources}, Figure~\ref{fig:source_grid}) and 14 fields (18 sources, Appendix~\ref{app:sources_fig}) containing multiple sources in the ALMA dust maps that are unlensed. In the latter fields, sources are close enough to each other to remain blended in the \vzgal data. Hence, these sources are not discussed further in this work, which focuses on the individual galaxies. Out of these DSFGs, 2/27 (HerBS-70W and HerBS-95W) fall on the edges of the available ALMA dust maps, and 3/27 have only tentative \coonezero detections (2$\sigma$--3$\sigma$, Table~\ref{tab:sources}). Further, HerBS-171 ($z=2.479$) is not robustly resolved with the available ALMA data (see Table~\ref{tab:sources}). After removing these 6/27 objects, our final sample consists of 21 unlensed \vzgal galaxies at $z\sim1-4$. These galaxies represent the largest homogeneous \coonezero sample of unlensed DSFGs in the early universe.

\setlength{\tabcolsep}{3pt} 

\begin{table}[h]
\scriptsize{
    \centering
    \caption{Identified isolated, unlensed sources from the \vzgal fields. See $\S$~\ref{subsec:sampleselection} for details.}
    \begin{tabular}{l cc c cc cc cc ccc}
        \toprule
Source & $z_{\mathrm{spec}}$ & $\lambda_{\rm rest}$ & {$S_{\lambda}$} & S/N & $^{a}$scale 
& \multicolumn{2}{c}{$^{b}$$r_{\mathrm{major}}$} 
& \multicolumn{2}{c}{$^{b}$$r_{\mathrm{minor}}$} 
& resolution & PA & CO(1--0) \\

 &  & [$\mu$m] & {[mJy]} &   & [kpc/${}^{\prime\prime}$]  & [mas] & [kpc] & [mas] & [kpc] & [${}^{\prime\prime} \times {}^{\prime\prime}$] & [deg] & significance\\ \midrule  

          HeLMS-26E & 2.690 &     274.61    &   2.99 $\pm$ 0.29    &   10.33 & 8.13  & 312 $\pm$ 55 & 2.54 $\pm$ 0.45   &   257 $\pm$ 48  & 2.09 $\pm$ 0.39 &   0.15$\times$0.10 &   41 $\pm$ 74 & 4$\sigma$\\
$^{c}${HeLMS-26W }& 2.688  &   274.76    &   3.93 $\pm$ 0.23    & 17.04   & 8.13  & 214 $\pm$ 20  & 1.74 $\pm$ 0.16  &  167 $\pm$ 16  &  1.36 $\pm$ 0.13  &   0.15$\times$0.10 & 41 $\pm$ 23 & $<$3$\sigma$\\
         HeLMS-32C & 1.715 &   366.49  &    3.06 $\pm$ 0.20   &  15.14  &  8.69  & 205 $\pm$ 20 & 1.78 $\pm$ 0.17  & 160 $\pm$ 17  & 1.39 $\pm$ 0.15    & 0.18$\times$0.16  & 48 $\pm$ 26 & 3$\sigma$\\ 
         HeLMS-39 & 2.766 &   264.43    &   12.14 $\pm$ 0.30    &  40.79  &  8.07 & 275 $\pm$ 17 & 2.22 $\pm$ 0.14  & 168 $\pm$ 13  & 1.36 $\pm$ 0.10   &  0.15$\times$0.11  & 178 $\pm$ 5 & 6$\sigma$\\
         HeLMS-47N & 2.223 &   309.66  &   6.90 $\pm$ 0.29    &  23.52  & 8.46   & 184 $\pm$ 15 & 1.56 $\pm$ 0.13  & 103 $\pm$ 21  & 0.87 $\pm$ 0.18   &  0.20$\times$0.15  & 8 $\pm$ 9 & 5$\sigma$\\
         HeLMS-47S & 2.223 &  309.66  &    3.81 $\pm$ 0.25   &   14.96 &  8.46   & 227 $\pm$ 20 & 1.92 $\pm$ 0.17  & 141 $\pm$ 22  & 1.19 $\pm$ 0.19    & 0.20$\times$0.15  & 32 $\pm$ 13 & 4$\sigma$\\
         HeLMS-50  & 2.053 &  326.90    &    6.74 $\pm$ 0.34   &  20.02  & 8.56   & 454 $\pm$ 34 & 3.88 $\pm$ 0.29  & 230 $\pm$ 20  & 1.97 $\pm$ 0.17    &  0.19$\times$0.15 & 163 $\pm$ 4 & 5$\sigma$\\
         HeLMS-54  & 2.707 &  268.64  &    4.55 $\pm$ 0.26   &   17.19 &  8.11   & 290 $\pm$ 22  & 2.35 $\pm$ 0.18 & 173 $\pm$ 18  & 1.40 $\pm$ 0.15    & 0.15$\times$0.11  & 138 $\pm$ 7 & 4$\sigma$\\
         HeLMS-57E & 1.982 &  339.81    &   2.25 $\pm$ 0.18    &  12.78  &  8.59  & 162 $\pm$ 20 & 1.39 $\pm$ 0.17 & 70 $\pm$ 16 & 0.60 $\pm$ 0.14    &  0.15$\times$0.10 & 76 $\pm$ 8 & 5$\sigma$\\
 $^{c}${HeLMS-57W} & 1.982 &   339.81   &   3.04 $\pm$ 0.23    &   13.36 &  8.59  & 245 $\pm$ 26 &  2.11 $\pm$ 0.22 & 104 $\pm$ 22  &  0.89 $\pm$ 0.19   &  0.15$\times$0.10  & 106 $\pm$ 7 & $<$3$\sigma$\\  
        HerBS-43a & 3.212 &   207.86   &   21.11 $\pm$ 0.53    &  40.11  & 7.72   & 500 $\pm$ 22 & 3.86 $\pm$ 0.17 & 324 $\pm$ 15 & 2.50 $\pm$ 0.12    & 0.20$\times$0.15  & 171 $\pm$ 4 & 5$\sigma$ \\
        HerBS-43b & 4.054 &  173.23   &   14.66 $\pm$ 0.52    &   28.12 &  7.07  &  488 $\pm$ 31 & 3.45 $\pm$ 0.22 & 312 $\pm$ 20 & 2.20 $\pm$ 0.14   &  0.20$\times$0.15  & 172 $\pm$ 6 & 5$\sigma$ \\     
        HerBS-51 & 2.183 &   332.05    &    5.84 $\pm$ 0.25   &   23.75 & 8.48  & 223 $\pm$ 15 & 1.89 $\pm$ 0.13 & 125 $\pm$ 10 & 1.06 $\pm$ 0.08    & 0.18$\times$0.15  & 110 $\pm$ 4 & 5$\sigma$ \\     
        HerBS-70E & 2.308 &  313.29   &   6.00 $\pm$ 0.27    & 22.24   & 8.40  &  403 $\pm$ 26 & 3.39 $\pm$ 0.22 & 157 $\pm$ 11 & 1.32 $\pm$ 0.09   &  0.14$\times$0.12  & 174 $\pm$ 2 & 4$\sigma$ \\
        $^{d}$HerBS-70W & 2.311 &  ...   &   ...    & ...  & ... &  ... & ... & ... & ...   &  ...  & ... & ...\\        
        HerBS-95E & 2.972 &  256.16    &     11.25 $\pm$ 0.32  &  34.88  & 7.91  & 287 $\pm$ 14 & 2.27 $\pm$ 0.11 & 165 $\pm$ 9 & 1.30 $\pm$ 0.07    & 0.18$\times$0.13  & 155 $\pm$ 4 & 5$\sigma$ \\
        $^{d}$HerBS-95W & 2.973 &  ...   &   ...    & ...  & ... &  ... & ... & ... & ...   &  ...  & ... & ...\\         
         HerBS-162SW & 2.474 &  304.93    &   4.08 $\pm$ 0.23    &  17.56  & 8.29 & 165 $\pm$ 16 & 1.37 $\pm$ 0.13  & 116 $\pm$ 14  & 0.96 $\pm$ 0.12   &  0.21$\times$0.16  & 136 $\pm$ 16 & 3$\sigma$\\
{HerBS-162NE} & 2.474 &   304.93   &    2.09 $\pm$ 0.28    &  7.54  & 8.29  & 258 $\pm$ 61 &  2.14 $\pm$ 0.51 &  145 $\pm$ 75 &  1.20 $\pm$ 0.62  &  0.21$\times$0.16  &  60 $\pm$ 28 & 3$\sigma$\\
        HerBS-165 & 2.225 & 328.42   &   3.27 $\pm$ 0.22    & 14.61   &  8.46  & 301 $\pm$ 32 & 2.54 $\pm$ 0.27  & 136 $\pm$ 18 & 1.15 $\pm$ 0.15   &  0.21$\times$0.14 & 95 $\pm$ 6 & 5$\sigma$ \\     
        HerBS-169 & 2.698 &    262.27   &    9.21 $\pm$ 0.36   &  25.22  & 8.12  & 295 $\pm$ 18 & 2.40 $\pm$ 0.15 & 171 $\pm$ 11 & 1.39 $\pm$ 0.09    &  0.18$\times$0.13 & 97 $\pm$ 5 & 5$\sigma$ \\
$^{e}$HerBS-171 & 2.479 &  278.78   &   3.57 $\pm$ 0.28    &   12.79 & 8.28  & 163 $\pm$ 24 & 1.35 $\pm$ 0.20 & 105 $\pm$ 35 & 0.87 $\pm$ 0.29    &  0.18$\times$0.13 & 24 $\pm$ 22 & 5$\sigma$ \\     
$^{c}${HerBS-187E}  & 1.828 &   342.95   &  2.17 $\pm$ 0.27     &  8.09  &  8.66  & 208 $\pm$ 61 &  1.80 $\pm$ 0.51 & 88 $\pm$ 66  &  0.73 $\pm$ 0.55   & 0.18$\times$0.13  & 58 $\pm$ 25 & $<$3$\sigma$\\           
         HerBS-187W & 1.827 &   342.95   &   1.31 $\pm$ 0.27    &  4.83  & 8.66   & 228 $\pm$ 35 & 1.97 $\pm$ 0.30  & 112 $\pm$ 45  & 0.97 $\pm$ 0.39    & 0.18$\times$0.13  & 153 $\pm$ 14 & 3$\sigma$\\
         HerBS-188 & 2.768 &   257.40    &    6.01 $\pm$ 0.31   &  19.42  & 8.07  & 215 $\pm$ 18 & 1.73 $\pm$ 0.14  & 98 $\pm$ 24 & 0.79 $\pm$ 0.19  &  0.18$\times$0.13 & 163 $\pm$ 6 & 4$\sigma$\\
         HerBS-191 & 3.443 &  233.26  &    8.54 $\pm$ 0.34   & 25.47   &  7.53  & 874 $\pm$ 63 & 6.58 $\pm$ 0.48  & 339 $\pm$ 26  & 2.55 $\pm$ 0.20    & 0.15$\times$0.12  & 48 $\pm$ 3 & 3$\sigma$\\
        HerS-8 & 2.243 &   316.88   &   3.05 $\pm$ 0.22    &  13.80  &  8.44 & 180 $\pm$ 21 & 1.52 $\pm$ 0.18 & 162 $\pm$ 22 & 1.37 $\pm$ 0.19    &  0.12$\times$0.10 & 120 $\pm$ 50  & 6$\sigma$ \\
\bottomrule
\end{tabular}

\parbox{\textwidth}{\footnotesize
\textbf{Notes.} \\
$^{a}$Angular scale at the source redshift in units of kiloparsec [kpc] per arcsecond [$^{\prime\prime}$]. \\
$^{b}$$r_{\mathrm{major}}$ and $r_{\mathrm{minor}}$ are the semi-major and semi-minor axes derived from 2D Gaussian fitting of the ALMA 1~mm dust maps in \texttt{CASA IMFIT}. These are beam-deconvolved sizes. \\
$^{c}${These sources with the \vzgal \coonezero line detection below 3$\sigma$ significance are not considered in our main analysis.} \\
$^{d}$Sources on the edge of the ALMA 1~mm maps. Therefore, they are excluded from further analyses in this work. \\
$^{e}${Not robustly resolved with the ALMA 1~mm data, as the deconvolved $r_{\mathrm{major}}$ is smaller than the beam. We also exclude this target.} 
}
\label{tab:sources}
}
\end{table}

\subsection{ALMA Continuum Observations} \label{subsec:almadustcont_dust}

Observed-frame 1~mm continuum in our target sample was observed using ALMA Band 7 across the years 2023 to 2025 (2022.1.00145.S, PI: Bakx; 2024.1.01570.S, PI: Algera; 2025.1.00326.S, PI: Bendo). The maximum baselines range from 2.5 to 3.7~km, with typically achieved angular resolutions across the 1~mm Band between ${0.1}^{\prime\prime}-{0.2}^{\prime\prime}$ or about 0.84--1.68~kpc at median $z\sim2.3$ (Table~\ref{tab:sources}). Each source was observed under excellent observing conditions, for a total on-source integration time of $\sim3$~min. Depending on the redshift of the source, these observations probe different rest-frame wavelengths (see Table~\ref{tab:sources}). The Common Astronomy Software Application \citep[\texttt{CASA},][]{McMullin+2007casa,casa2022} has been used to reduce and image the ALMA data; more details on this will be provided by Bakx et al. (in prep.). We show the resulting maps in Figure~\ref{fig:source_grid}. Further, Table~\ref{tab:sources} provides primary-beam-corrected continuum flux densities and signal-to-noise (S/N) ratios at the respective rest-frame wavelengths of each target. All the detections have peak S/N ratio of $\gtrsim$10, with HerBS-187W ($z=1.827$; S/N $\sim$ 5) as an exception.

\subsection{\vzgal \coonezero Measurements} \label{subsec:co10obs}

Unresolved measurements of the \coonezero emission-line fluxes for our targets were presented in the \vzgal studies \citep{stanley+2023,prajapati2026}. All the 21 DSFGs selected here have $>3\sigma$ detection of the \coonezero line. Table~\ref{tab:dynamicalmasses} summarizes their FWHM and \coonezero line luminosities adopted from the \vzgal papers. The overall spread in \coonezero line luminosities of our sources is consistent with that found in the literature for high-\textit{z} DSFGs or submillimeter galaxies (SMGs) \citep[e.g.,][]{aravena+2016,friascastillo23}. \coonezero emission for most of the 21 targets can be well described by a single Gaussian profile with an FWHM of 300 to 700~km~$\rm{s}^{-1}$. Similar linewidths have also been observed in the literature for other unlensed SMGs/DSFGs \citep[e.g.,][]{carilli10,friascastillo23}.  
A few of our systems, however, show more complex \coonezero line profiles (Appendix $\S$~\ref{app:lineprofiles}). 

\begin{itemize}[noitemsep]
    
    \item HeLMS-32C, HeLMS-50, HerBS-43a, and HerBS-43b show double-peaked \coonezero line profiles. All except HeLMS-50 are nearly symmetric, and the spatial distribution of dust emission in all of them appears compact. 

    \item HeLMS-50, HerBS-43a, HerBS-188, and HerBS-191 show broad (FWHM~$\gtrsim$~1000~km~${\rm s}^{-1}$) \coonezero line profiles. Their resolved dust emission does not show any clear merger signatures.
    
\end{itemize}

Moreover, HeLMS-50, HerBS-43a, HerBS-43b, and HerBS-191 exhibit dust radii larger than 3 kpc ($\S$~\ref{subsec:dustsize}), leading to their total dust extents spanning beyond $\sim$6~kpc. For all systems displaying complex \coonezero line profiles, the lack of obvious merger signatures in the current ALMA dust-continuum maps, however, does not rule out ongoing or past merger activity, particularly if such features are obscured by projection effects.\footnote{We also expect that deeper, multi-array-configuration dust observations would further help reveal more extended dust emission in all sources.}

\section{Analysis and Results} \label{sec:analysis}

\subsection{Dust Morphology} \label{subsec:dustmorphology} 

As mentioned in $\S$~\ref{subsec:sampleselection}, all the selected targets show compact dust morphology in the ALMA maps (Figure~\ref{fig:source_grid}). Their measured beam-deconvolved major- and minor-axis sizes, $r_{\mathrm{major}}$ and $r_{\mathrm{minor}}$, are reported in Table~\ref{tab:sources}, which are derived using 2D Gaussian fitting with \texttt{CASA IMFIT}. The beam position angles (PA) are also tabulated.

\subsubsection{Dust Radii and Axis Ratios} \label{subsec:dustsize}

We assume that the dust is distributed in a circular, inclined disk-like morphology, whose size ($R_{\rm dust}$) is thus best represented by the major axis size of the dust emission; i.e., ${R}_{\mathrm{dust}} = r_{\mathrm{major}}$. Most sources in our sample have dust-emitting radii between 1--3~kpc (Figure~\ref{fig:rdust_inclination}), with a sample median radius of $2.2 \pm 0.4$~kpc. Only 5/21 sources (HeLMS-50, HerBS-43a, HerBS-43b, HerBS-70E, and HerBS-191) exhibit radii above 3~kpc. Among these, HerBS-191 is the most extended source with a dust-emitting radius $\sim$7~kpc. Based on the minor and major axis sizes listed in Table~\ref{tab:sources}, we derive the observed axis ratios $q=r_{\rm minor}/r_{\rm major}$ spanning 0.4--0.9, with a median value of $0.6 \pm 0.1$. The median axis ratio and overall range are comparable between 6/21 sources ($\S$~\ref{subsec:co10obs}) with complex \coonezero line profiles and the rest with a single Gaussian-like line profile. The observed range of axis ratios reveals a broad diversity in projected dust morphologies, from nearly circular to highly elongated.

\subsubsection{Dust versus Cold Molecular Gas Sizes} \label{subsubsec:discussion_radius}

The median dust radius of 2.2~kpc for our sample is about $3\times$ lower than previously assumed \citep{bothwell2013} or previously observed \citep{riechers+2011sled2,ivison+2011,hodge2012} low-\textit{J} CO radii of 7~kpc in the literature. For a given galaxy, the dust continuum size is expected to be lower than the cold gas size due to the radiative-transfer effects \citep[e.g.,][]{rivera2018,boogard2026} involving spatial variations in gas column density, temperature, and optical depth (see Appendix $\S$~\ref{app:sizeRT}). This is true even in the absence of plausible biases coming from the observational constraints.

Therefore, dust sizes may not necessarily trace the entire cold molecular gas disk; instead, they may represent a relatively compact region near the center, e.g., a bulge or a bar. James Webb Space Telescope (JWST) observations of high-\textit{z} unlensed DSFGs already confirm that their active star formation extends beyond the central bulges traced by dust \citep{hodge2025}. Since dynamical masses ($M_{\rm dyn}\propto {\rm radius}$; see $\S$~\ref{subsec:mdynisovir} and $\S$~\ref{subsec:mdynmixed}), adopting dust continuum sizes may therefore yield systematically lower dynamical masses than would be inferred using the full CO(1--0) extent.\footnote{We emphasize here that although CO(1--0) is generally optically thick and therefore does not linearly trace molecular column density, its luminosity remains correlated with molecular gas mass via \aco because, in the galaxy-scale observations, the velocity-integrated emission from ensembles of turbulent, self-gravitating molecular clouds scales approximately with the total molecular mass \citep[e.g.,][]{dickman1986,solomon1987,bolatto2013}.}

\subsubsection{Dust Inclination Angles} \label{subsec:almadustcont_incl}

As our source selection (Section~\ref{sec:obs}) excludes regions with double or multiple dust sources --- plausibly hosting complex velocity fields --- we assume that the remaining 21 unlensed DSFGs selected here broadly follow the inclination angle prescription\footnote{originally derived by \citet{hubble1926}} by \cite{natascha2020} for star-forming galaxies (i.e., main-sequence disks) at cosmic noon. \cite{tacconi2013phibss} observe that most of the PHIBSS ($z\sim1-3$) massive, star-forming galaxies selected from the same parent sample are rotationally supported, turbulent disks in CO. Further, \citet{lee2025} observe that about 50\% of their massive star-forming main-sequence galaxies at $4<z<6$ are consistent with rotating disks. Therefore, our assumption in $\S$~\ref{subsec:dustsize} that the observed dust emission in our sources comes from disk-like morphologies remains consistent here, and we derive the sky-projected inclination angles ($i$) using:

\begin{equation}
      {\mathrm{sin}}^{2}(i) = \frac{(1-{q^2})}{(1-{{q_0}^2})} \, ; \, q=\frac{{r}_{\rm minor}}{{r}_{\rm major}}
         \label{eqn:inclination}
\end{equation}

where ${q_0}$ is a finite intrinsic thickness for a galactic disk, which we assume to be 0.2 based on \cite{natascha2020}.

\begin{figure}[h]
\centering
\includegraphics[width=0.45\textwidth]{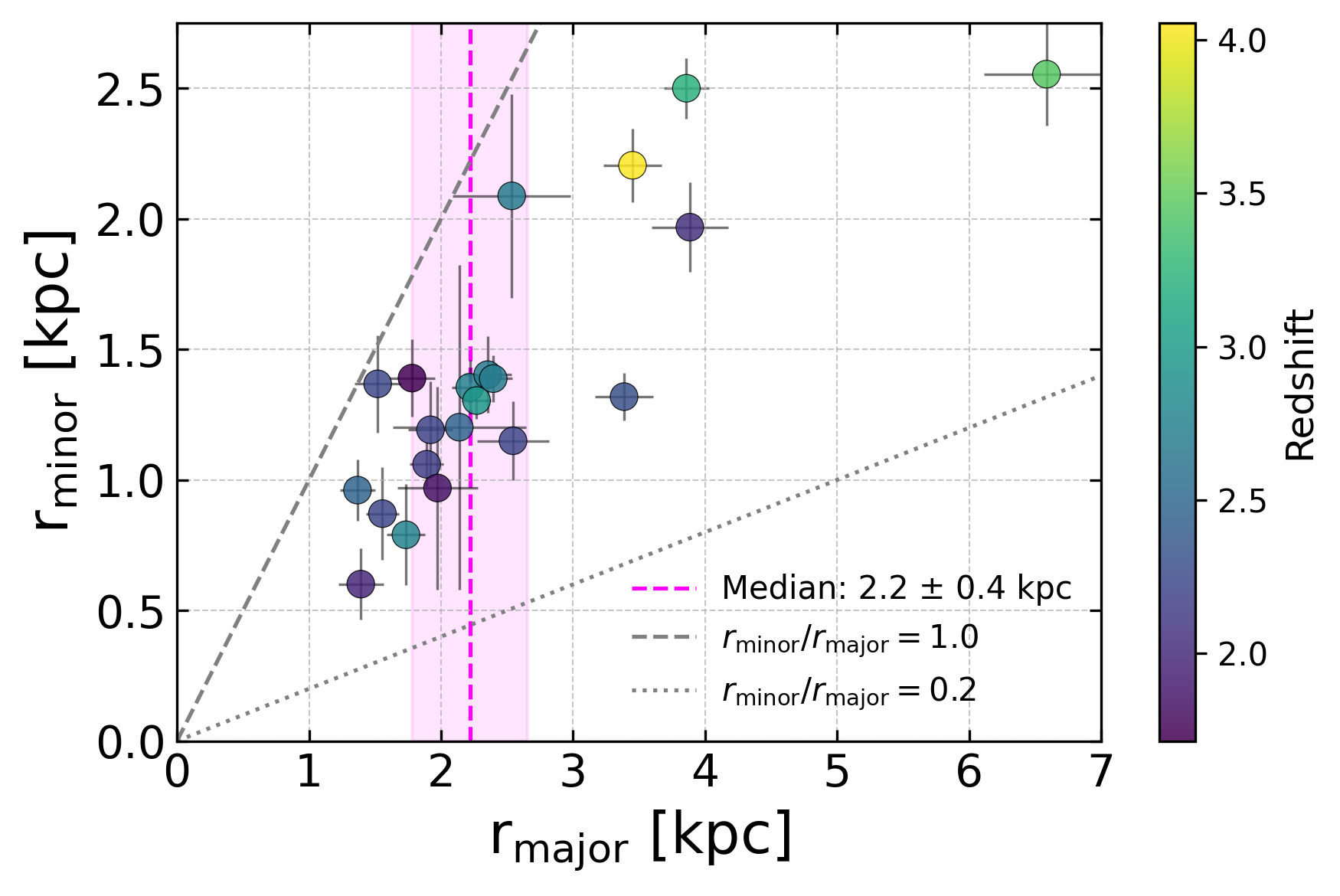}
\includegraphics[width=0.45\textwidth]{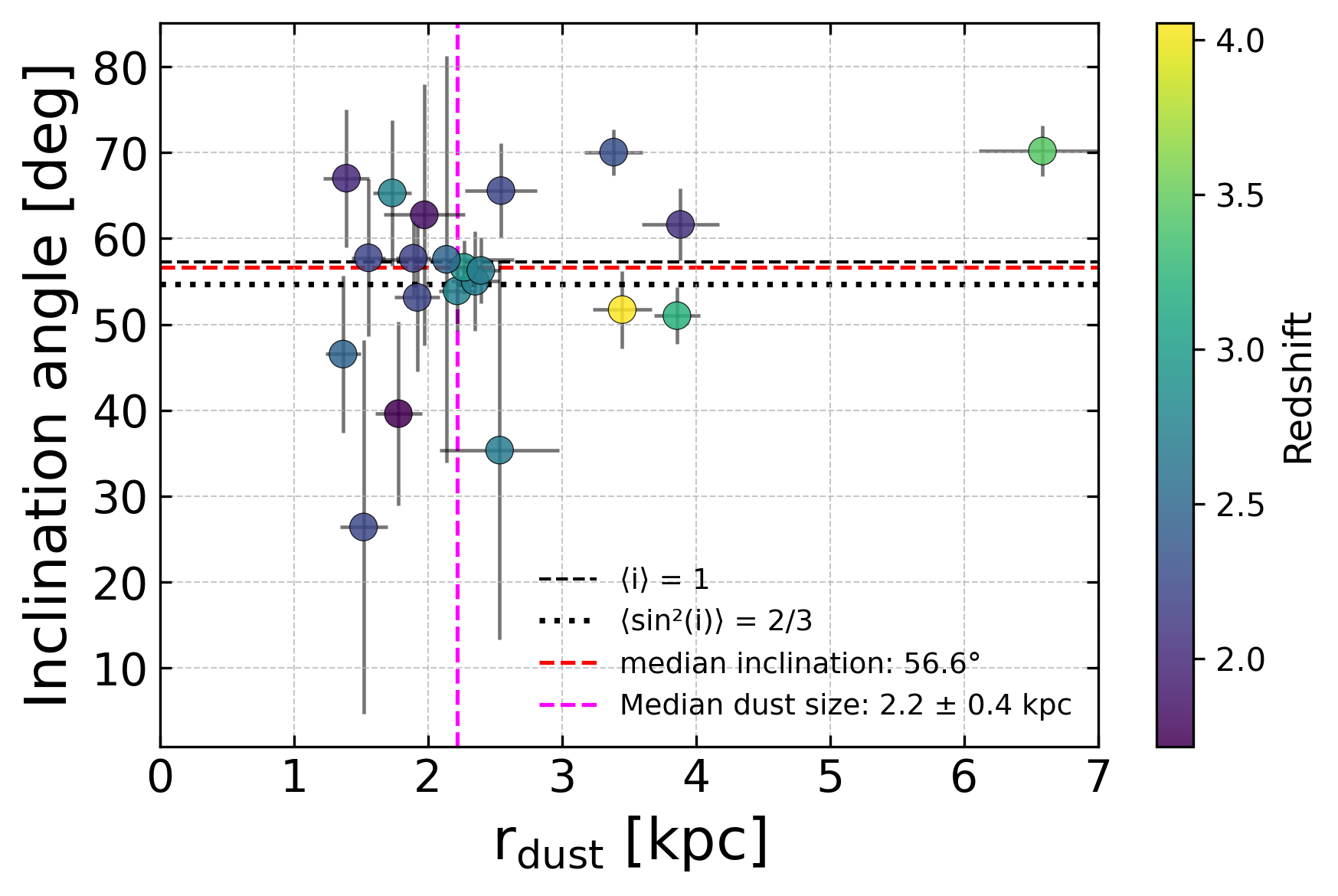}
\caption{Left panel: Observed major and minor axes of the targets using ALMA 1~mm observations. Right panel: Dust inclination angle distribution of the sources as compared to their derived dust sizes. {The median inclination angle is consistent with that expected for randomly oriented disks in the sky, corresponding to $\langle{i}\rangle = 1$ ($i = {57.3}^{\circ}$) or $\langle{\mathrm{sin}^{2}(i)}\rangle =  2/3$ ($i = {54.7}^{\circ}$).} These values should not be confused with the commonly adopted average inclination factor using $\langle{\mathrm{sin}(i)}\rangle = \pi/4 \simeq 0.79$, used in earlier studies \citep[e.g.,][]{neri2003,bothwell2013}; see Appendix~\ref{app:s-rand} for further discussion.}
\label{fig:rdust_inclination}
\end{figure}

\begin{figure}[h]
\centering
\includegraphics[width=0.45\textwidth]{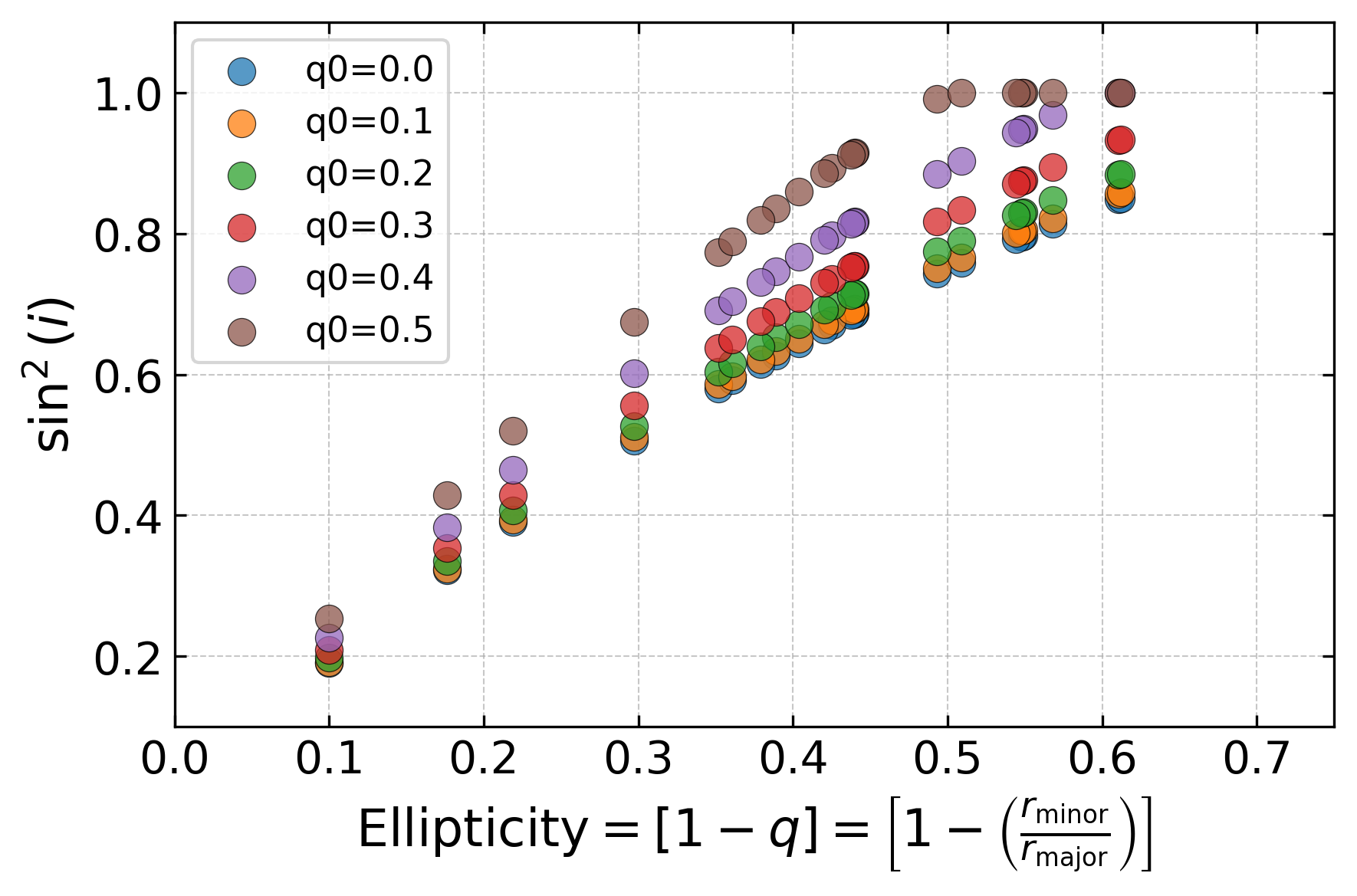}
\includegraphics[width=0.45\textwidth]{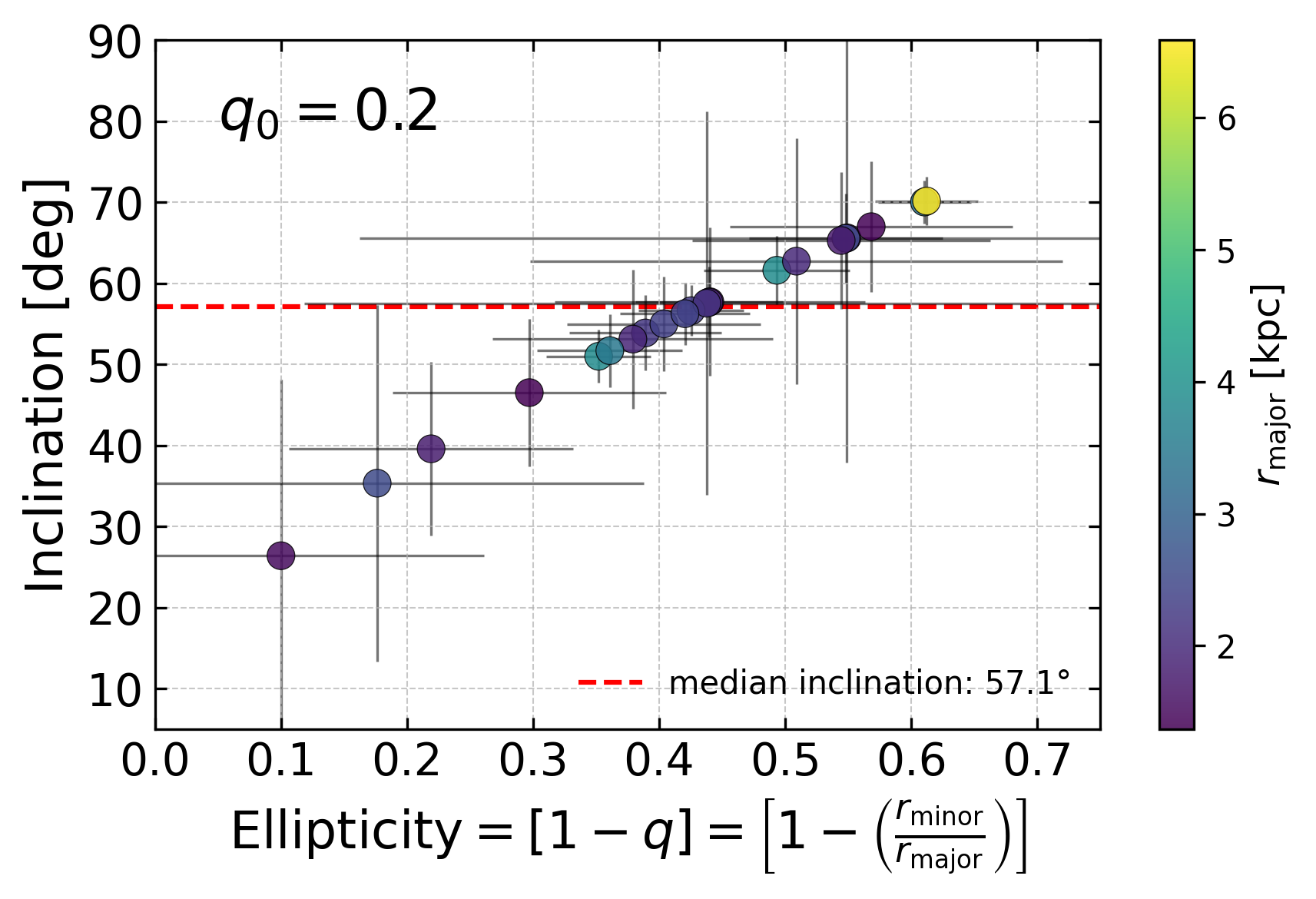}
\caption{Correlation between inclination angle and ellipticity of a disk. Both parameters depend on the observed axis ratio `q'. Left panel: Variation in inclination-angle contribution to the dynamical mass ($M_{\rm dyn}$) --- via ${\rm sin}^{-2}(i)$ factor \citep[][Eqn.~\ref{eqn:Mdyn}]{bothwell2013} --- with assumed intrinsic thickness ($q_0$) of the disk. For all $q_0$ values and near face-on systems, a steeper increase in observed inclination angles is expected with a smaller rise in ellipticity. Right panel: A specific case of $q_0=0.2$, as assumed in this work. Derived inclination angles are sensitive to the measured dust axis ratio, not the major axis alone.}
\label{fig:ellips}
\end{figure}

Dust inclination angles range from about 25$^{\circ}$ to 70$^{\circ}$ (Figure~\ref{fig:rdust_inclination}, Table~\ref{tab:dynamicalmasses}), with a median of 57.1$^{\circ}$. When using ${q_0}=0.2$, we recover a distribution of inclination angles that is consistent with randomly oriented disks ($p(i) \propto {\rm sin}\,i$; Appendix~\ref{app:s-rand}). However, the intrinsic thickness parameter ${q_0}$ is not directly constrained for our sources. Adopting ${q_0}$ to 0.1 (0.3) yields a median inclination of about 56$^{\circ}$ (60$^{\circ}$), which still remains consistent with random disk orientations within the measurement uncertainties. In contrast, assuming a substantially thicker disk with ${q_0}=0.5$ increases the median inclination to about 72$^{\circ}$. In Figure~\ref{fig:ellips}, we illustrate that adopting a larger intrinsic thickness ($q_0$) increases the inferred inclination, with the effect becoming progressively stronger for systems with higher projected ellipticity than for those with more circular dust emission. As a consequence, the same observed projected axis ratio can be reproduced by either a moderately inclined, intrinsically thin disk (low $q_0$) or a highly inclined, intrinsically thicker disk (high $q_0$), giving rise to a degeneracy in the inferred disk geometry.

By construction in Eqn.~\ref{eqn:inclination}, the inferred inclination is a monotonic, but non-linear, function of the observed axis ratio `$q$' and assumed intrinsic thickness `$\rm q_0$', leading to an intrinsic correlation between ellipticity $\rm e=(1-q)$ and $i$ (Figure~\ref{fig:ellips}). 
It is a direct consequence of both quantities sharing the same underlying observable, namely the axis ratio for dust emission, which in itself may not truly represent the axis ratio of the cold molecular gas disk \citep[see e.g.,][]{hodge2015,rowland2024}.\footnote{{A recent JWST study of such unlensed, high-\textit{z} DSFGs \citep{hodge2025} finds no correlation between the ellipticity derived from the rest-frame near-infrared and ALMA dust continuum emission, despite a match in their emission peaks. In the absence of resolved data of cold gas and stars for our sample, the geometries of gas and stellar disks remain unknown. Their future comparison with the dust data presented here will be crucial for mapping all the baryons in these systems.}} For example, a recent study by \citet{boogard2026b} identified a stellar bar with luminous dust emission in the central region of GN20, which would make the galaxy appear nearly edge-on if its inclination were inferred solely from the dust morphology. In contrast, both the cold molecular gas \citep{hodge2012} and stellar kinematics \citep{colina2023} indicate that the galaxy is viewed close to face-on. Therefore, in the absence of corrections based on resolved cold-gas data and dynamical modeling \citep[e.g., \texttt{$^{3{\rm D}}$BAROLO},][]{teodoro20153dbarolo}, dust/gas inclinations inferred from such a prescription should be used with caution as tracers of galactic disk orientation and geometry.

\begin{table}[h]
{
    \centering
    \caption{Dust sizes, inclination angles, assumed velocity dispersion, and integrated \coonezero line properties of the targeted DSFGs.}
    \begin{tabular}{l c c c c c c c c}
        \toprule
Category & Source & $z_{\mathrm{spec}}$ & $^{a}$$\rm R_{{dust}} = r_{{major}}$ & $^{b}$${\sigma}_0$ & $^{c}$Inclination &  $^{d}$FWHM & $^{d}$$\rm {L}^{\prime}_{CO(1-0),obs}$ & $^{e}$SFR\\ 
& &  & [kpc] & [km $\rm s^{-1}$]  & [deg] & [km $\rm s^{-1}$] & [${10}^{11}$ $L_l$] & [$M_{\odot}$ $\rm {yr}^{-1}$] \\ \midrule

$^f$Group-1 & HeLMS-39 & 2.766 &  2.22 $\pm$ 0.14  & {34.2} &53.9 $\pm$ 4.6  &  575 ± 58  &1.75 ± 0.25   & 2720 ± 200 \\
& HeLMS-54  & 2.707 & 2.35 $\pm$ 0.18 &  {33.9} & 55.0 $\pm$ 5.8  &  564 ± 130  &0.84 ± 0.24  & 2170 ± 260 \\
& HerS-8 & 2.243 &  1.52 $\pm$ 0.18  &  {32.1} & 26.4 $\pm$ 21.7  &  687 ± 89  & 0.89 ± 0.15 & 2660 ± 230 \\
& HerBS-51 & 2.183 &  1.89 $\pm$ 0.13 &  {31.8} &57.7 $\pm$ 4.3 &  448 ± 87   &0.71 ± 0.17  & 2445 ± 200 \\
& HerBS-70E & 2.308 &  3.39 $\pm$ 0.22 &  {32.3} & 70.1 $\pm$ 2.7  &  622 ± 130  & 0.90 ± 0.20&   1960 ± 90 \\
& HerBS-165 & 2.225 &   2.54 $\pm$ 0.27  &  {32.0} &65.6 $\pm$ 5.5  &  516 ± 83   &0.78 ± 0.18 & 1370 ± 130 \\
& HerBS-169 & 2.698 & 2.40 $\pm$ 0.15 &  {33.9} &56.3 $\pm$ 3.8  &  535 ± 99  & 1.13 ± 0.28 & 2630 ± 240 \\
\midrule

$^f$Group-2 & HeLMS-50  & 2.053 &  3.88 $\pm$ 0.29  & {31.2} &61.6 $\pm$ 4.2 &  1387 ± 139   &1.44 ± 0.22  & 1740 ± 150 \\
& HerBS-43a & 3.212 & 3.86 $\pm$ 0.17  &  {35.7} & 51.0 $\pm$ 3.3  &  1166 ± 249  &1.70 ± 0.50  & 3080 ± 120 \\
& HerBS-43b & 4.054 &  3.45 $\pm$ 0.22 & {38.1} &51.7 $\pm$ 4.5  &  744 ± 290  & 0.40 ± 0.20& 1560 ± 200 \\
& HerBS-188 & 2.768 & 1.73 $\pm$ 0.14 &  {34.2} & 65.3 $\pm$ 8.5  &  987 ± 192  &0.91 ± 0.22   &  2570 ± 220   \\
& HerBS-191 & 3.443 & 6.58 $\pm$ 0.48 &  {36.5} &70.2 $\pm$ 3.0  &  1751 ± 490  &1.93 ± 0.72   & 3130 ± 240 \\
& ... & ... & ... & ... & ... &  $^g$761 ± 36  & ...   &  ...\\
\midrule

Others & HeLMS-26E & 2.690 &  2.54 $\pm$ 0.45   & {33.9} & 35.4 $\pm$ 22.0 &  216 ± 46  & 0.52 ± 0.05  & $<$ 2910 (930) \\
& HeLMS-32C & 1.715 & 1.78 $\pm$ 0.17 & {29.7} & 39.6 $\pm$ 10.7  & 493 ± 49  & 0.40 ± 0.11 & $<$ 1000 (140) \\
& HeLMS-47N & 2.223 &  1.56 $\pm$ 0.13   & {32.0} &57.8 $\pm$ 9.1 &  834 ± 135   &1.42 ± 0.14 & $<$ 3250 (325) \\
& HeLMS-47S & 2.223 & 1.92 $\pm$ 0.17 & {32.0} &53.1 $\pm$ 8.6  &  439 ± 100  &0.82 ± 0.08 & $<$ 3250 (325) \\
& $^h$HeLMS-57E & 1.982 & 1.39 $\pm$ 0.17  &  {30.9} &67.0 $\pm$ 8.0  & 649 ± 108  & 0.67 ± 0.10  &  $<$ 1780 (160) \\
& HerBS-95E & 2.972 & 2.27 $\pm$ 0.11  &  {34.9} &56.6 $\pm$ 3.1  &  658 ± 137   & 0.60 ± 0.20 & 1210 ± 110 \\
& HerBS-162SW & 2.474 &  1.37 $\pm$ 0.13 &  {33.0} &46.5 $\pm$ 9.1  &  305 ± 94  &0.45 ± 0.05 & ... \\
& {HerBS-162NE} & 2.474 &   2.14 $\pm$ 0.51  &  {33.0} & 57.6 $\pm$ 23.7  & 408 ± 116  & 0.54 ± 0.05  & ... \\
& HerBS-187W & 1.827 & 1.97 $\pm$ 0.30 &  {30.2} &62.8 $\pm$ 15.2 &  517 ± 106   &0.62 ± 0.16   & $<$ 1180 (140)  \\

\bottomrule
\end{tabular}

\parbox{\textwidth}{\footnotesize
\textbf{Notes.} \\
$^{a}$Effective dust radius (${R}_{\mathrm{dust}} = r_{\mathrm{major}}$). \\
$^{b}${Velocity dispersion based on \cite{rizzo2024}.} \\
$^{c}$Dust inclination angle derived using Eqn.~\ref{eqn:inclination}. \\
$^{d}$From the \vzgal \coonezero catalog \citep{prajapati2026}. Here, \coonezero line luminosity is in the units of ${L}_{l}$=$\rm K~km~{s}^{-1}~{pc}^{2}$. \\
$^{e}$Star formation rates (SFRs) from the NOEMA \zgal catalog \citep[see][]{berta+2023}. For targets with unresolved spatial components, we provide upper limits (with error bars) based on the summed far-infrared luminosity. \\
$^{f}$Group-1 an -2 categories defined for the sources as discussed in $\S$~\ref{subsec:gasmasses}. \\
$^{g}$Linewidth of CO(5--4) from \zgal results \citep{cox+2023}, as the \coonezero line has an extremely broad line profile with large errors in \coonezero FWHM (see Appendix~\ref{app:lineprofiles}). We use this linewidth of HerBS-191 for the LVG modeling and dynamical mass estimates presented in this work. \\
$^{h}$We separately extract the \coonezero parameters for the eastern component that were not reported by \cite{prajapati2026}.
}

\label{tab:dynamicalmasses}
}
\end{table}

\subsection{Molecular Gas Masses} \label{subsec:gasmasses}

A value of \aco= 0.8~$\rm M_{\odot}~{(K~km~{s}^{-1}~{pc}^{2})}^{-1}$ (units omitted hereafter) in Eqn.~\ref{eqn:acodef} is frequently assumed for high-\textit{z} DSFGs, following studies of local ULIRGs \citep{downes_solomon_1998}. This choice is largely motivated by dynamical mass arguments, as larger conversion factors would often imply molecular gas masses exceeding the available dynamical mass budget. With a dataset of galaxies that does not require corrections for either gravitational lensing or CO excitation, our aim here is to derive molecular gas masses using two complementary methods, which can then be compared with dynamical mass estimates.

First, in $\S$~\ref{subsec:gasmass_gdmr}, we fit the far-infrared-to-millimeter dust continuum with a general modified blackbody (GMBB; Eqn.~\ref{eq:general_mbb}) model, using the resolved dust size ($R_{\rm dust}$) to constrain the dust surface density and optical depth, thereby reducing the otherwise strong degeneracy between $T_{\rm dust}$, $\tau_\nu$, and $M_{\rm dust}$. The resulting dust masses are converted to gas masses assuming a solar-metallicity gas-to-dust ratio of G/D = 100 \citep[e.g.,][]{leroy2011,dunne21,Gururanjan2023}. Second, we use the TUNER large velocity gradient (LVG) radiative transfer framework \citep{boogard2026} to jointly model the CO line and dust continuum SEDs within a turbulent, multi-density interstellar medium or ISM ($\S$~\ref{subsec:gasmass_tuner}). In this approach, the molecular gas mass is constrained primarily by CO excitation and radiative transfer, while the dust emission is modeled self-consistently using the full GMBB method. Unlike the dust-only method, TUNER treats the characteristic emitting radius as a free parameter and therefore does not require the resolved dust size to define the emitting area. Both approaches are thus complementary, drawing on different observables and subject to different dominant uncertainties: one is driven primarily by the continuum SED and resolved dust geometry, while the other is constrained by the combined dust and CO excitation properties on galaxy-integrated scales. However, they are not fully independent, as both adopt the same dust opacity normalization ($\kappa_\nu$) and G/D = 100 and rely on simplified global descriptions of the dust emission. Consequently, differences in the inferred gas masses (Tables~\ref{tab:dustmass_gdmr} and \ref{tab:tunerlvg}) primarily reflect the distinct constraints on dust temperature, $\beta_{\rm dust}$, optical depth, and effective emitting size arising from continuum-only versus joint dust-and-gas radiative transfer modeling.

To compare these gas-mass estimators, we further restrict our analysis to a subset of \vzgal systems for which reliable LVG radiative transfer modeling can be carefully performed. From 21 DSFGs, we exclude sources containing multiple DSFGs that were blended within the {Herschel}/SPIRE beam but resolved in subsequent VLA \vzgal and ALMA observations (e.g., HeLMS-47N and HeLMS-47S, HerBS-162SW and HerBS-162NE). For these systems, the peak of the far-infrared dust SED is poorly constrained because the Herschel photometry blends emission from multiple components. Three systems (HerBS-43a, HerBS-43b, and HerBS-70E), however, are retained because their robustly de-blended dust continuum flux densities are available from \citet{stanley+2023}. After applying these selection criteria, the final sample comprises 12 galaxies with sufficiently well-constrained photometric and spectroscopic data to enable robust TUNER LVG modeling and a meaningful comparison with the other gas-mass estimation method. These 12 systems comprise two distinct subsets (Table~\ref{tab:dynamicalmasses}, Group-1 and Group-2). The first consists of seven sources whose \coonezero line profiles appear more consistent with a single Gaussian with a FWHM $< 700$~km~s$^{-1}$: HeLMS-39, HeLMS-54, HerS-8, HerBS-51, HerBS-70E, HerBS-165, and HerBS-169. The second includes five sources whose broad or complex \coonezero line profiles (FWHM $\gtrsim 700$~km~s$^{-1}$, see also $\S$~\ref{subsec:co10obs}): HeLMS-50, HerBS-43a, HerBS-43b, HerBS-188, and HerBS-191. Hereafter, we refer to these two subsets as Group-1 and Group-2, respectively.

Integrated dust continuum SED data points used in the following two sections are adopted from the \zgal studies \citet{neri+2020} and \citet{ismail+2023}, including all the $>2\sigma$ detections. These have been restricted within the rest-frame $50-1000~\mu$m frequency range to cover the far-infrared dust continuum.

\subsubsection{{Gas Masses Derived using Dust Continuum and Galactic Gas-to-Dust Mass Ratio}} \label{subsec:gasmass_gdmr}

Most local and high-\textit{z} studies typically use the Galactic calibrations of the gas-to-dust mass ratio (G/D) $\sim$ 100 for deriving gas masses in solar metallicity environments \citep[e.g.,][]{leroy2011}. A similar value has also been used while cross-calibrating cold gas tracers, namely CO, [CI], and dust, in extragalactic observations \citep{dunne+2022,Gururanjan2023}. Dunne et al. argue that high-\textit{z} DSFGs/SMGs are likely to have near-solar or even super-solar metallicities, given their extreme far-infrared luminosities, substantial dust reservoirs, and the intense star formation activity required to sustain them. However, direct metallicity measurements remain unreliable in such systems due to heavy extinction at rest-frame optical/near-infrared wavelengths \citep[high $\rm A_{V}$; see][]{bik2024,hodge2025}. A few other studies argue for variations in chemical abundances and metallicity with redshift in DSFGs \citep[e.g.,][]{eales2024}. In our analysis, however, we have fixed the G/D ratio to 100.

All the down-selected 12 sources have dust masses derived using modified blackbody (MBB) fits in their optically thin ($\tau_{\rm \nu} \ll 1$) approximation to the far-infrared SED \citep{neri+2020, ismail+2023,stanley+2023}. However, Ismail et al. demonstrated through mock SED fitting that combining the MBB in its general form (GMBB) with resolved dust sizes reliably recovers the dust mass (within $\sim 10\%$) compared to systematic {overestimates} of a factor of $>1.5$ under an optically thin approximation. Here, we adopt the GMBB following the form below \citep[see][for a detailed description]{ismail+2023} to fit the dust continuum data, which are tabulated in Appendix~\ref{herbs165_corner}.

\begin{equation}
S_{\nu} \propto (1-e^{-\tau_{\nu}}) \ B_{\nu}(T_{\rm dust});
\label{eq:general_mbb}
\end{equation}

where $B_{\nu}(T_{\rm dust})$ is the Planck function at frequency $\nu$ and dust temperature $T_{\rm dust}$ and $(1-e^{-\tau_{\nu}})$ is the emissivity coefficient. The optical depth: 

\begin{equation}
    \tau_\nu \propto \nu^{\beta} M_{\rm dust} R_{\rm dust}^{-2}; 
    \label{eqn:taugdmr}
\end{equation}

\begin{figure}[h]
\centering
\includegraphics[width=0.75\textwidth]{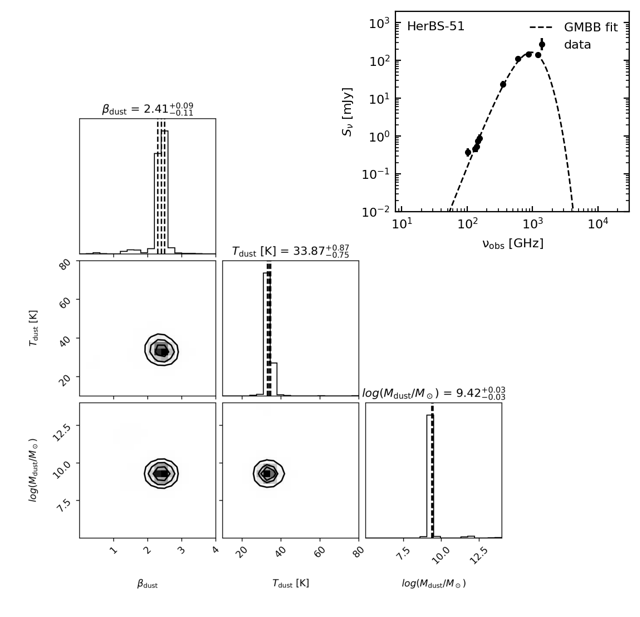}
\caption{HerBS-51 ($z=2.183$) as an example of the GMBB approach to fit the dust SED.}
\label{fig:gmbb_herbs51}
\end{figure}

$\beta$ being the dust emissivity index, $M_{\rm dust}$ is the dust mass, and $R_{\rm dust}$ is the effective dust emission radius (see Table~\ref{tab:dynamicalmasses}). Our calibration here is based on \cite{draine2014} with corrections for $M_{\rm dust}$ following \citet{berta2016} and for the cosmic microwave background (CMB) effects following the prescriptions of \citet{dacunha2013}. In Figure~\ref{fig:gmbb_herbs51}, we show an example of HerBS-51 ($z=2.183$) with an SED fit and the corresponding corner plot of the posterior distributions of the derived parameters.

The GMBB approach naturally yields warmer dust temperatures for larger optical depths and lower dust masses than those reported by \zgal studies under the optically thin approximation (see Table~\ref{tab:dustmass_gdmr}). We find median dust masses of (3.24 $\pm$ 0.15)$\times {10}^{9}$~${\rm M}_{\odot}$ and (3.98 $\pm$ 0.59)$\times {10}^{9}$~${\rm M}_{\odot}$ for Groups-1 and -2, respectively. Using these dust masses scaled with G/D = 100, we derive the \aco conversion factor having:

\begin{equation}
    M_{\rm gas,G/D} = 100 \times M_{\rm dust,G/D} = \alpha_{\rm CO,G/D} \times L^{\prime}_{\rm CO(1-0),obs}
    \label{eqn:acogdmr}
\end{equation}

The values of $\alpha_{\rm CO,G/D}$ vary from 1.45 to 4.55 (median = $3.06 \pm 1.21$) for Group-1 and from 2.45 to 8.47 (median = $3.73 \pm 0.51$) for Group-2 sources.

\begin{table}[h]
    \centering
    \caption{Far-infrared SED-derived dust masses of the down-selected 12 sources. These results are derived using the GMBB approach combined with the resolved dust sizes (see $\S$~\ref{subsec:gasmass_gdmr}).}
    \begin{tabular}{l c c c c c}
        \toprule
Category & Source  & $\rm {M}_{dust,G/D}$ & $^{a}$$\rm {M}_{gas,G/D}$ &$\rm T_{\mathrm{dust,G/D}}$ &   $\beta_{\mathrm{dust,G/D}}$   \\ 

&   & [${10}^{9}$ $\rm M_{\odot}$]  &[${10}^{11}$ $\rm M_{\odot}$]  &  [K]  &     \\ \midrule

Group-1 &  HeLMS-39  &  $3.24^{+1.29}_{-1.26}$  & $3.24^{+1.29}_{-1.26}$ & ${33.64}^{+3.10}_{-1.24}$    &   ${2.39}^{+0.20}_{-0.32}$    \\

&  HeLMS-54  &  $2.57^{+1.45}_{-1.41}$   & $2.57^{+1.45}_{-1.41}$& ${32.08}^{+6.55}_{-1.92}$    &   ${2.52}^{+0.35}_{-0.47}$     \\

 &  HerS-8   &  $1.29^{+2.04}_{-1.82}$  & $1.29^{+2.04}_{-1.82}$ & ${36.28}^{+3.70}_{-1.60}$    &   ${3.09}^{+0.50}_{-0.52}$   \\

 & HerBS-51  &  $3.09^{+1.07}_{-1.07}$   & $3.09^{+1.07}_{-1.07}$ & ${33.87}^{+0.87}_{-0.75}$     &   ${2.41}^{+0.09}_{-0.11}$     \\

& HerBS-70E  &  $3.24^{+1.07}_{-1.07}$  & $3.24^{+1.07}_{-1.07}$ & ${29.81}^{+1.25}_{-1.07}$    &   ${2.40}^{+0.10}_{-0.11}$    \\

& HerBS-165  &  $3.55^{+1.32}_{-1.23}$   &  $3.55^{+1.32}_{-1.23}$ & ${28.82}^{+2.37}_{-1.49}$     &   ${2.32}^{+0.23}_{-0.25}$    \\

& HerBS-169  & $3.39^{+1.15}_{-1.15}$   & $3.39^{+1.15}_{-1.15}$  & ${35.46}^{+1.85}_{-1.57}$     &   ${2.04}^{+0.15}_{-0.15}$    \\

\midrule

Group-2 & HeLMS-50  &   $5.37^{+1.32}_{-1.20}$  & $5.37^{+1.32}_{-1.20}$ & ${27.54}^{+4.94}_{-1.58}$     &   ${2.28}^{+0.20}_{-0.36}$    \\

& HerBS-43a  &   $4.17^{+1.12}_{-1.12}$   & $4.17^{+1.12}_{-1.12}$ & ${36.03}^{+3.03}_{-2.34}$    &   ${1.87}^{+0.16}_{-0.17}$     \\

& HerBS-43b  &   $3.39^{+1.29}_{-1.23}$   & $3.39^{+1.29}_{-1.23}$ & ${33.39}^{+9.81}_{-4.83}$    &   ${1.76}^{+0.36}_{-0.45}$   \\

& HerBS-188 &   $3.24^{+1.12}_{-1.10}$  &  $3.24^{+1.12}_{-1.10}$ & ${38.52}^{+2.03}_{-1.61}$    &   ${1.86}^{+0.14}_{-0.15}$   \\

& HerBS-191  &  $3.98^{+1.15}_{-1.15}$  & $3.98^{+1.15}_{-1.15}$ & ${36.52}^{+2.96}_{-2.63}$    &   ${1.81}^{+0.18}_{-0.18}$    \\

\bottomrule
\end{tabular}

\parbox{\textwidth}{\footnotesize
\textbf{Notes.} \\
$^{a}$ using gas-to-dust mass ratio: G/D ratio = 100.  \\
}

\label{tab:dustmass_gdmr}
\end{table}

\subsubsection{Gas Masses from TUNER LVG Modeling} \label{subsec:gasmass_tuner}

All the galaxies in our sample have robust \coonezero detections from \vzgal \citep{prajapati2026} and CO spectral line energy distribution (SLED) coverage up to CO(5--4) from the NOEMA \zgal \citep[Table~\ref{tab:tuner_data};][]{neri+2020,cox+2023}. We here model their integrated dust and CO emission using TUrbulent Non-Equilibrium Radiative transfer (TUNER) framework \citep{jarugula2021,harrington2021,boogard2026}, a semi-empirical LVG model. TUNER uses a 50-component density PDF to represent the turbulent ISM density structure within an unresolved DSFG \citep[see e.g.,][]{krumholz2005,glover2011}. Its Bayesian implementation \citep[\texttt{EMCEE};][]{foreman2013} jointly fits the observed dust SED and CO SLED using this density PDF to infer global ISM properties by determining their posterior distributions.

\begin{table}[h]
    \centering
    \caption{CO line fluxes used for the TUNER LVG modeling. These are adopted from the \vzgal and \zgal papers \citep{neri+2020,cox+2023,stanley+2023,prajapati2026}.}
    \begin{tabular}{l c c ccc ccc}
        \toprule
Category & Source  & \coonezero & \cotwoone  & CO(3--2)  & CO(4--3)  & CO(5--4)    \\ 

&   & [Jy~km~${s}^{-1}$]  & [Jy~km~${s}^{-1}$] & [Jy~km~${s}^{-1}$] & [Jy~km~${s}^{-1}$] & [Jy~km~${s}^{-1}$]  \\ \midrule

Group-1 &  HeLMS-39  & 0.48 $\pm$ 0.07 &... & 2.65 $\pm$ 0.36 &... & 3.98 $\pm$ 0.55   \\

&  HeLMS-54  &  0.24 $\pm$ 0.07 & ...& 0.72 $\pm$ 0.23 &...&... \\

 &  HerS-8  &  0.35 $\pm$ 0.06 &...& 1.14 $\pm$ 0.30 & 2.75 $\pm$ 0.56 &...\\ 
 
 & HerBS-51  &   0.30 $\pm$ 0.07 &...& 1.33 $\pm$ 0.14 & 2.12 $\pm$ 0.22 &...  \\
 
& HerBS-70E  &  0.32 $\pm$ 0.09 &...& 1.80 $\pm$ 0.50 & 3.40 $\pm$ 0.35 &...   \\ 
 
& HerBS-165  & 0.31 $\pm$ 0.07 &...& 0.97 $\pm$ 0.25 & 1.49 $\pm$ 0.30 &...   \\

& HerBS-169 & 0.32 $\pm$ 0.08 &...& 1.73 $\pm$ 0.18 &...&...  \\

\midrule

Group-2 & HeLMS-50  &   0.66 $\pm$ 0.10 &...& 4.41 $\pm$ 0.83 & 3.86 $\pm$ 0.39 &... \\

& HerBS-43a & 0.37 $\pm$ 0.10 &...&...& 5.50 $\pm$ 0.80 & 6.70 $\pm$ 0.80    \\

& HerBS-43b  &  0.06 $\pm$ 0.03 &...&...& 1.70 $\pm$ 0.34 &1.50 $\pm$ 0.34    \\ 

& HerBS-188 & 0.25 $\pm$ 0.06 &...& 1.94 $\pm$ 0.38 &...& 2.54 $\pm$ 0.53    \\

&  HerBS-191  & $^{a}$0.18 $\pm$ 0.06 &...&...& 3.23 $\pm$ 0.55 & 2.52 $\pm$ 0.39   \\

\bottomrule
\end{tabular}

\parbox{\textwidth}{\footnotesize
\textbf{Notes.} \\
$^{a}$Line flux integrated from the moment-0 map created using the CO(5--4) linewidth ($761 \pm 36$~km~s$^{-1}$) from \zgal results \citep{cox+2023}, as the \coonezero line has an extremely broad line profile with large errors in \coonezero FWHM (see Appendix~\ref{app:lineprofiles}).  \\
}

\label{tab:tuner_data}
\end{table}

TUNER has a variety of radiative, physical, and chemical parameters (Table~\ref{tab:tuner_priors}). Among these, we fix the dust opacity coefficient (${\kappa}_{\nu}$) to be 0.047~$\rm m^2~{kg}^{-1}$ at a reference frequency of 352.7~GHz (850~$\mu$m), following \cite{draine2014}. We have also fixed the G/D ratio to 100, i.e., the same as that used in $\S$~\ref{subsec:gasmass_gdmr}. We further tested TUNER LVG models with the G/D ratio left as a free parameter. The resulting fits produced median G/D ratios of $\sim 150$--380, with 9/12 sources favoring values above 200. These solutions imply $\alpha_{\rm CO}$ values of $\sim$7--22, with many sources requiring $\alpha_{\rm CO} \gtrsim 10$. Given that such large G/D ratios are not generally expected for dust-rich, far-infrared- and submillimeter-selected, highly star-forming DSFGs in the early universe \citep[see][]{dunne21,dunne+2022}, we chose to adopt a fixed G/D ratio of 100. 

Our models assume a virial parameter of ${\kappa}_{\rm vir}=0.9$, consistent with gravitationally bound, near-virial molecular clouds typical of star-forming environments. The turbulent velocity width ($\Delta v_{\rm turb}$) is set to $2 \sigma_0\sqrt{2~ln(2)}$, using their velocity dispersion ($\sigma_0$) values derived using the redshift evolution explored by \cite{rizzo2024} (see Table~\ref{tab:dynamicalmasses}). 

Additionally, we treat several ISM properties as free parameters with physical priors as given in Table~\ref{tab:tuner_priors}. These include the molecular gas density ($n_{\rm H_2}$), kinetic and dust temperatures ($T_{\rm kin}$, $T_{\rm dust}$), dust emissivity index ($\beta_{\rm dust}$), abundance $\left[{\rm CO}\right]/\left[{\rm H_2}\right]$, and the kinetic temperature and gas density power-law index ($\rm \gamma_{T}$) with $T_{\rm kin} \propto {\rm log(n_{H_2})}^{\gamma_{T}}$. We set $\rm \gamma_{T}\leq0$ such that denser gas has lower kinetic temperatures \citep[e.g.,][]{glover2007}.

\begin{table}[h]
    \centering
    \caption{Parameters used in the TUNER LVG modeling and their adopted priors.}
    \begin{tabular}{l c c c c}
        \toprule
Category & Parameter & Fixed value & \multicolumn{2}{c}{Explored range}  \\
\cmidrule(lr){4-5}
    &      &   & Minimum    & Maximum  \\ 
\midrule

Physical parameters & Radius [kpc] &... & 0.0001 & 10.0 \\
&log$_{10}$($n_{\rm H_2}$) [${cm}^{-3}$] &... & 1.0 & 7.0\\
&$T_{\rm kin}$ [K] & ...& 2.7 & 1000.0 \\
&$\Delta v_{\rm turb}$ [km~$\rm {s}^{-1}$] & (fixed to $2\sigma_0 \sqrt{2~ln(2)}$) &... &... \\
&$\kappa_{\rm vir}$ & 0.9 & ...& ...\\
&$\gamma_{\rm T}$ &... & -0.3 & -0.001 \\
&$\rm C_{ff}$ & 0.0 & ...&... \\

\midrule

Chemical parameters & log$_{10}$($[\rm {}^{12}CO]/[H_2]$) &... & -5.3 & -3.5 \\
&G/D ratio  & 100 &... &... \\
&$\nu_0$ for $\kappa_d$ & 352.7~GHz (850~$\mu$m) &... &... \\
&$\kappa_d$ [m$^2$ kg$^{-1}$] at $\nu_0$ & 0.047 & ...&... \\

\midrule

Radiation field &$T_{\rm dust}$ [K] &... & 5.0 & 125.0\\
&$\beta_{\rm dust}$ &... & 1.5 & 3.4 \\
&redshift ($z$) & (fixed as per Table~\ref{tab:dynamicalmasses}) & ...&... \\

\midrule

Other priors&$T_{\rm kin}$/$T_{\rm dust}$ &... & 0.5 & 6.5\\
\bottomrule
\end{tabular}

\label{tab:tuner_priors}
\end{table}

We parameterize the total emitting area in the TUNER LVG model through the characteristic radius ($R_{\rm LVG}$), which is equivalent to expressing the total source solid angle as ($\Omega_{\rm total} = \pi R_{\rm LVG}^2$). For the unresolved dust and gas observations used in this modeling, $R_{\rm LVG}$, therefore, represents the effective projected radius of the emitting region and is expected to remain smaller than the observed beam(s). Because TUNER is sensitive only to the total emitting solid angle of each density component, rather than its detailed spatial distribution, $R_{\rm LVG}$ should provide a reasonable estimate of the physical extent of the dominant emitting component when the molecular gas is distributed relatively smoothly. However, if multiple spatially separated compact clumps dominate the observed line luminosity, the inferred $R_{\rm LVG}$ represents only the combined projected area of those clumps and may therefore underestimate the true spatial extent of the molecular gas distribution. Since our targets exhibit relatively compact and smooth dust continuum morphologies, we assume that the molecular gas follows a similarly smooth distribution. Under this assumption, the inferred LVG sizes are expected to provide a reasonable approximation to the physical extent of the molecular gas in these galaxies. 

\begin{figure}[h]
\centering
\includegraphics[width=0.45\textwidth]{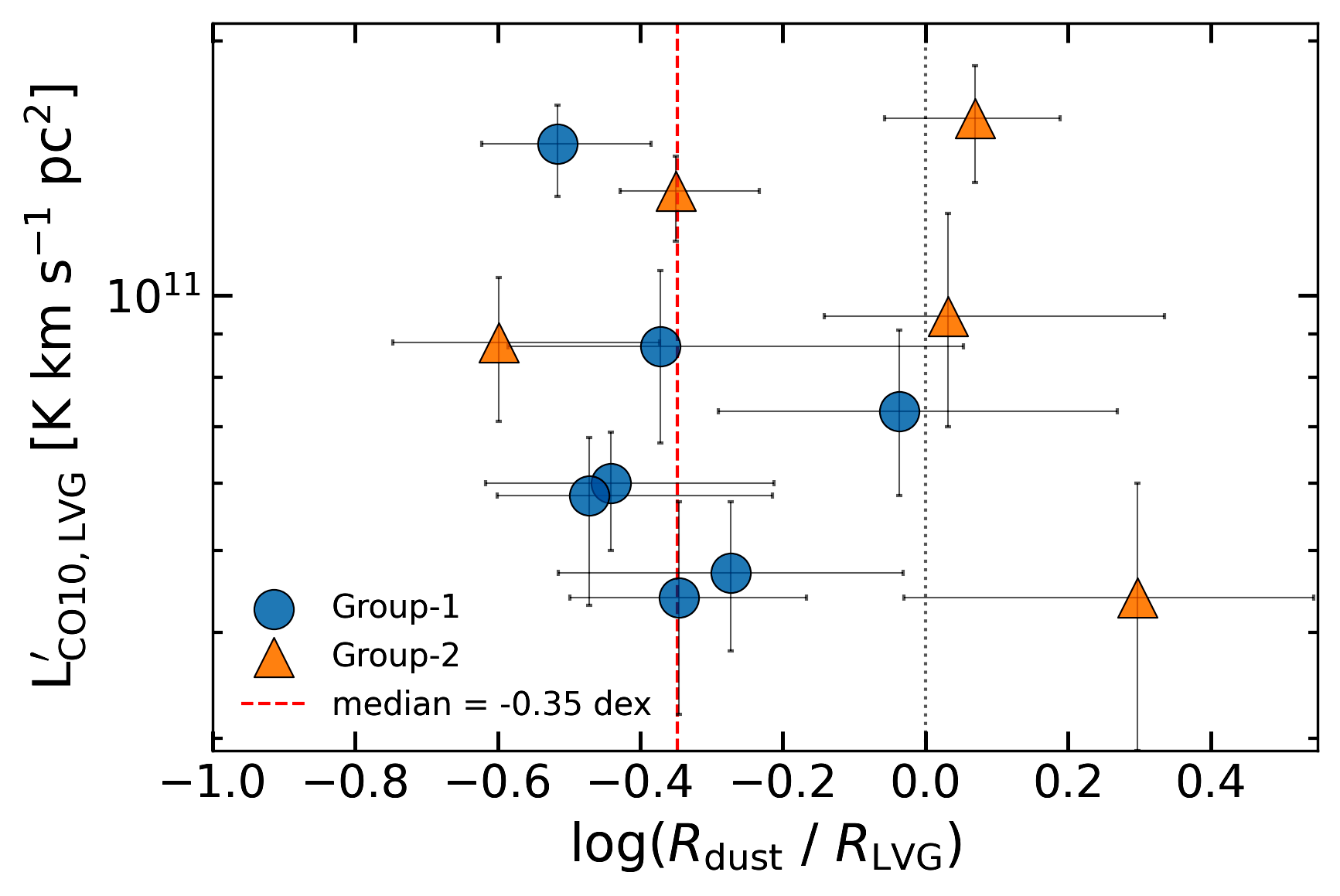}
\includegraphics[width=0.455\textwidth]{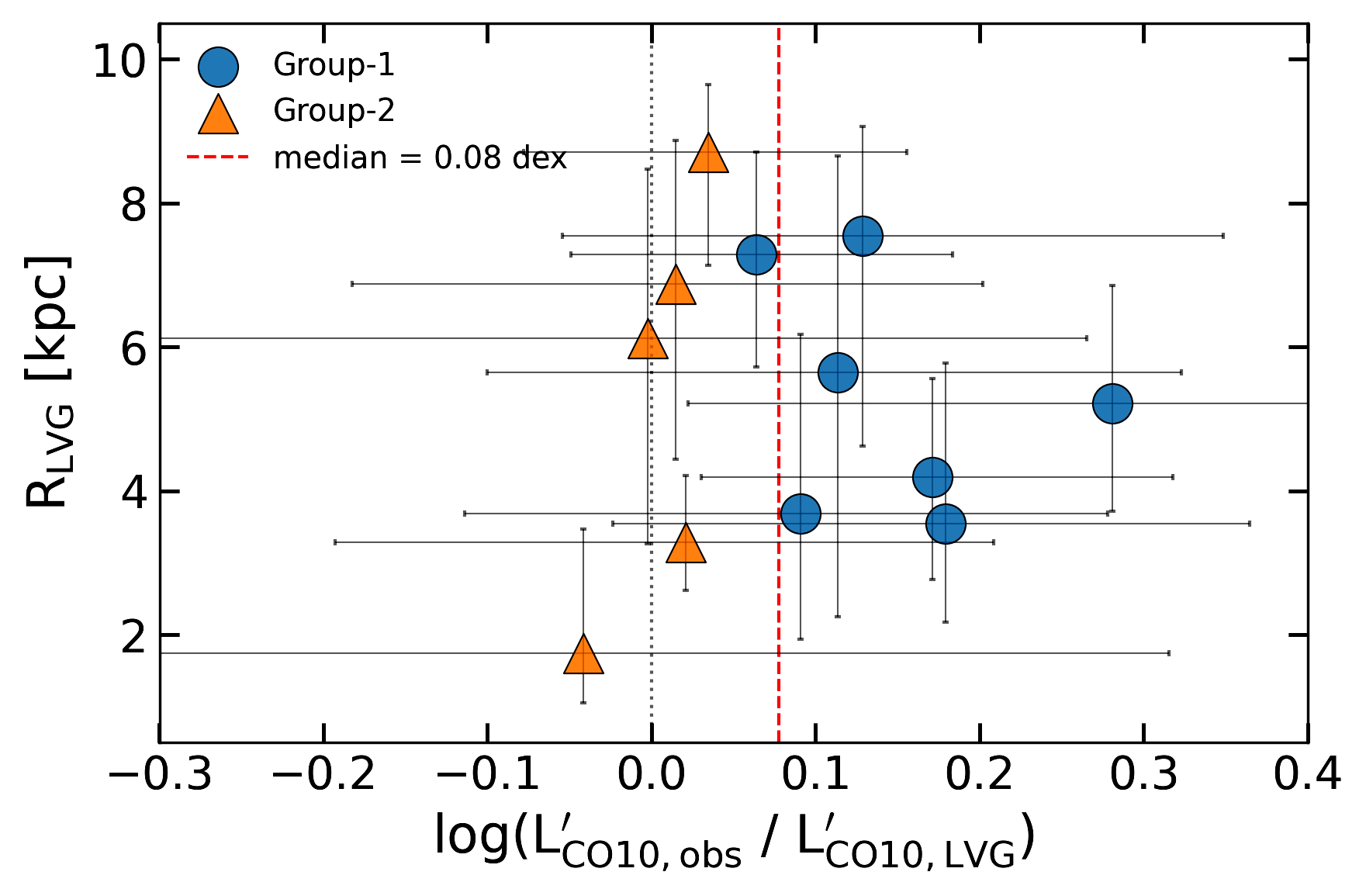}
\caption{(Left panel) Comparison of the ALMA-observed dust sizes to the characteristic radius inferred from the TUNER LVG modeling. As expected due to radiative transfer effects \citep{rivera2018,boogard2026}, the LVG-derived radii are systematically larger than the observed dust sizes in most targets. (Right panel) Observed \coonezero line luminosities compared with those fitted by TUNER. Our best-fit models fit the observed \coonezero line luminosities within the error bars, with a median underestimation of $\sim$0.10~dex. HeLMS-54 is an exception with TUNER underestimating its $L^{\prime}_{\rm CO(1-0),obs}$ by $\sim$0.3~dex. In both panels, blue markers represent Group-1 targets, and orange triangles represent Group-2 sources.}
\label{fig:acoratio_rratio}
\end{figure}

Further, radiative transfer arguments \citep[e.g.,][]{rivera2018,boogard2026} suggest that the dust continuum can have a more compact appearance compared to molecular gas, implying $R_{\rm dust} \lesssim R_{\rm LVG}$. Figure~\ref{fig:acoratio_rratio}~(left panel) shows that most sources are consistent with this expectation. In contrast, two Group-2 DSFGs (HerBS-43a and HerBS-43b) show median LVG-derived radii that are marginally smaller than their observed dust radii, albeit within the uncertainties. 
Figure~\ref{fig:acoratio_rratio} (right panel) demonstrates how TUNER-fits to CO SLEDs of our targets reproduce their observed \coonezero line luminosities. We note HeLMS-54 as an exception; however, it has only galaxy-integrated CO(3--2) besides \coonezero data, suggesting a plausible improvement with more CO line observations. Overall, Group-1 targets are less distributed in both these parameter spaces than the Group-2 DSFGs.

\begin{figure}[h]
\centering
\includegraphics[width=\textwidth]{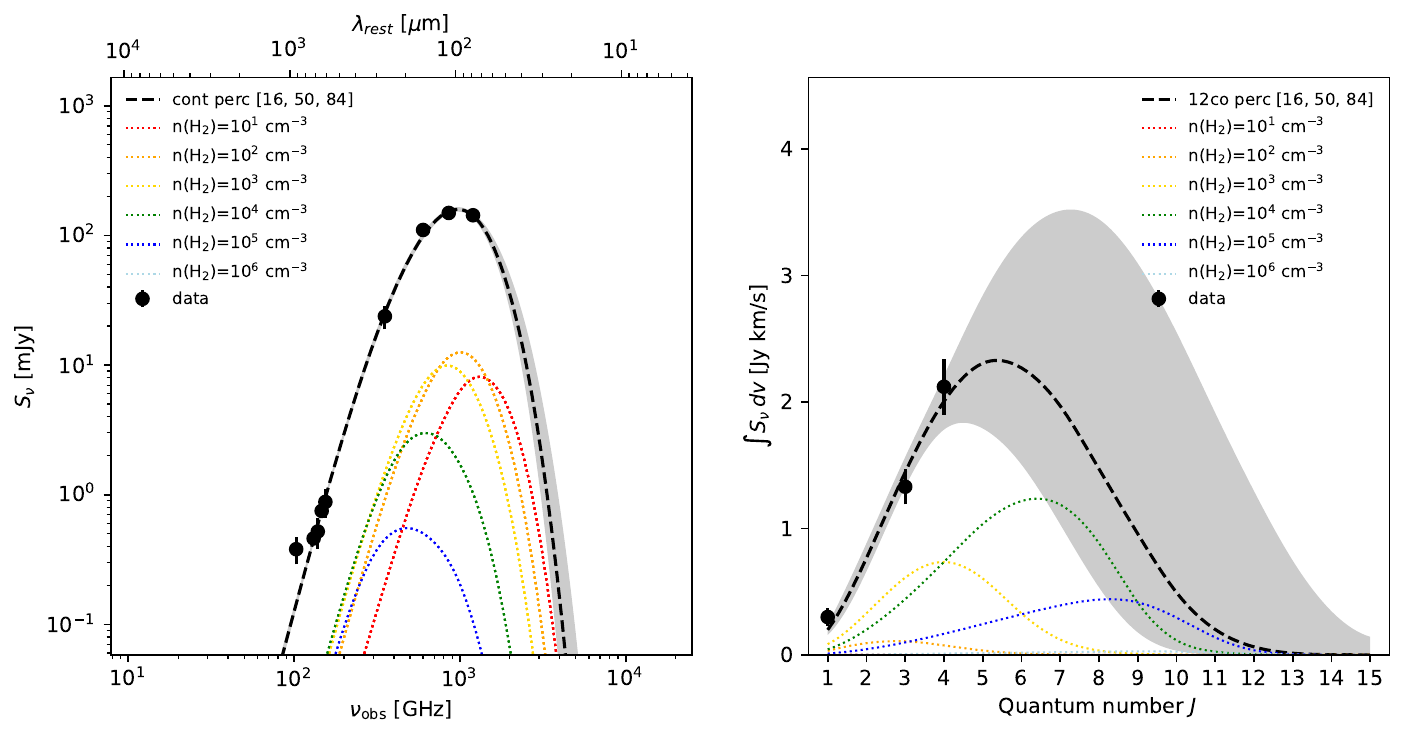}
\includegraphics[width=0.67\textwidth]{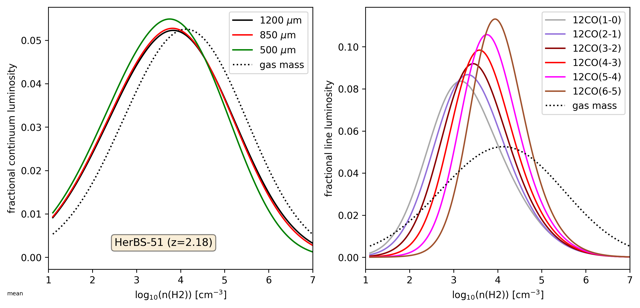}
\includegraphics[width=0.32\textwidth]{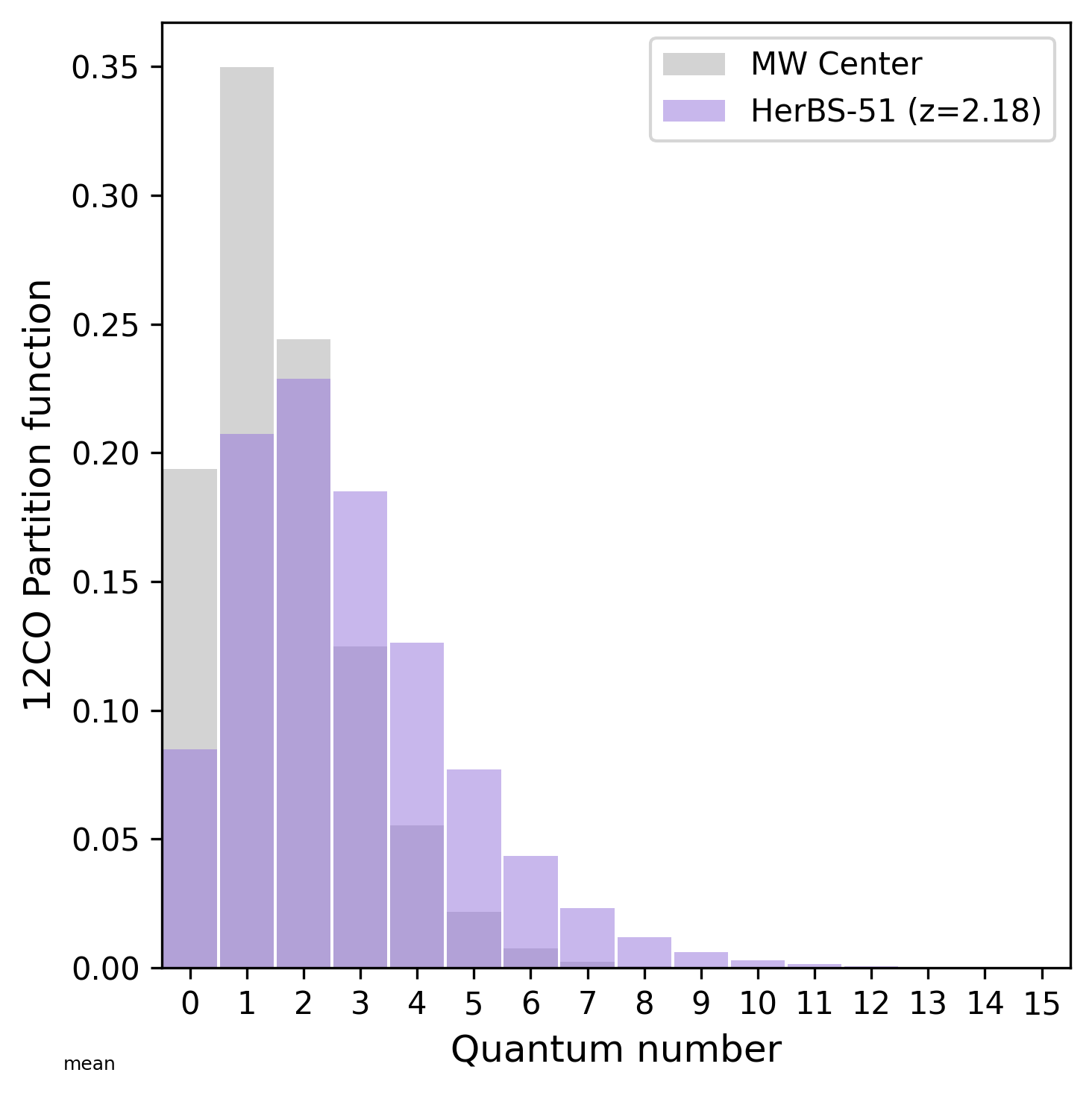}
\caption{HerBS-51 ($z=2.183$) is shown as a representative example of the TUNER LVG modeling results. Upper panel: The black dashed curve represents the median best-fit model to the combined dust and CO SEDs, while the gray shaded region indicates the 1$\sigma$ confidence interval. Black points show the observed fluxes, and the colored dotted curves correspond to individual density components, scaled up for clarity. Comparable fits are obtained for the remaining 11 DSFGs (Appendix $\S$~\ref{herbs165_corner}). Lower panel: (Left, Center) Fractional far-infrared multiwavelength dust luminosities, CO line luminosities, and molecular gas mass distributions as a function of gas density. (Right) A comparison of CO-level populations between HerBS-51 and the Milky Way center.}
\label{fig:tuner_herbs51}
\end{figure}

We show an example of HerBS-51 ($z=2.183$) in Figure~\ref{fig:tuner_herbs51} to illustrate the TUNER LVG results, with the corresponding corner plot shown in Appendix~\ref{herbs165_corner}. Similar to this representative case, the median best-fit dust SEDs for all sources are consistent with the observed far-infrared continuum data. However, 4/12 sources exhibit one of their CO transitions with measured flux lying marginally outside the 1$\sigma$ spread of the median CO SLED model (Appendix~\ref{herbs165_corner}). While TUNER underestimates \coonezero line flux for HeLMS-54 (only 2 CO line-SLED) and HerS-8 (undersampled Rayleigh-Jeans tail in dust SED), CO(4--3) is also underpredicted for HerBS-43b and HerBS-191. In addition, the CO(3--2) flux is marginally overpredicted for HerS-8. Comparing the TUNER-derived gas masses with those inferred from the dust-SED approach ($\S$~\ref{subsec:gasmass_gdmr}) for these objects, however, we find agreement within the quoted uncertainties (Tables~\ref{tab:dustmass_gdmr} and \ref{tab:tunerlvg}) for all sources.

In the entire sample, HerBS-43a is the only source for which the TUNER analysis yields a higher gas mass (${8.44}^{+0.72}_{-1.20}~\times{10}^{11}~{\rm M}_{\odot}$) compared to the dust-based estimate (${4.17}^{+1.12}_{-1.12}~\times{10}^{11}~{\rm M}_{\odot}$). At the same time, the sizes agree within the error bars (see Figure~\ref{fig:acoratio_rratio}). The noted mass difference does not arise from the adopted G/D ratio, which is fixed at 100 in both methods, but instead reflects the different ways in which the gas reservoir is constrained. In the TUNER LVG framework, the molecular gas mass is primarily linked to the CO excitation and the column density required to reproduce the observed CO line luminosities over the spatially resolved emitting region. With the emitting area independently constrained, the CO SLED effectively restricts the allowed range of column densities, leading to a plausibly higher inferred gas mass solution. For example, SLED of HerBS-43a (Appendix~\ref{herbs165_corner}) has a significantly unconstrained peak, allowing denser density components to dominate the CO energy distribution. In contrast, the GMBB approach infers the dust mass directly from the far-infrared continuum without incorporating constraints from molecular line excitation or optical-depth effects in CO. For HerBS-43a, the joint dust and CO fit, therefore, suggests higher column densities, resulting in a molecular gas mass approximately a factor of two higher than that derived from the continuum-only analysis.

Furthermore, TUNER-inferred CO abundances for our sources are consistent with Galactic values within the uncertainties.\footnote{The TUNER models yield $[\rm {}^{12}CO]/[H_2]$ abundance ratios that are generally consistent with near-solar or solar values (Table~\ref{tab:tunerlvg}). These estimates, however, are derived assuming fixed values of the G/D ratio and dust opacity normalization, $\kappa_\nu$ \citep{draine2014}.} Finally, we summarize the LVG-derived median values (50th percentile of the \textit{MCMC} chains) and $1\sigma$ uncertainties (16th-84th percentiles) for the global gas and dust properties of our targets in Table~\ref{tab:tunerlvg}. Among these, the reported characteristic radius of ISM ($R_{\rm LVG}$) serves as a lower limit to the effective extent of the cold molecular gas. The table also contains the fitted \coonezero luminosity, total molecular gas (H$_2$) mass\footnote{including Helium contribution}, and the corresponding \aco conversion factor defined as:

\begin{equation}
    M_{\mathrm{H_2,LVG}} = \alpha_{\rm CO,LVG} \cdot L^{\prime}_{\rm CO(1-0),LVG}
    \label{eqn:acolvg}
\end{equation}

The inferred $\alpha_{\rm CO,LVG}$ values span 2.62--8.63 (median = $4.75 \pm 1.74$) for Group-1 sources and 5.14--11.48 (median = $5.50 \pm 0.12$) for Group-2 sources. {Although the median conversion factors inferred here are marginally higher than those derived in $\S$~\ref{subsec:gasmass_gdmr}, both methods give values consistent within their respective error bars.}

\begin{table}[h]
    \centering
    \caption{Integrated (global) gas and dust properties from the TUNER LVG modeling.}
    \begin{tabular}{l c c c c c c c c c}
        \toprule
Category & Source  & $\rm R_{\mathrm{LVG}}$ &  $^{a}$$\rm {L}^{\prime}_{CO(1-0),LVG}$ & $^{b}$$\rm M_{\mathrm{H_2,LVG}}$ & $^{c}$$\alpha_{\rm CO,LVG}$    & $^{d}$$\rm M_{\rm dust,LVG}$  & $^{d}$$\rm T_{\rm dust,LVG}$ & $\beta_{\rm dust,LVG}$ & $\rm [{}^{12}CO]/[H_2]$  \\ 

&   & [kpc] & [${10}^{11}$ $L_l$]  & [${10}^{11}$ $M_{\odot}$] & [$\alpha_0$]   & [${10}^{9}$ $M_{\odot}$] & [K] &  & \\ \midrule

Group-1 &  HeLMS-39  & $7.28^{+1.43}_{-1.57}$ & $1.51^{+0.17}_{-0.20}$ & $4.00^{+0.69}_{-0.83}$ & $2.62^{+0.50}_{-0.43}$  & $4.00^{+0.68}_{-0.83}$ & $32.40^{+2.04}_{-1.79}$ & $2.94^{+0.22}_{-0.22}$    & $-4.06^{+0.33}_{-0.25}$  \\

&  HeLMS-54  &  $5.21^{+1.64}_{-1.50}$ & $0.44^{+0.13}_{-0.12}$ & $2.09^{+0.77}_{-0.57}$ & $4.75^{+2.18}_{-1.41}$    &   $2.09^{+0.77}_{-0.57}$ & $32.14^{+2.65}_{-2.06}$ & $3.27^{+0.10}_{-0.14}$ & $-4.57^{+0.55}_{-0.39}$   \\

 &  HerS-8  &  $4.19^{+1.37}_{-1.43}$ & $0.60^{+0.09}_{-0.10}$ & $1.59^{+0.68}_{-0.53}$ & $2.65^{+1.09}_{-0.77}$   & $1.59^{+0.68}_{-0.53}$ & $37.34^{+5.13}_{-2.63}$ & $3.23^{+0.12}_{-0.22}$   & $-4.09^{+0.36}_{-0.38}$ \\ 
 
 & HerBS-51  &  $3.54^{+2.23}_{-1.37}$ & $0.47^{+0.10}_{-0.10}$ & $3.01^{+1.22}_{-0.82}$ & $6.49^{+2.11}_{-1.64}$  &   $3.01^{+1.21}_{-0.82}$ & $38.98^{+7.99}_{-5.01}$ & $2.97^{+0.27}_{-0.25}$  & $-4.41^{+0.55}_{-0.61}$ \\
 
 & HerBS-70E  &  $3.68^{+2.49}_{-1.74}$ & $0.73^{+0.18}_{-0.15}$ & $2.59^{+0.95}_{-0.71}$ & $3.52^{+1.10}_{-0.76}$  &  $2.59^{+0.95}_{-0.70}$ & $37.40^{+7.98}_{-3.86}$ & $2.82^{+0.25}_{-0.25}$   & $-4.25^{+0.43}_{-0.42}$  \\ 
 
& HerBS-165  &  $7.54^{+1.52}_{-2.92}$ & $0.58^{+0.10}_{-0.15}$ & $4.99^{+1.06}_{-1.41}$ & $8.63^{+2.10}_{-1.76}$  &   $4.99^{+1.06}_{-1.41}$ & $28.64^{+4.02}_{-2.25}$ & $2.94^{+0.26}_{-0.26}$ & $-4.50^{+0.62}_{-0.51}$  \\

& HerBS-169 &  $5.64^{+3.02}_{-3.40}$ & $0.87^{+0.20}_{-0.21}$ & $5.10^{+1.24}_{-1.86}$ & $5.59^{+1.70}_{-1.40}$  &   $5.09^{+1.24}_{-1.86}$ & $37.52^{+12.63}_{-4.81}$ & $2.73^{+0.37}_{-0.39}$   & $-4.35^{+0.53}_{-0.49}$   \\

\midrule

Group-2 & HeLMS-50  & $8.70^{+0.94}_{-1.57}$ & $1.33^{+0.14}_{-0.17}$ & $7.45^{+1.02}_{-1.29}$ & $5.50^{+1.17}_{-0.85}$  &    $7.45^{+1.02}_{-1.29}$ & $26.96^{+2.64}_{-1.38}$ & $3.06^{+0.17}_{-0.36}$   & $-4.42^{+0.45}_{-0.36}$ \\

& HerBS-43a & $3.28^{+0.92}_{-0.68}$ & $1.62^{+0.26}_{-0.26}$ & $8.44^{+0.72}_{-1.20}$ & $5.15^{+0.86}_{-0.93}$  &  $8.44^{+0.72}_{-1.20}$ & $44.87^{+6.40}_{-4.60}$ & $3.01^{+0.27}_{-0.38}$  & $-3.89^{+0.28}_{-0.36}$   \\ 

& HerBS-43b  &   $1.74^{+1.73}_{-0.69}$ & $0.44^{+0.16}_{-0.15}$ & $5.18^{+1.39}_{-2.00}$ & $11.48^{+6.26}_{-4.24}$  & $5.18^{+1.39}_{-2.00}$ & $53.51^{+19.76}_{-10.72}$ & $2.66^{+0.45}_{-0.48}$ & $-4.52^{+0.68}_{-0.55}$ \\ 

& HerBS-188 &  $6.88^{+2.00}_{-2.44}$ & $0.88^{+0.17}_{-0.18}$ & $4.76^{+1.67}_{-0.96}$ & $5.61^{+1.90}_{-1.41}$ &  $4.75^{+1.67}_{-0.96}$ & $39.76^{+3.33}_{-2.86}$ & $2.14^{+0.27}_{-0.22}$    & $-4.73^{+0.65}_{-0.31}$  \\ 
 
& HerBS-191  & $8.09^{+1.41}_{-2.37}$ & $1.48^{+0.32}_{-0.41}$ & $6.03^{+1.58}_{-1.35}$ & $4.19^{+1.40}_{-0.85}$  &  $6.02^{+1.58}_{-1.35}$ & $39.94^{+3.78}_{-3.22}$ & $2.17^{+0.29}_{-0.26}$    & $-4.36^{+0.49}_{-0.41}$  \\

\bottomrule
\end{tabular}

\parbox{\textwidth}{\footnotesize
\textbf{Notes.} \\
$^{a}$\coonezero line luminosity from the CO SLED fitting provided by the LVG model. Here, ${L}_{l}$=$\rm K~km~{s}^{-1}~{pc}^{2}$. \\
$^{b}$LVG-derived gas mass, $\rm M_{\mathrm{H_2,LVG}}$, includes Helium contribution. \\
$^{c}$$M_{\mathrm{H_2,LVG}} = \alpha_{\rm CO,LVG} \cdot L^{\prime}_{\rm CO(1-0),LVG}$. The LVG-derived CO--H$_2$ conversion factor given here is in the units of ${\alpha}_{0}$=$\rm M_{\odot}~{(K~km~{s}^{-1}~{pc}^{2})}^{-1}$. \\
$^{d}$Dust masses and temperatures derived from the TUNER LVG modeling, which simultaneously fits the dust (GMBB) and CO SEDs assuming a fixed G/D ratio, $M_{\rm dust,LVG} \, / \, M_{\mathrm{H_2,LVG}} = 100$.  \\
}

\label{tab:tunerlvg}
\end{table}

\subsubsection{Comparison of Gas Mass Estimates} \label{subsec:gasmass_comparison}
Our targets have median \coonezero line luminosities of (0.89 $\pm$ 0.11) $\rm \times {10}^{11}$ and (0.94 $\pm$ 0.50) $\rm \times {10}^{11} K~km~{s}^{-1}~{pc}^{2}$ observed for Group-1 and Group-2 DSFGs, respectively. Under the traditional bimodal framework, \aco= 0.8 is commonly adopted for local ULIRGs and high-\textit{z} DSFGs, SMGs, and quasars \citep[e.g.,][]{downes_solomon_1998,carilliwalter2013,bolatto2013}, whereas \aco= 4.3 --- comparable to the Milky Way value\footnote{also with Helium contribution} --- is typically assumed for local star-forming main-sequence galaxies. Applying \aco= 0.8 to our sample yields median molecular gas masses of $0.7 \times {10}^{11}$~${\rm M}_{\odot}$ and $0.8 \times {10}^{11}$~${\rm M}_{\odot}$ for Groups-1 and -2, respectively. In contrast, adopting \aco= 4.3 increases these estimates by a factor of $\sim 5$, giving median gas masses of $3.8 \times {10}^{11}$~${\rm M}_{\odot}$ and $4.0 \times {10}^{11}$~${\rm M}_{\odot}$.

Independently, the two gas-mass estimation methods described in $\S$~\ref{subsec:gasmass_gdmr} and $\S$~\ref{subsec:gasmass_tuner} produce mutually consistent results for each source, except HerBS-43a, within their uncertainties. With the dust continuum-based method, we find median gas masses of (3.24 $\pm$ 0.15)$\times {10}^{11}$~${\rm M}_{\odot}$ and (3.98 $\pm$ 0.59)$\times {10}^{11}$~${\rm M}_{\odot}$ for Groups-1 and -2, respectively. These are broadly in agreement with those derived using the TUNER LVG method: (3.01 $\pm$ 0.99)$\times {10}^{11}$~${\rm M}_{\odot}$ and (5.18 $\pm$ 0.42)$\times {10}^{11}$~${\rm M}_{\odot}$ for Group-1 and Group-2.

Results from the dust continuum-based method have only 2/12 DSFGs with inferred \aco values below 2, namely HeLMS-39 ($\alpha_{\rm CO,G/D}={1.85}^{+0.31}_{-0.23}$) and HerS-8 ($\alpha_{\rm CO,G/D}={1.45}^{+0.29}_{-0.21}$) from Group-1. In this method, we do not find any object with \aco $\sim$ 0.8, which is typically assumed in the literature for local ULIRGs and high-\textit{z} SMGs/DSFGs. Similarly, none of the galaxies in Groups-1 and -2 show $\alpha_{\rm CO,LVG}\lesssim 2$ using the TUNER, suggesting that the latest LVG model does not support \aco $\sim$ 0.8 on galaxy-integrated scales for our sample of high-\textit{z} DSFGs\footnote{Similar trend has also been seen for high-\textit{z} dust-obscured galaxies \citep[e.g., W2246-0526;][]{harrington2025} and quasars (e.g., Cloverleaf; Prajapati et al., in prep.).}, unlike the inference derived for local ULIRGs by \citet{downes_solomon_1998}.  

Our derived gas masses in $\S$~\ref{subsec:gasmass_gdmr} and \ref{subsec:gasmass_tuner} are broadly consistent with each other. The median values lie significantly closer to those inferred using $\alpha_{\rm CO}=4.3$ than to those obtained with $\alpha_{\rm CO}=0.8$, while the overall spread in \aco ($\sim$~1.5--11.5) is similarly recovered across both approaches. Together, these results statistically favor a Milky Way-like CO--H$_2$ conversion factor for the DSFG sample in this work. 

Understanding such higher than \aco= 0.8 values in the context of the TUNER LVG radiative transfer model with CO excitation is important. \citet{papadopoulos2012b} analyzed a large sample of luminous infrared galaxies (LIRGs) and found that single-phase radiative transfer models reproduce the low-\textit{J} CO emission with a typical $\alpha_{\rm CO} \sim 0.6$, consistent with the canonical ULIRG value. They argued that the low \aco is driven primarily by high gas velocity dispersion, with gas temperature playing a secondary role. However, they proposed that a cold, dense, gravitationally-bound gas phase, traced by high-\textit{J} CO and dense-gas tracers (e.g., HCN, HCO$^+$), could contain a substantial fraction of the molecular mass and may increase \aco to 2--6. This highlighted the possibility that unresolved starbursts contain multiple gas density components, making a single global \aco an oversimplification. 

The TUNER modeling used in our work accounts for such a turbulent gas density distribution (see Figure~\ref{fig:tuner_herbs51}). The lower-left and lower-center panels of Figure~\ref{fig:tuner_herbs51} for HerBS-51 show that although no single CO transition or dust continuum measurement can recover the full molecular gas reservoir in high-\textit{z} DSFGs, the gas mass distribution indeed peaks around densities that are better probed by mid-\textit{J} CO transitions ($J\sim4-7$) than by low-\textit{J} CO lines. This is also in agreement with their CO level population showing an extended tail towards mid/high-\textit{J} rotational levels (Figure~\ref{fig:tuner_herbs51} lower-right panel).\footnote{Such shifts toward higher rotational states are expected in high-\textit{z} DSFGs owing to their elevated gas excitation and, potentially, the influence of increasing CMB temperature with redshift \citep[see][]{harrington2025,prajapati2026,cescon2026}.} In Appendix~\ref{herbs165_corner}, the remaining 11 DSFGs from our sample also show a similar trend. This highlights that \coonezero and other low-\textit{J} lines are useful for tracing more diffuse ISM, while dense gas tracers remain important for revealing a significant fraction of the total gas mass in such high-\textit{z} objects with high star formation rates. 
Most of our targets have at least 3 CO lines observed up to mid-\textit{J} levels (see Table~\ref{tab:tuner_data}), showing consistency between observed and modeled CO luminosities. In the future, observing galaxy-integrated $J>6$ CO transitions for our sample will be of further use to test the presented TUNER results. This is a limitation based on the currently available data, which will be addressed in future work.

\section{Dynamical Tests of the Molecular Gas-Mass Estimates}  \label{sec:mdyn}

\coonezero emission is easily excited across a broad range of physical conditions, from cold and diffuse molecular gas to warmer and denser regions, making it the best-calibrated tracer of the total molecular gas reservoir in the local universe \citep[e.g.,][]{Saintonge+2017}. For this reason, \coonezero is commonly used in integrated dynamical mass estimates, as both its linewidth and spatial extent are expected to trace the bulk of the gravitationally bound molecular medium. In contrast, higher-$J$ CO transitions and dust continuum emission preferentially probe warmer and denser gas and might appear more compact in DSFGs \citep{rivera2018,boogard2026}. As a result, these tracers may not fully sample the system's spatial extent, potentially biasing dynamical mass estimates. The \coonezero linewidth and radius, therefore, may provide a more representative characterization of the galaxy-scale gas dynamics via dynamical mass ($M_{\rm dyn}$) calculations. Central to the ``\aco tension" is a discrepancy between dynamical masses and molecular gas masses. In this section, we examine whether or not the molecular gas masses inferred in $\S$~\ref{subsec:gasmasses} are consistent with the available dynamical mass budget --- inferred from global \coonezero line properties --- in high-\textit{z} DSFGs. Specifically, we test whether the condition $M_{\rm gas} \leq M_{\rm dyn}$ is satisfied, as expected for systems that are gravitationally bound.\footnote{We use the inequality sign to reflect the lack of independent constraints on stellar masses and dark matter contributions in our dust-obscured, high-\textit{z} sources. In this comparison, we use only the molecular gas and dynamical masses and assume the dust mass contribution is negligible.}

\subsection{Dynamical Masses from the Isotropic Virial Estimator} \label{subsec:mdynisovir}

Assuming the cold molecular gas is gravitationally supported, a first-order consistency check on the gas masses can be obtained using an isotropic virial estimator for the dynamical mass (Eqn.~\ref{eqn:virialestimator}). This approach assumes a spherically symmetric gas distribution and characterizes the kinematics via a one-dimensional velocity dispersion, $\sigma_0={\rm FWHM}/2.35$ \citep[e.g.,][]{tacconi+2008,natascha2009,bothwell2013}\footnote{We note that because of the unresolved \coonezero FWHM used here for demonstrating a single pressure-support term, we cannot distinguish between the individual contributions from rotation, turbulence, mergers/accretion/outflows, and plausible beam-smeared velocity gradients. However, this provides a robust first test of whether or not the high-\textit{z} DSFGs have an obviously unbound nature with $M_{\rm gas} > M_{\rm dyn,vir}$.}. It is commonly applied to SMGs and DSFGs in the early universe, using an effective radius, $R_{\rm eff}$ [kpc].

\begin{equation}
    M_{\rm dyn,vir} \, \, [\rm M_{\odot}] \, \, = \, \, 6.7  \, {\sigma_0}^{2} \, \, R_{\rm eff}/G \, \, = \, \, 2.82 \, \times \, {10}^{5} \, \, {\rm (FWHM)}^{2} \, \, R_{\rm eff}; 
    \label{eqn:virialestimator}
\end{equation}

where G is the gravitational constant in the units of [kpc~km$^2$~s$^{-2}$~$\rm {M_{\odot}^{-1}}$]. Based on \cite{binney2008}, \cite{natascha2009} argue that the factor 6.7 is more appropriate for a variety of galactic mass distributions observed in star-forming galaxies at cosmic noon. Using this ``virial estimator" with $R_{\rm eff}=R_{\rm dust}$ and \coonezero FWHM for our sources, the derived median dynamical masses are (2.02 $\pm$ 0.09)$\times {10}^{11}$~${\rm M}_{\odot}$ and (10.7 $\pm$ 0.54)$\times {10}^{11}$~${\rm M}_{\odot}$ for Group-1 and Group-2 DSFGs, respectively. Group-2 sources not only show a $5\times$ higher median $M_{\rm dyn,vir}$ than that of Group-1, but also a $\sim 6 \times$ broader distribution. This larger spread is primarily driven by the wider ranges of dust sizes and \coonezero linewidths observed in Group-2 (Table~\ref{tab:dynamicalmasses}). Such behavior may reflect significant diversity in the underlying kinematics and/or geometries of the Group-2 systems.

\begin{figure}[h]
\centering
\includegraphics[width=\textwidth]{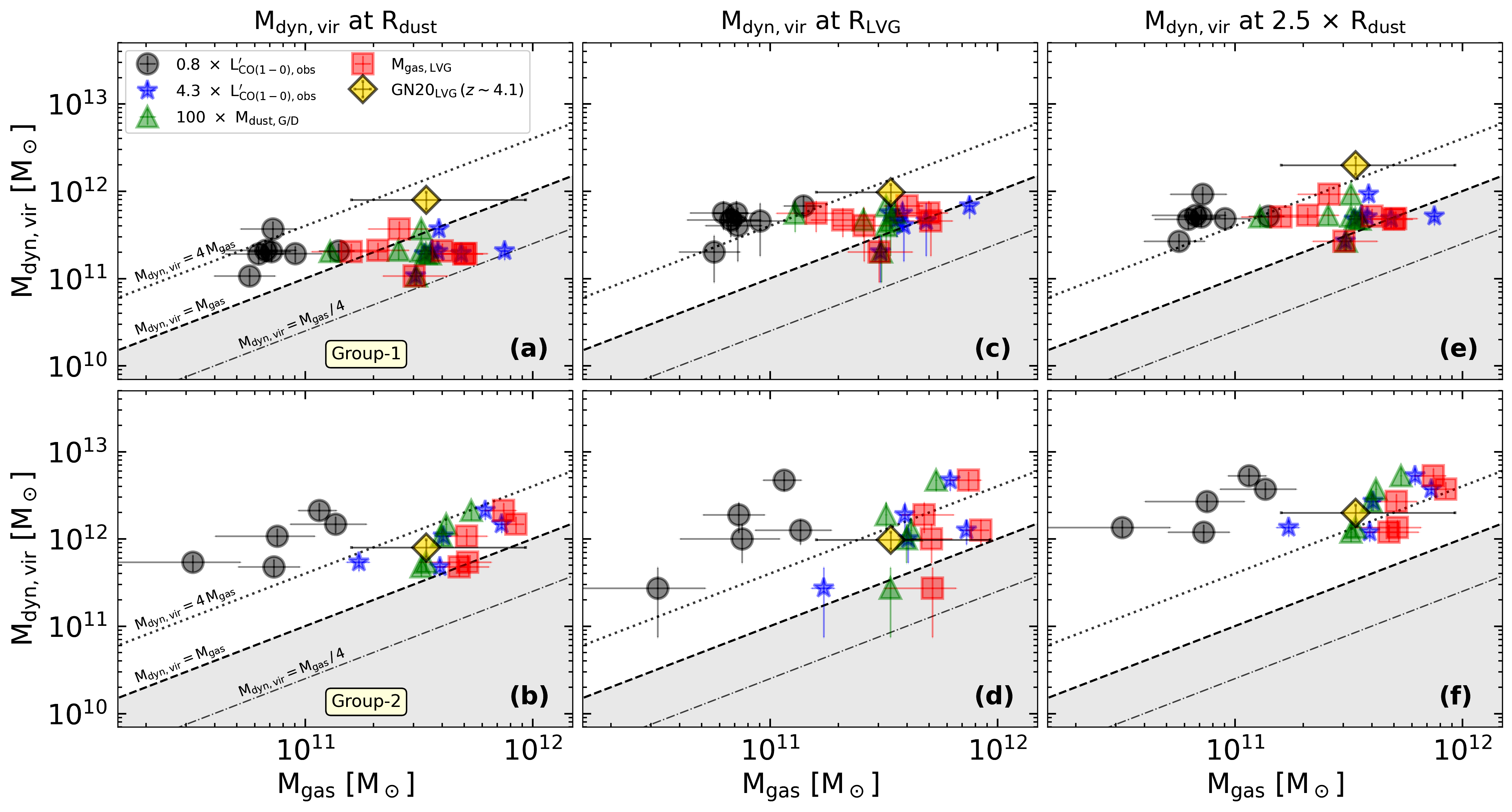}
\caption{Comparison between dynamical masses and gas masses for the explored 12 sources. The upper panel and lower panel are for Group-1 and Group-2 targets, respectively. We explore three different radii to derive dynamical masses, namely dust size ($R_{\rm dust}$, panels-(a,b)), LVG radius ($R_{\rm LVG}$, panels-(c,d)), and a fiducial cold gas disk of 2.5$\times$ the dust size (panels-(e,f)). Different colored symbols represent various gas mass derivation methods, including \aco= 0.8- and \aco= 4.3-scaled \coonezero line luminosities (black and blue), and two methods discussed in $\S$~\ref{subsec:gasmass_gdmr} (green) and $\S$~\ref{subsec:gasmass_tuner} (red). Using diagonal black lines, we show three different relations between the dynamical and gas masses, and also shade the physically forbidden regime ($M_{\rm gas} > M_{\rm dyn,vir}$) in gray. All the plots here use the isotropic ``virial estimator" (Eqn.~\ref{eqn:virialestimator}) to derive dynamical masses. We also show GN20 \citep{boogard2026} as an example compared to our Group-1 and -2 sources.}
\label{fig:virial_mdyn}
\end{figure}

In Figure~\ref{fig:virial_mdyn}(a,b), we compare these isotropic dynamical masses derived from dust-based sizes with molecular gas masses estimated using the different methods described in $\S$~\ref{subsec:gasmass_comparison}. For the Group-1 sources, the inferred gas masses systematically exceed the dynamical masses when adopting either of our two gas-mass derivation approaches: the dust continuum-based method (method-1; $\S$~\ref{subsec:gasmass_gdmr}) and the TUNER LVG-based method (method-2; $\S$~\ref{subsec:gasmass_tuner}). This discrepancy persists even under a Galactic \aco= 4.3, whereas \aco= 0.8 would largely alleviate the tension.

\citet{rivera2018} found that the cold molecular gas in high-\textit{z} SMGs/DSFGs is typically at least twice as extended as the rest-frame 250~$\mu$m dust continuum emission, owing to radiative transfer effects on galaxy-integrated scales. Similar gas-to-dust size offsets have been observed in GN20 \citep[$z\sim4.1$;][]{boogard2026}, and several [CII] 158~$\mu$m studies; e.g., CRISTAL survey of $z\sim5$ main-sequence galaxies \citep{ikeda2025} and REBEL-25 \citep[$z\sim7$;][]{rowland2024}. We therefore explore the consequences of adopting larger effective radii, specifically (i) the LVG-based radii derived using TUNER (Figure~\ref{fig:virial_mdyn}c,\ref{fig:virial_mdyn}d) and (ii) a fiducial radius\footnote{The \coonezero observations from the VLA in \vzgal have typical beam sizes of $\sim$2--3$^{\prime\prime}$, corresponding to spatial scales of $\sim$17--25~kpc ($z\sim2.3$), or radii of $\sim$9--13~kpc, encompassing the unresolved \coonezero emission. Assuming the CO emission is 2.5 times more extended than the dust continuum, the median dust radius of 2.2~kpc (Figure~\ref{fig:rdust_inclination}) implies a fiducial CO radius of $\sim$11~kpc, comparable to the VLA beam; i.e., beam-filling factor close to unity for cold gas.} of $2.5\times R_{\rm dust}$ (Figure~\ref{fig:virial_mdyn}e,\ref{fig:virial_mdyn}f). The median LVG-derived radii are $\sim 2\times$ larger than the dust radii, increasing from $2.4 \pm 0.2$ kpc to $5.2 \pm 1.5$ kpc for Group-1 sources and from $3.9 \pm 0.4$ kpc to $6.1 \pm 2.6$ kpc for Group-2 sources.

For both radius choices, each of which exceeds the observed dust sizes, nearly all targets in Groups-1 and -2 lie within the physically allowed regime, $M_{\rm gas} \leq M_{\rm dyn,vir}$, independent of the gas mass estimation method. HerBS-51 from Group-1 is an exception, with marginally higher LVG-derived gas masses than the dynamical limits. On the other hand, HerBS-43b from Group-2 also shows higher gas masses when compared to $M_{\rm dyn,vir}$ using LVG sizes, which can be attributed to its smaller LVG-derived size than the observed dust radius. Given that the presented LVG modeling is physically motivated, which will have improvements when new data of mid/high-\textit{J} CO transitions and far-infrared continuum are added, we argue that the LVG-based gas masses ($M_{\rm gas,LVG}$) --- mutually consistent with dust continuum-derived gas masses ($M_{\rm gas,G/D}$) --- are overall consistent with the dynamical masses of the sources as expected from a simple isotropic virial estimator. 

In summary, Figure~\ref{fig:virial_mdyn} shows that the gas masses derived using both methods-1 and -2 can be reconciled with the dynamical mass constraints once realistic estimates of the cold molecular gas extent are adopted. Our TUNER modeling further reveals substantial source-to-source variation in the inferred gas masses, reflecting a variety in intrinsic ISM conditions and the galaxies' excitation properties. This is evident from the considerably larger spread in the LVG-based gas masses relative to those obtained from method-1. 

Together, these results argue against the assumption of a universal CO-to-H$_2$ conversion factor on galaxy-integrated scales for our sample of high-\textit{z} DSFGs. Instead, the appropriate value of \aco appears to vary from system to system according to the physical state and excitation conditions of the molecular gas. Consequently, our analysis does not require adoption of the canonical ULIRG value of \aco = 0.8 to reconcile the inferred gas masses with the available dynamical mass constraints (see also $\S$~\ref{subsec:gasmass_comparison}), highlighting the dynamical viability of intermediate (e.g., \aco= 2--3) or near-Galactic conversion factors. Moreover, if applied indiscriminately, \aco = 0.8 may systematically underestimate the molecular gas content of many high-\textit{z} dusty galaxies -- see also a recent theoretical work by \citet{anirudh2025}. We therefore suggest constraining \aco on a source-by-source basis wherever possible, using detailed radiative-transfer modeling in combination with complementary observational/dynamical diagnostics.

\subsection{Dynamical Masses from Rotation and Pressure Support} \label{subsec:mdynmixed}

When the underlying geometry and kinematics of a galaxy are unknown, the isotropic virial estimator provides a useful first-order approach for assessing the dynamical masses of unresolved systems. This is particularly relevant for many dusty galaxies in the early universe, whose dynamical structures are often poorly constrained. However, the isotropic virial estimator effectively averages over the source geometry and folds together the contributions of ordered rotation, turbulence, accretion, and other dynamical processes into the observed \coonezero linewidths. 

With time, the growing capabilities of modern radio and submillimeter interferometers, including ALMA, NOEMA, and the VLA, have enabled an increasing number of high-\textit{z} systems to be spatially resolved. These studies reveal a diverse range of dynamical states and corresponding geometries. Some of the most luminous dusty galaxies appear to be associated with major merger events \citep[e.g.,][]{tacconi+2008,riechers+2017}. In contrast, several others, including both gravitationally lensed and unlensed sources, show ordered rotation indicative of cold molecular gas disks \citep[e.g.,][]{swinbank2015,rivera2018,sharon2019}. The distinction, however, is not always clear, except in cases of ongoing mergers where multiple components are both spatially and kinematically distinct and not aligned along the same line of sight. High-\textit{z} DSFGs, therefore, appear to comprise a heterogeneous population, spanning coalescing systems, late-stage mergers transitioning toward more disk-like configurations, and galaxies with well-ordered rotating disks. For example, GN20 ($z\sim 4.1$) exhibits a regularly rotating \cotwoone disk \citep{hodge2012}, although \cite{boogard2026} suggest it may have experienced an interaction in the recent past.\footnote{This diversity also raises the possibility that dynamically disturbed systems in the early universe can settle into disk-like configurations on relatively short timescales.}

Therefore, in this section, we propose to use a ``mixed estimator" of dynamical masses for high-\textit{z} dusty galaxies (Eqn.~\ref{eqn:Mdyn}). This approach is consistent with the studies of star-forming main-sequence disks and starburst galaxies up to $z\sim4$ \citep[e.g.,][]{genzel2006,natascha2009,rizzo2024} using various components of the ISM\footnote{These studies span from stellar and ionized gas tracers to cold gas resolved data using [CI], [CII], and mid-\textit{J} CO transitions \citep[see also][]{rizzo2023}.}, which often model the resolved kinematics with both rotational and velocity-dispersion contribution.

\begin{equation}
    M_{\rm dyn} = \left[\frac{{{v}^{2}_{\rm rot}}+{k {\sigma}^{2}_{0}}}{G} \right] \cdot R_{\rm eff}
     \label{eqn:Mdyn}
\end{equation}

We fix the coefficient $k=3$ for all our targets --- assuming rotating (thick) disks with significant pressure support --- while acknowledging that this choice is not empirically constrained. $G$ denotes the gravitational constant (in units of kpc~${\rm km}^2$~$\rm s^{-2}~M_{\odot}^{-1}$) and $\sigma_0$ is the velocity dispersion. The dispersion of cold molecular gas spanning $\sim$30--40~km~$\rm {s}^{-1}$ with a typical 10\% uncertainty is assumed to evolve with redshift (see Table~\ref{tab:dynamicalmasses}) as suggested by \cite{rizzo2024}. As expected, these values are significantly lower than those observed for ionized/warm gas traced by H$\alpha$ \citep[e.g.,][]{wisnioski2015,ubler2019}. The current data do not allow us to separate the rotational velocity ($v_{\rm rot}$) from the velocity dispersion using the integrated \coonezero FWHM of our targets. Thus, we assume $v_{\rm rot}$ similar to that used in a pure rotational estimator for DSFGs\footnote{Rotating disk estimator: $M_{\rm dyn,rot} \, \, {\rm sin}^{2} \, i \, \, [\rm M_{\odot}] = 4 \, \times \, {10}^{4} \, \, {(FWHM)}^{2} \, \, R_{\rm eff}$.} \citep{neri2003,bothwell2013}. We reiterate that the integrated FWHM from unresolved CO data in DSFGs may not have only the rotational contribution. So, this is just an assumption.

\begin{equation}
    v_{\rm rot} = 0.5 \left[\frac{\rm {FWHM}}{{\rm sin}(i)}\right]
     \label{eqn:vrot}
\end{equation}

In $\S$~\ref{subsec:mdynisovir}, we showed that dynamical masses derived using the isotropic virial estimator are generally consistent with the molecular gas masses obtained from methods-1 and -2, provided that realistic cold-gas extents are assumed. We therefore compare the mixed estimator (Eqn.~\ref{eqn:Mdyn}) to the isotropic virial estimator for a representative range of observed properties. Assuming identical sizes, a median velocity dispersion of $\sim34$~km~s$^{-1}$ (Table~\ref{tab:dynamicalmasses}), and characteristic CO(1--0) linewidths of $\sim500$~km~s$^{-1}$ (Group-1) and $\sim1000$~km~s$^{-1}$ (Group-2), we find that matching the dynamical masses derived from the mixed and isotropic virial estimators requires inclination angles of only $\sim26^\circ$--$28^\circ$.\footnote{Similarly, approaches discussed by \citet{bothwell2013}, namely virial and rotating disk estimators, also require an inclination of $\sim22^\circ$ to be mutually compared.} Varying the pressure-support coefficient over the range $k=1$--10 yields a similar result, with only a weak dependence on the adopted linewidth. Such systematically low inclination angles are unlikely to be ubiquitous among DSFGs, suggesting that generic assumptions may not adequately capture the diversity of their underlying dynamical structures.

This outcome highlights a key limitation of applying Eqn.~\ref{eqn:Mdyn} to unresolved observations, despite its greater flexibility in accounting for turbulent and dispersion-supported gas motions compared to a pure rotational disk estimator. The observed CO FWHM reflects a combination of ordered rotation, random motions, and potentially other sources of kinematic broadening, whereas unresolved data for most high-\textit{z} DSFGs do not allow these contributions to be disentangled. Consequently, the same integrated linewidth can arise from different combinations of $v_{\rm rot}$ and $\sigma_0$, leading to substantially different inferred dynamical masses. In addition, DSFGs are not spherically symmetric; the observed linewidth therefore depends on viewing angle, such that galaxies with identical intrinsic kinematics but different inclinations can exhibit different FWHM values.\footnote{Additional velocity components, such as inflows, outflows, or merger-induced motions, may further contribute to the observed linewidth.}

The requirement for nearly face-on inclinations when equating the two estimators thus reflects the fundamentally different assumptions underlying an arguably more physically motivated pressure-supported rotating disk geometry and a spherically symmetric virialized system. The latter does not account for geometric and inclination-dependent effects that are explicitly relevant in disk-like systems. As a result, unresolved CO linewidth measurements alone do not provide sufficient information to uniquely constrain both the rotational velocity ($v_{\rm rot}$) and intrinsic velocity dispersion ($\sigma_0$). We therefore rely on the apparent consistency between gas and dynamical masses inferred from the isotropic virial estimator, given that it remains the most practical first-order method for unresolved DSFG populations. Nonetheless, we argue that the mixed estimator (Eqn.~\ref{eqn:Mdyn}) is likely to provide a more realistic representation of DSFG dynamics once independent constraints on the underlying geometry and kinematics are obtained from resolved observations and detailed dynamical modeling \citep[e.g., with \texttt{GalPaK$^{\rm 3D}$, {$^{3{\rm D}}$BAROLO}, DysmalPy$^{\rm 3D}$};][]{bouch2015galpak3d,teodoro20153dbarolo,lee2025dysmalpy}.

\subsection{GN20, an Archetype Dusty Galaxy} \label{subsec:gn20}

GN20, at a redshift of $z=4.055$, is an unlensed dusty galaxy in the GOODS-North field. Its integrated CO emission was first reported by \cite{carilli10}, while subsequent \cotwoone observations spatially resolved its cold molecular gas disk (size: $7 \pm 2$~kpc) and kinematics \citep{hodge2012}. Resolved dust continuum imaging later revealed a dust-emitting radius of $R_{\rm dust} \sim5.3$~kpc at $880~\mu$m \citep{hodge2015}. As one of the very few unlensed high-\textit{z} SMGs with spatially resolved \cotwoone kinematics --- the lowest-\textit{J} CO transition resolved in such a system --- GN20 provides a unique benchmark for testing how dynamical masses inferred from unresolved CO linewidths compare with those derived from resolved observations. We therefore use GN20 as a case study to investigate the impact of assumptions underlying the use of galaxy-integrated data in the proposed dynamical mass estimator (Eqn.~\ref{eqn:Mdyn}).

We first investigate the conventional isotropic virial estimator (Eqn.~\ref{eqn:virialestimator}) for GN20, using the integrated \cotwoone linewidth ($730 \pm 140$~km~${\rm s}^{-1}$) and compare the resulting dynamical masses with the TUNER LVG-derived gas mass\footnote{Assuming that the self-consistent TUNER LVG modeling provides the most robust molecular gas mass estimates of GN20 currently available.} from \citealt{boogard2026} (Figure~\ref{fig:virial_mdyn}). We evaluate three choices of effective radius: the resolved 880~$\mu$m dust radius ($R_{\rm dust}$), the median LVG-derived size ($R_{\rm LVG} \sim 6.5 \pm 1.2$~kpc)\footnote{Comparable to the resolved \cotwoone gas size of $\sim 7$~kpc measured by \cite{hodge2012}.}, and a fiducial value $2.5 \times R_{\rm dust}$. In all three cases, the inferred molecular gas mass satisfies $M_{\rm gas,LVG} \leq M_{\rm dyn,vir}$, placing GN20 comfortably within the distribution of our Group-2 DSFGs, consistent with its broad $>$ 700~km~${\rm s}^{-1}$ wide CO profile. Notably, \citet{boogard2026} derive an $\alpha_{\rm CO,LVG}$ of ${3.6}^{+7.3}_{-1.4}$ from unresolved, joint dust and CO SED modeling using TUNER\footnote{The modeled CO SLED spans \coonezero to CO(7--6), while the G/D ratio is left free, yielding a median value of ${57}^{+116}_{-28}$.}, consistent with both Galactic values and the dynamical estimate of $\alpha_{\rm CO} \sim 3.3$ reported by \citet{hodge2012} assuming that GN20 is composed of 100\% molecular gas.

Resolved dynamical modeling further provides independent measurements of the rotational velocity ($v_{\rm rot} = 575 \pm 100$~km~${\rm s}^{-1}$), intrinsic velocity dispersion ($\sigma_0 = 100 \pm 30$~km~${\rm s}^{-1}$), and cold-gas disk inclination ($i_{\rm gas} = {30}^{\circ} \pm {15}^{\circ}$). These quantities cannot be disentangled from an integrated CO linewidth alone and therefore offer a substantially more physical characterization of the galaxy's dynamical structure. Interestingly, the resolved dust morphology implies a much larger inclination ($i_{\rm dust} \sim {67}^{\circ}$), more than twice that inferred for the cold gas disk. Although the uncertainty on the gas inclination is significant, this discrepancy suggests that dust continuum morphology may not always provide a reliable proxy for the underlying dynamical geometry of high-\textit{z} DSFGs.

\begin{figure}[h]
\centering
\includegraphics[width=\textwidth]{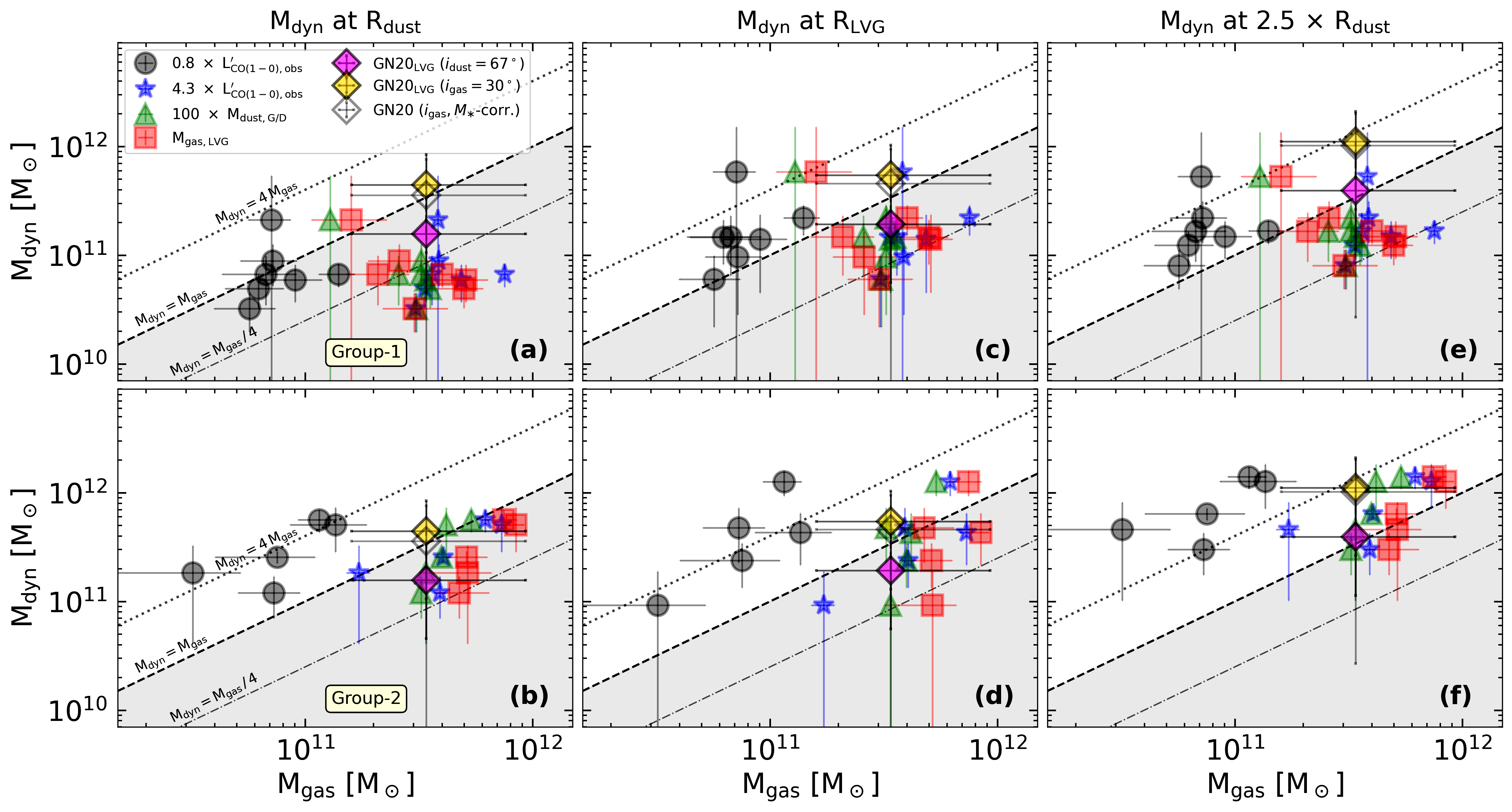}
\caption{Same as Figure~\ref{fig:virial_mdyn} but with the proposed ``mixed estimator" of dynamical masses (Eqn.~\ref{eqn:Mdyn}), using dust-inferred inclination angles for our targets. Definitions of colored symbols for the \vzgal sources are as per Figure~\ref{fig:virial_mdyn}. In addition, we overplot the values for GN20 using dust inclination (magenta diamond), gas disk inclination from resolved \cotwoone data (gold), and the latter corrected for stellar masses (white).}
\label{fig:grid_mass3methods}
\end{figure}

To explore the implications of this difference for GN20, Figure~\ref{fig:grid_mass3methods} compares dynamical masses inferred from the proposed ``mixed estimator" with both the gas- and dust-based inclination angles, $i_{\rm gas}$ and $i_{\rm dust}$. We use the resolved values of $v_{\rm rot}$ (i.e., minimum FWHM of 575~km~$\rm s^{-1}$ in Eqn.~\ref{eqn:vrot}) and $\sigma_0$, together with the same three characteristic radii considered above. We find that the LVG-derived gas mass is consistent within the limit set by the mixed dynamical mass estimator only when the resolved cold-gas inclination is adopted. In this case, Eqn.~\ref{eqn:Mdyn} yields $M_{\rm dyn} \sim 5.9\times {10}^{11}~{\rm M}_{\odot}$, in excellent agreement with the resolved dynamical modeling result of $(5.4 \pm 2.4) \times {10}^{11}~{\rm M}_{\odot}$ reported by \citet{hodge2012}.

In \citet{boogard2026}, the authors also present the TED (Turbulent Extended Distribution) model, which resolves radial profiles of the CO and dust continuum by combining individual TUNER models to build a spatially resolved profile. The derived $\alpha_{\rm CO,LVG} = {2.8}^{+0.5}_{-0.3}$ of GN20 from this analysis challenges yet the conventional bimodal assumption of \aco= 0.8 commonly adopted for SMGs/DSFGs. This \aco$\sim$ 2.8 is also in agreement with the dynamical upper limit for GN20, derived using Eqn.~\ref{eqn:Mdyn} with the resolved cold gas data and also its stellar mass-corrected version (Table~\ref{tab:gn20}, Figure~\ref{fig:grid_mass3methods}).

Nevertheless, the uncertainties associated with current LVG- and dynamical-modeling approaches remain considerable for high-\textit{z} dusty objects. Improving the latter will require higher-resolution and more sensitive observations plausibly using low/mid-\textit{J} CO rotational transitions (or [CI] and [CII]\footnote{[CII] 158~$\mu$m measurements are not possible from ground for GN20 at its redshift of $z=4.055$.}) together with multi-frequency dust continuum data. Although such measurements for larger samples may not entirely resolve the ``\aco tension" --- since even Galactic \aco calibrations exhibit large scatter and systematic differences depending on the adopted methodology --- we argue that they would significantly reduce \aco uncertainties in dusty galaxies across redshift and help determine whether any remaining \aco discrepancy has a genuinely physical origin.

\begin{table}[h]
    \centering
    \caption{{A comparison of dynamical upper limits on \aco derived from galaxy-integrated versus resolved \cotwoone measurements of GN20, an unlensed SMG at $z=4.055$.}}
    \begin{tabular}{l c c c}
        \toprule
Contributor  & Assumed or &  Value from the & Difference [dex] \\ 

   & integrated value &  resolved kinematics &  log($\alpha_{\rm CO,resolved} / \alpha_{\rm CO,dyn}$) \\ \midrule

 FWHM [km~$\mathrm{s}^{-1}$] & 730  & $^{a}$575  &  $-$0.200   \\ 

 Dispersion $\sigma_0$ [km~$\mathrm{s}^{-1}$]  & $^{b}$38  & 100  &  0.064  \\ 

 Radius $R_{\rm eff}$ [kpc]  & 5.3  &  7.0 &  0.121   \\ 

 Inclination [deg]  & 67  &   30 & 0.522  \\ 

$L^{\prime}_{\rm CO(1-0),obs}$ [K~km~$\mathrm{s}^{-1}$~$\mathrm{pc}^2$] &  $^{c}$$L^{\prime}_{\rm CO(2-1)}$ & $^{d}(1/0.88) \cdot L^{\prime}_{\rm CO(2-1)}$  & $-$0.056 \\ \midrule

 $^{e}$$\sum(\Delta\alpha_{\rm CO,contributors})$  & ... & ... & 0.451 \\
 $\alpha_{\rm CO}$ & $\leq 1.24$ & $\leq 3.22$ ($\leq {2.78}^{f}$) & 0.414 (${0.351}^{f}$) \\
 $^{g}$$\alpha_{\rm CO,LVG}$ & ${3.6}^{+7.3}_{-1.4}$ & ${2.8}^{+0.5}_{-0.3}$ & ... \\

\bottomrule
\end{tabular}

\parbox{\textwidth}{\footnotesize
\textbf{Notes.} \\
$^{a}$Rotational speed ($v_{\rm rot}$). \\
$^{b}$Assumed based on the redshift evolution suggested by \citet{rizzo2024}. \\
$^{c}$Assuming thermalized gas. \\
$^{d}$Adopting a (median) brightness temperature ratio of $r_{21}= L^{\prime}_{\rm CO(2-1)}/L^{\prime}_{\rm CO(1-0)} =0.88$ \citep{prajapati2026}. The observed $L^{\prime}_{\rm CO(2-1)}$ is $1.6 \times {10}^{11}$ K~km~$\mathrm{s}^{-1}$~$\mathrm{pc}^2$ \citep{hodge2012}. \\
$^{e}$For this order-of-magnitude comparison, uncertainties on the individual parameters are not considered. \\
$^{f}$Corrected for the stellar component with mass $(8.6 \pm 4.3) \times {10}^{10}~{\rm M}_{\odot}$ \citep{boogard2026}. \\
$^{g}$\cite{boogard2026}.
}

\label{tab:gn20}
\end{table}

\begin{figure}[h]
\centering
\includegraphics[width=0.55\textwidth]{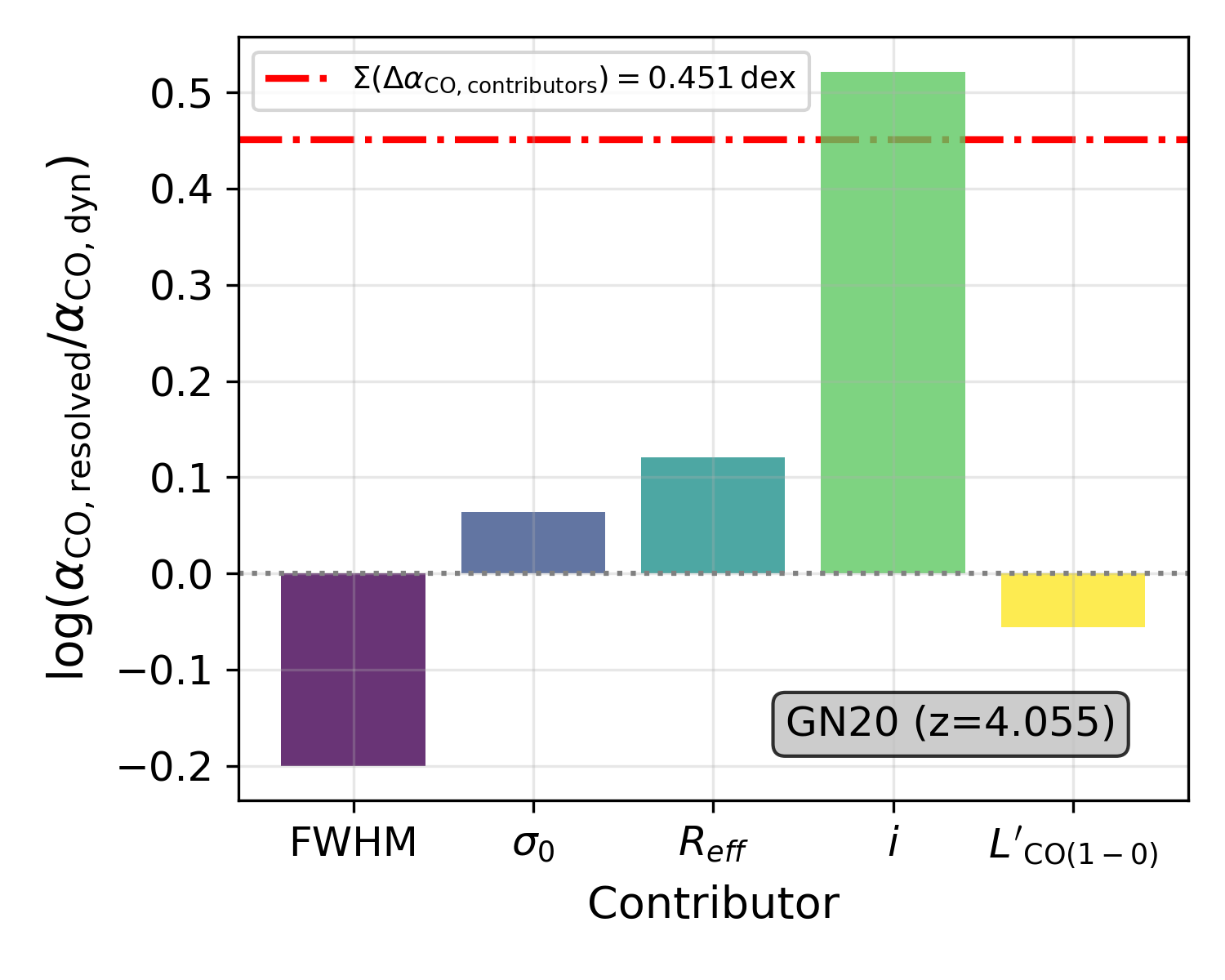}
\caption{GN20 as an example to show collective effects of various contributors, uncertainties/assumptions that contribute to plausible changes in the dynamical upper limits of $\alpha_{\rm CO}$. We note that CO excitation correction also contributes to the \aco discrepancy in the absence of integrated \coonezero measurements.}
\label{fig:gn20}
\end{figure}

Finally, we compare GN20 with the broader DSFG sample in Figure~\ref{fig:grid_mass3methods} by overplotting the distributions of our Group-1 and Group-2 galaxies, for which rotational velocities are estimated from unresolved CO linewidths corrected using dust-based inclinations, and velocity dispersions (Table~\ref{tab:dynamicalmasses}) are adopted from the empirical redshift-dependent relation of \citet{rizzo2024}. While isotropic virial estimators yield gas masses consistent with dynamical constraints for our sample, a mixed model with rotating turbulent disk produces inconsistencies for a subset of sources, implying that either (i) dust morphology does not fully trace the kinematic plane, (ii) CO linewidths are significantly contaminated by non-rotational motions, or (iii) a significant fraction of DSFGs are not well described by single-component rotating disks. This reflects that getting a robust cold gas radius does not solve the \aco tension when a sophisticated dynamical mass estimator such as Eqn.~\ref{eqn:Mdyn} is considered, and the unavoidable degeneracies associated with unresolved linewidth measurements and inclination estimates based solely on dust morphology remain important. That said, the GN20 case demonstrates that, once the underlying geometry and kinematics are independently constrained, the proposed mixed estimator reproduces the results of full dynamical modeling remarkably well. Therefore, we argue that it may provide a physically motivated framework for interpreting DSFGs with unresolved low-\textit{J} CO lines and a practical bridge between the simple virial estimator and resolved dynamical analyses using brighter CO, [CI], or [CII] lines until the challenging, lowest-\textit{J} CO transitions are resolved.

In Appendix~\ref{subsec:allcontriaco}, we summarize relative contributions to the \aco conversion factor from various parameters involved in dynamical mass estimation using the ``mixed estimator". The \aco estimates are more sensitive to inclination and integrated FWHM --- both contributing to the rotational speeds (Eqn.~\ref{eqn:vrot}) --- radius being the second largest contributor. Results from the GN20 case study (Figure~\ref{fig:gn20}) show agreement with this inference.

\section{Discussion} \label{sec:discussion} 

We investigate the dynamical masses of 21 \vzgal DSFGs that show no clear evidence of gravitational lensing and whose dust continuum emission is spatially resolved by ALMA at observed-frame 1~mm. The ALMA-derived dust sizes and inclination angles (Figure~\ref{fig:rdust_inclination}), together with the integrated CO(1--0) linewidths from the \vzgal survey \citep[Table~\ref{tab:dynamicalmasses};][]{prajapati2026}, are used to estimate dynamical masses using two complementary approaches: an isotropic ``virial estimator" ($\S$~\ref{subsec:mdynisovir}) and a ``mixed estimator" that assumes rotating thick disks with significant pressure support ($\S$~\ref{subsec:mdynmixed}).

\begin{figure}[h]
\centering
\includegraphics[width=0.45\textwidth]{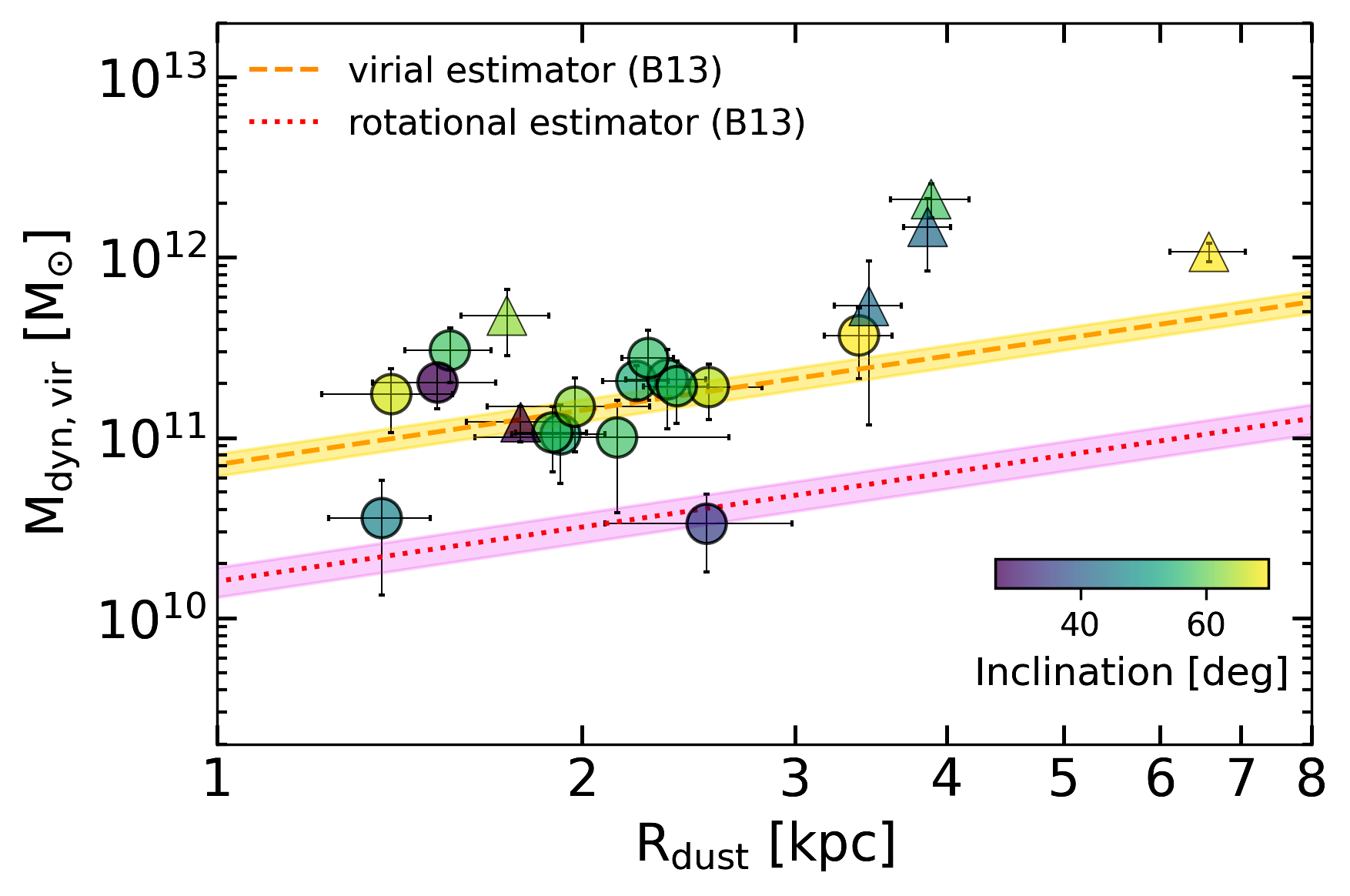}
\includegraphics[width=0.45\textwidth]{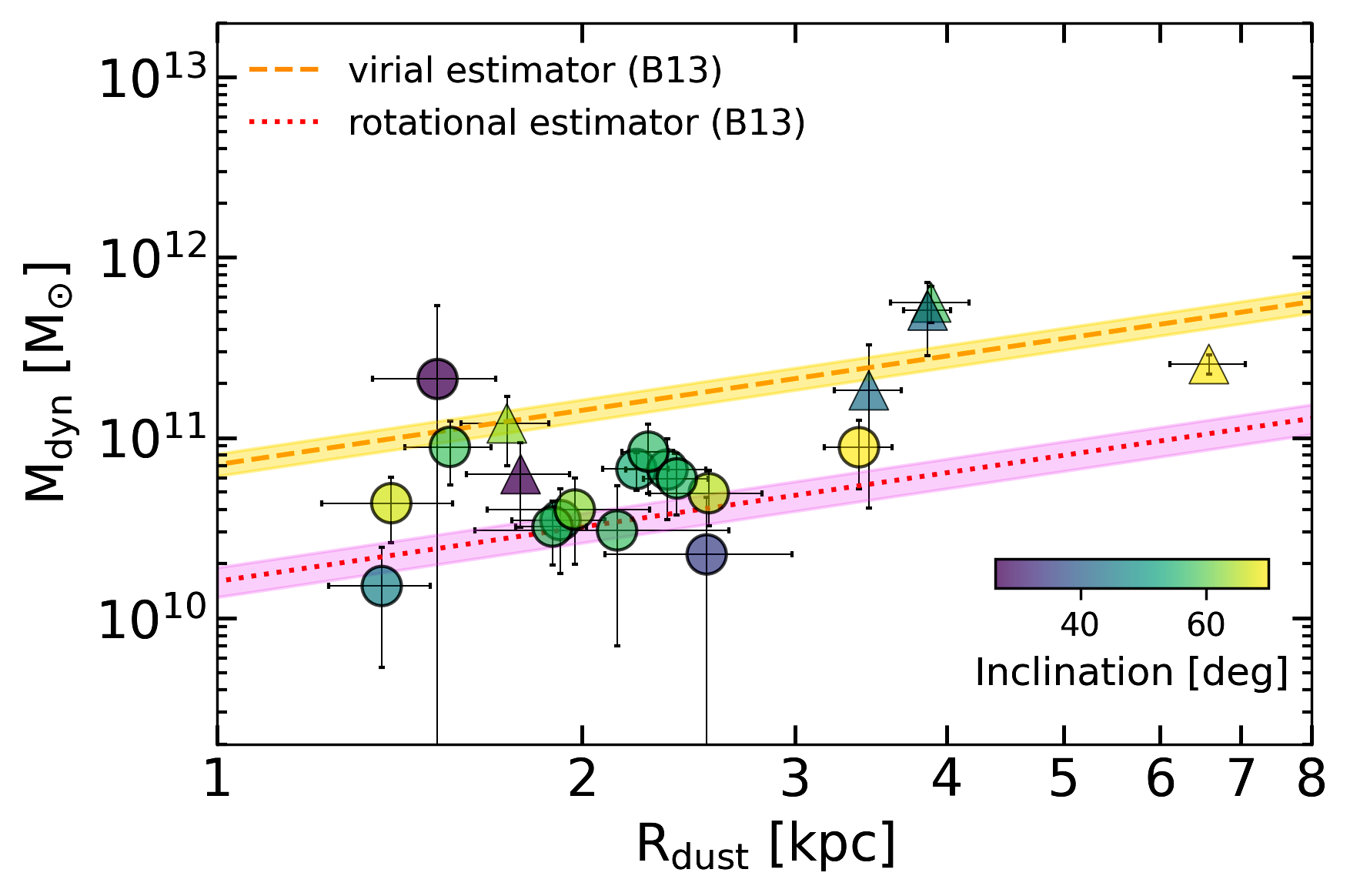}
\caption{Dynamical mass as a function of dust size for 21 \vzgal unlensed DSFGs. All the data points are colored using their dust-inferred inclination angles ($i$; Table~\ref{tab:dynamicalmasses}). The dashed yellow and dotted red lines show the empirical $M_{\rm dyn}$--R relations from \cite{bothwell2013}. Left panel: Dynamical masses shown here are using the isotropic virial estimator discussed in $\S$~\ref{subsec:mdynisovir}. Right panel: Here we use the ``mixed estimator" assumptions to derive the dynamical masses (see $\S$~\ref{subsec:mdynmixed}). Systems with broad, complex \coonezero line profiles discussed in $\S$~\ref{subsec:co10obs} are shown using triangle markers in both plots.}
\label{fig:mdyn_dustsize}
\end{figure}

Figure~\ref{fig:mdyn_dustsize} compares the resulting dynamical masses with the empirical mass--size scaling relations derived by \citet{bothwell2013}. Using the isotropic virial estimator, our galaxies closely follow the virial relation reported by \citet{bothwell2013}, with $(7.1 \, \pm \, 1.0) \, \times \, {10}^{10} \, R_{\rm kpc} \, [\rm M_{\odot}]$. In contrast, the mixed estimator places our sample between the virial and pure rotational relations ($(1.6 \, \pm \, 0.3) \, \times \, {10}^{10} \, R_{\rm kpc} \, [\rm M_{\odot}]$) presented by Bothwell et al. This intermediate behavior is expected, as the mixed estimator explicitly accounts for both rotational support and a non-negligible pressure-support term ($k{\sigma^2_0}$, Eqn.~\ref{eqn:Mdyn}), naturally bridging the two limiting cases.

The inferred isotropic dynamical masses of our sample span a wide range, from ${10}^{10.4}-{10}^{12.2}~{\rm M}_{\odot}$ with a majority of the sources spanning $(0.3-5)\times{10}^{11}~{\rm M}_{\odot}$. Similar values have also been derived for ultraviolet/optical-selected main-sequence galaxies at cosmic noon \citep[e.g., SINS sample][]{natascha2006} and mixed/heterogeneous DSFG/SMG populations across redshifts \citep{neri2003,swinbank2004,swinbank2006,tacconi+2008,aravena+2016,rivera2018}. A few notable outliers (HeLMS-50, HerBS-43a, and HerBS-191), which also exhibit the broadest \coonezero line profiles ($\S$~\ref{subsec:co10obs}, Group-2 sources), may indicate strong candidates for a post-merger coalescence phase, or mergers viewed at unfavorable projection angles. They are distinguished by their extended dust sizes (Figure~\ref{fig:mdyn_dustsize}) and large isotropic virial dynamical masses (${M}_{\rm dyn}\gtrsim {10}^{12}~{\rm M}_{\odot}$), corresponding to halo masses of at least ${M}_{\rm halo}\sim {10}^{13}~{\rm M}_{\odot}$. Notably, their gas masses are also the largest among our sample (Tables~\ref{tab:dustmass_gdmr} and \ref{tab:tunerlvg}). Although we do not have stellar masses of these DSFGs, this could plausibly suggest their large gas fractions \citep[see theoretical predictions by e.g.,][]{behroozi2019}. 

Among these 21 galaxies, we further selected a subset of 12 ($\S$~\ref{subsec:gasmasses}) to derive independent molecular gas masses and corresponding \aco values, which are subsequently compared to the dynamical upper limits on the $\alpha_{\rm CO}$. We therefore restrict the following discussion to the results obtained for this sub-sample of 12 DSFGs.

\subsection{Integrated molecular gas masses and \aco of dusty star-forming systems across cosmic time} \label{subsec:unresolvedlimits}

There has been considerable debate regarding whether or not the \aco conversion factor is bimodal across galaxy populations \citep[e.g.,][and references therein]{daddi2010,genzel2015,scoville2016,dunne+2022,berta+2023}. Under the traditional bimodal framework, \aco= 0.8 is commonly adopted for local ULIRGs and high-\textit{z} DSFGs, SMGs, and quasars \citep[e.g.,][]{downes_solomon_1998,carilliwalter2013,bolatto2013,Xu2026qsos}, whereas a value of Galactic \aco= 4.3 is typically assumed for local star-forming main-sequence galaxies. Theoretical studies have long discussed that \aco is not adequately described by bimodality, but instead it varies continuously with the physical conditions of the ISM \citep{narayanan2011,narayanan2012}. Although many observational studies have discussed this, \cite{harrington2021} and \cite{prajapati2026} attempt to quantify the trend better using larger sample sizes. However, unresolved dust and gas observations provide only limited constraints on the internal ISM properties of high-\textit{z} galaxies; methods such as the TUNER LVG radiative transfer modeling explored in our work offer a valuable means to infer these properties.

The commonly adopted \aco of 0.8 for local ULIRGs (with a reported range spanning 0.3--1.3) originated from the study by \cite{downes_solomon_1998}, which was subsequently adopted for high-\textit{z} DSFGs, SMGs, and quasars. This conversion factor gained widespread acceptance because it yielded molecular gas masses broadly consistent with the available dynamical mass estimates. In that work, \aco was derived using radiative transfer modeling; however, the analysis was limited primarily to observations of the \coonezero and \cotwoone transitions. Based on these data, the authors inferred relatively low gas densities, a condition that leads to a low $\alpha_{\rm CO}$, as $\alpha_{\rm CO} \propto \sqrt{n_{\rm H_2}}\,/\,{T_{\rm b}}
$ \citep{solomon2005}. However, observations of higher-\textit{J} CO transitions, particularly after the Herschel era, have revealed that local ULIRGs are generally highly excited systems \citep[see e.g.,][]{krumholz2014,rosenberg2014}. Modern radiative transfer analyses of these sources (Weiss et al., in prep.) no longer support the low gas densities inferred from the limited low-\textit{J} datasets available to \cite{downes_solomon_1998}. Consequently, the latest LVG and other excitation models often favor substantially higher \aco values -- see also the discussion by \citet{papadopoulos2012b} for LIRGs. Additionally, a recent theoretical study of massive star-forming galaxies (on and above the main sequence) at cosmic noon by \citet{anirudh2025} also suggests that the \aco is $\sim1-5$ in the central regions of their massive simulated galaxies, but these values rise sharply beyond the central few kiloparsecs. 

Leading to a range in \aco of about 1.5--8.5 for our sample, the dust SED-based method-1 ($\S$~\ref{subsec:gasmass_gdmr}) support this interpretation, for which the inferred large gas masses are consistent, within the uncertainties, with those derived from the detailed radiative transfer modeling performed with TUNER (\aco= 2.6--11.5, method-2; $\S$~\ref{subsec:gasmass_tuner}). In particular, TUNER simultaneously models the integrated dust and CO SEDs, providing a self-consistent framework for constraining the gas excitation conditions and molecular gas content of these sources. It is important, however, to note the current limitations of the TUNER LVG modeling. It relies on the currently available galaxy-integrated measurements, which are robustly anchored using \coonezero but lack constraints on CO transitions above $J=6$ for all of our targets. Consequently, the inferred physical conditions can be refined with improved measurements, providing stronger constraints on the warmer and denser molecular gas components. It will also enable a more robust characterization of the range of gas excitation conditions present in these systems.

The tension between gas-based and dynamical estimates of \aco is also not unique to CO or dust as molecular gas tracers. Similar discrepancies have been reported using alternative probes such as [CI] and [CII], when their Galactic calibrations are adopted \citep{dye2022}. More broadly, studies of dust-obscured star-forming galaxies across cosmic time consistently find that molecular gas masses inferred from ISM tracers exceed those implied by unresolved dynamical analyses \citep[e.g.,][]{yang2017,dye2022,hagimoto2023,eales2024}, leading to the requirement for their lower \aco conversion factors. While \cite{eales2024} propose metallicity evolution as one of the plausible explanations, this is unlikely to dominate in the most luminous high-$z$ DSFGs, which are expected to have near-solar or super-solar metallicities \citep{dunne+2022,Gururanjan2023}.

\subsection{Are large molecular gas masses and Galactic \aco values dynamically allowed?} \label{subsec:largeaco}

The \coonezero line luminosities of our high-$z$ \vzgal DSFGs are consistent with those of the broader DSFG population \citep[see Figure 2 of][]{prajapati2026}, indicating that these galaxies are representative rather than exceptional systems. Their gas-mass estimates based on the dust SED and the TUNER LVG analysis ($\S$~\ref{subsec:gasmass_gdmr} and $\S$~\ref{subsec:gasmass_tuner}) consistently imply large molecular gas reservoirs, encompassing near-Galactic conversion factors within the inferred broad range of $\alpha_{\rm CO} \sim1.5-11.5$. The central question is, therefore, whether such large gas masses are compatible with the observed dynamics, or whether the kinematics instead necessarily require the commonly adopted ULIRG value of $\alpha_{\rm CO} = 0.8$.

\begin{table}[h]
\scriptsize{
    \centering
    \caption{\aco constraints compared from three different methods: two gas mass estimators, and dynamical upper limits.}
    \begin{tabular}{l|c|c|c c c|c c c}
        \toprule
 & \multicolumn{8}{c}{Median $\alpha_{\rm CO}$ value/limit}\\ 
\cmidrule(lr){2-9}
Sources & Dust SED & TUNER LVG & \multicolumn{3}{c}{Mixed estimator of $M_{\rm dyn}$ (Eqn.~\ref{eqn:Mdyn})} & \multicolumn{3}{c}{Isotropic virial estimator of $M_{\rm dyn,vir}$ (Eqn.~\ref{eqn:virialestimator})}\\ 
 &  (a) & (b) & (c) & (d) & (e) & (f) & (g) & (h)  \\
\midrule
Group-1 & 3.06 ± 1.21 & 4.75 ± 1.74 & $\leq$ 0.63 ± 0.18 &  $\leq$ 1.26 ± 0.41 &  $\leq$ 1.58 ± 0.45 
& $\leq$ 2.27 ± 0.56 &  $\leq$ 4.46 ± 1.10   &   $\leq$ 5.68 ± 1.41   \\
Group-2 & 3.73 ± 0.51 & 5.50 ± 0.12 & $\leq$ 2.99 ± 0.93 &  $\leq$ 2.54 ± 0.23  &  $\leq$ 7.47 ± 2.31 
& $\leq$ 11.44 ± 2.74 &    $\leq$ 10.64 ± 3.86    &   $\leq$ 28.60 ± 6.85    \\
\bottomrule
\end{tabular}

\parbox{\textwidth}{\footnotesize
\textbf{Notes.} \\
(a) $\alpha_{\rm CO,G/D}$ from method-1 ($\S$~\ref{subsec:gasmass_gdmr}). \\
(b) $\alpha_{\rm CO,LVG}$ from method-2 ($\S$~\ref{subsec:gasmass_tuner}). \\
(c) $\alpha_{\rm CO,dyn}$ using a ``mixed estimator" (Eqn.~\ref{eqn:Mdyn}) with dust sizes ($R_{\rm dust}$) and the observed, galaxy-integrated \coonezero line luminosity; i.e., $\alpha_{\rm CO,dyn}  \leq M_{\rm dyn} / L^{\prime}_{\rm CO(1-0),obs}$. \\
(d) Same as above, but with LVG-derived sizes ($R_{\rm LVG}$). \\\
(e) Same as above, but with a fiducial size of $2.5 \times R_{\rm dust}$. \\
(f) $\alpha_{\rm CO,dyn,vir}$ using an isotropic ``virial estimator" (Eqn.~\ref{eqn:virialestimator}) with dust sizes ($R_{\rm dust}$) and observed \coonezero line luminosity; i.e., $\alpha_{\rm CO,dyn}  \leq M_{\rm dyn,vir} / L^{\prime}_{\rm CO(1-0),obs}$. \\
(g) Same as above, but with LVG-derived sizes ($R_{\rm LVG}$). \\
(h) Same as above, but with a fiducial size of $2.5 \times R_{\rm dust}$. \\
}

\label{tab:aco3methods_table}
}
\end{table}

Table~\ref{tab:aco3methods_table} compares the gas mass-inferred \aco estimates with the maximum \aco values allowed by different dynamical mass prescriptions. When our proposed rotating thick-disk ``mixed estimator" (Eqn.~\ref{eqn:Mdyn}, Figure~\ref{fig:grid_mass3methods}) is combined with compact dust radii, the inferred dynamical masses are relatively small, particularly for Group-1, yielding upper limits that can fall below the gas-based estimates. However, using GN20 as a case study, we argue that these constraints would relax substantially once more realistic cold gas extents and geometry are adopted. Assuming a smoothly distributed cold gas component, the TUNER LVG-inferred sizes are expected to be broadly consistent with the radii inferred from resolved low-\textit{J} CO emission extents (e.g., GN20). 

The isotropic virial estimator (Eqn.~\ref{eqn:virialestimator}, Figure~\ref{fig:virial_mdyn}) provides a complementary test by avoiding explicit assumptions about disk inclination and rotational support. Using the LVG-derived gas radii in the virial estimator, the median dynamical limits increase to $\alpha_{\rm CO,dyn,vir}\leq4.46\pm1.10$ for Group-1 and $\leq10.64\pm3.86$ for Group-2, fully encompassing the values inferred from both gas-mass methods. 

The particularly large dynamical limits obtained for Group-2 should nevertheless be interpreted as conservative upper bounds. Because both dynamical estimators scale approximately with the square of the observed CO linewidth, galaxies with broad integrated profiles naturally yield much larger dynamical masses. As discussed in $\S$~\ref{subsec:mdynmixed} for unresolved systems, broad linewidths may contain a combination of ordered rotation, turbulence, mergers, or projection effects. Consequently, the Group-2 limits should be used with caution (see also Appendix $\S$~\ref{app:lineprofiles}).  

Overall, the comparison in Table~\ref{tab:aco3methods_table}, supported by Figures~\ref{fig:virial_mdyn} and \ref{fig:grid_mass3methods}, demonstrates that the inferred high \aco values cannot be completely ruled out by the available dynamical constraints. Instead, the apparent preference for \aco= 0.8 may arise mainly when compact effective sizes and inclination-dependent rotating-disk models are applied to unresolved observations. Once plausible uncertainties in the cold gas extent, geometry, and kinematic structure are taken into account, intermediate (e.g., \aco= 2--3) or near-Galactic \aco values remain dynamically viable across our sample (e.g., GN20, $\S$~\ref{subsec:gn20}). 

We therefore find no dynamical justification for universally adopting \aco= 0.8 in the presented high-\textit{z} DSFGs here. To avoid underestimating molecular gas masses systematically, \aco= 0.8 should not be assumed for DSFGs unless supported by ISM and/or dynamical modeling. Rather, the appropriate conversion factor is likely to vary from source to source. Robust estimates would require combining independent gas-mass tracers, resolved gas and dust distributions, and supporting dynamical models \citep[e.g., \texttt{GalPaK$^{\rm 3D}$};][]{bouch2015galpak3d} that overcome the limitations set by unresolved observations.

\subsection{Why do rotating thick-disk-based dynamical masses remain low?} \label{subsec:mixedestlow}

The dependence of the inferred \aco limits on the adopted dynamical estimator naturally raises the question of why the rotating thick-disk prescription --- although generic and flexible for accommodating various $v_{\rm rot}$/$\sigma_0$ values --- systematically yields lower masses than the spherically symmetric virial estimator. We argue that this primarily reflects the application of a geometrically specific model to unresolved observations, where several key quantities cannot be directly constrained.

\begin{itemize}[noitemsep]

\item \textit{Cold gas extents:} Both dynamical mass estimators scale linearly with the effective radius of the cold gas distribution, which remains uncertain. Our TUNER LVG radiative transfer analysis, together with resolved studies of DSFGs \citep[e.g.,][]{rivera2018,boogard2026}, indicates that molecular gas may significantly extend beyond the dust continuum. This suggests that dust-based sizes likely underestimate the true spatial extent over which the observed linewidths are produced. 

\item \textit{Source geometry:} While the isotropic virial estimator performs well for unresolved data --- and, in our case, yields dynamical masses broadly encompassing the inferred gas masses --- more sophisticated approaches require spatially and spectrally resolved observations. If DSFGs are assumed to be rotating thick disks, the poorly constrained inclination angles introduce substantial uncertainties in the derived dynamical masses. Moreover, the dust morphology may not accurately trace the more extended distribution of cold molecular gas in high-\textit{z} DSFGs. Consequently, resolved cold gas observations are essential to fully characterize the source geometry and to assess its impact on dynamical mass estimates. 

\item \textit{CO linewidth:} The integrated FWHM of CO lines cannot be uniquely decomposed into ordered rotation and turbulent dispersive medium in unresolved galaxies. This is extremely relevant for bright dusty sources across redshifts, many of which are proposed to be involved in ongoing/past merger activities. Turbulence, mergers, beam smearing, cold mode accretion, outflows, and multiple unresolved components may all broaden the observed CO line profile. Consequently, the integrated FWHM used in any dynamical mass estimator introduces a fundamental degeneracy into geometry-dependent dynamical models. 

\end{itemize}

Ultimately, resolving these uncertainties requires spatially and spectrally resolved observations of the cold gas reservoir, particularly with low-\textit{J} CO emissions, which directly traces the bulk molecular gas distribution and its kinematics in high-\textit{z} DSFGs. Such observations are essential for constraining the gas extent, source geometry, and the relative contributions of ordered rotation and turbulent motions (i.e., $v_{\rm rot}$/$\sigma_0$). While high-resolution dust continuum observations provide a more accessible and observationally efficient means of obtaining initial size estimates, they are unlikely to capture the full extent of the cold molecular gas reservoir and its dynamics. Resolved low-\textit{J} CO observations with ALMA, NOEMA, and future facilities such as the ngVLA will therefore be essential. At the same time, resolved observations of alternative tracers, including [CII], [CI], and mid-\textit{J} CO transitions, can provide complementary, intermediate constraints on galaxy kinematics; e.g., the recent NOEMA$^{\rm 3D}$ survey for high-\textit{z} main-sequence galaxies by \cite{chen2026noema3d,jolly2026noema3d}, and several [CII] studies \citep[e.g.,][]{leung2019,rowland2024}. Together, these data will enable more physically motivated dynamical mass estimates and, consequently, more robust upper limits on $\alpha_{\rm CO}$ than those obtained under the isotropic virial assumption.

\subsection{Group-1 versus Group-2 DSFGs: Different kinematic classes?} \label{subsec:groups}

The distinction between Group-1 and Group-2 sources suggests that the \vzgal DSFG population is unlikely to be kinematically uniform. Group-1 galaxies, characterized by comparatively simple CO(1--0) line profiles, occupy a relatively homogeneous region of parameter space in which the unresolved kinematics can be reproduced by similar combinations of inclination and dynamical support. In contrast, Group-2 galaxies exhibit broader and/or more complex CO(1--0) profiles and span a substantially wider range of acceptable dynamical solutions. Their larger integrated linewidths and, in some cases, larger dust sizes naturally yield higher dynamical masses and consequently weaker upper limits on $\alpha_{\rm CO}$. This broader distribution may reflect greater diversity in the underlying gas dynamics, arising from enhanced turbulence, non-circular motions, mergers, or multiple unresolved kinematic components. For example, GN20 --- with plausible recent interaction \citep{colina2023,ubler2024,boogard2026} --- occupies a similar region of parameter space to the Group-2 sources in Figures~\ref{fig:virial_mdyn} and \ref{fig:grid_mass3methods}.

Although we note that marginal differences in the viewing angle of the system's geometry can change the FWHM shapes, large differences in linewidths are more likely to arise because of dynamics. We therefore interpret the two groups as representing different degrees of dynamical complexity rather than a strict disk--merger division. While the available unresolved observations do not uniquely distinguish between the physical mechanisms responsible for the observed linewidths, they suggest that our high-\textit{z} DSFGs likely span a continuum of dynamical states, with Group-1 representing the more homogeneous end of the distribution and Group-2 encompassing systems with increasingly diverse kinematic configurations.

This interpretation is broadly consistent with recent theoretical studies, which suggest that unresolved kinematic measurements may not fully capture the dynamical complexity of massive high-\textit{z} star-forming galaxies. Cosmological simulations indicate that the molecular gas is preferentially confined to the rotating disk, with expected $v_{\rm rot}/\sigma_0$ ratios approximately 2--3 times higher than those of the HI component \citep{kretschmer2022_mdynsimulations}. Furthermore, FIRE-2 simulations show that the mid-plane pressure is well described by the weight of the gas disk and scales approximately linearly with the star formation rate surface density, a quantity that can be inferred from resolved dust continuum observations \citep{gurvich2020_mdynsimulations}. Consistent with this picture, simulation-based studies of $z\gtrsim1$ galaxies with stellar masses $M_{\ast} > {10}^{11}~{\rm M}_{\odot}$ find that dynamical masses can be underestimated by up to 40\% if turbulent pressure gradients are neglected in disk models \citep{wellons2020_mdynsimulations}. Together, these results provide theoretical motivation for modeling DSFGs as rotating, pressure-supported systems while highlighting the importance of spatially resolved gas and dust observations for disentangling rotational support, turbulence, and merger-driven motions, and for testing the validity of such dynamical models.

\begin{figure}[h]
\centering
\includegraphics[width=0.329\textwidth]{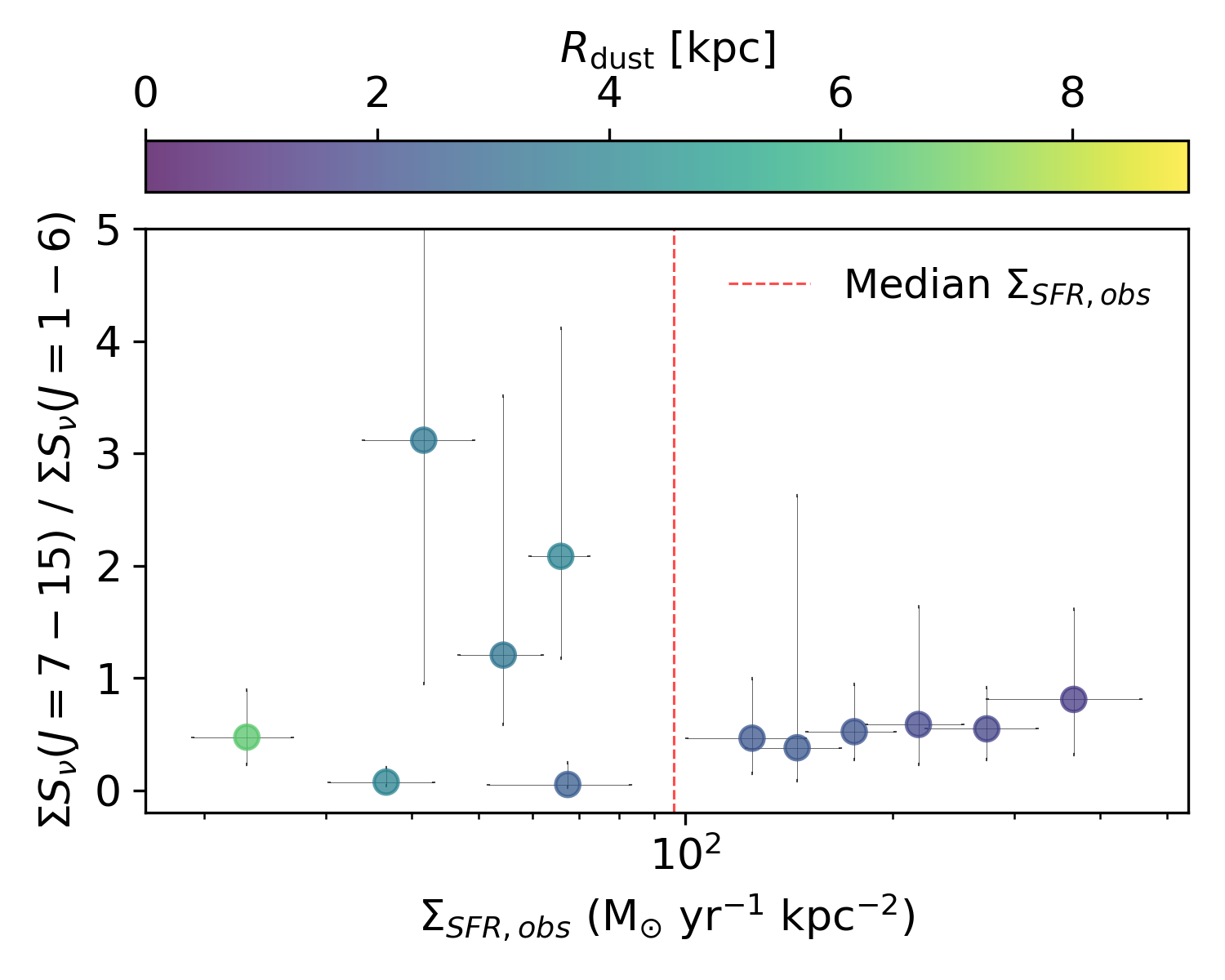}
\includegraphics[width=0.329\textwidth]{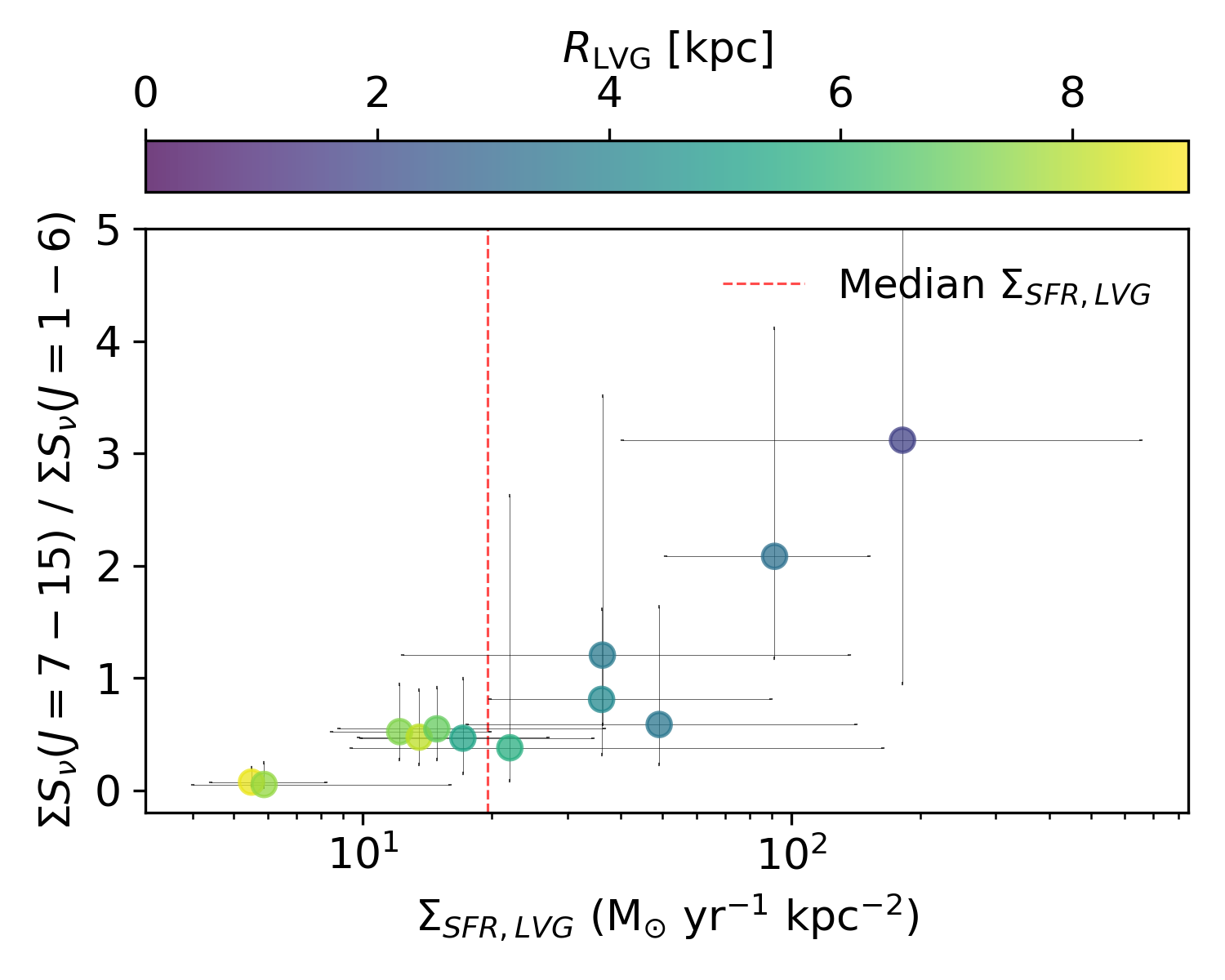}
\includegraphics[width=0.329\textwidth]{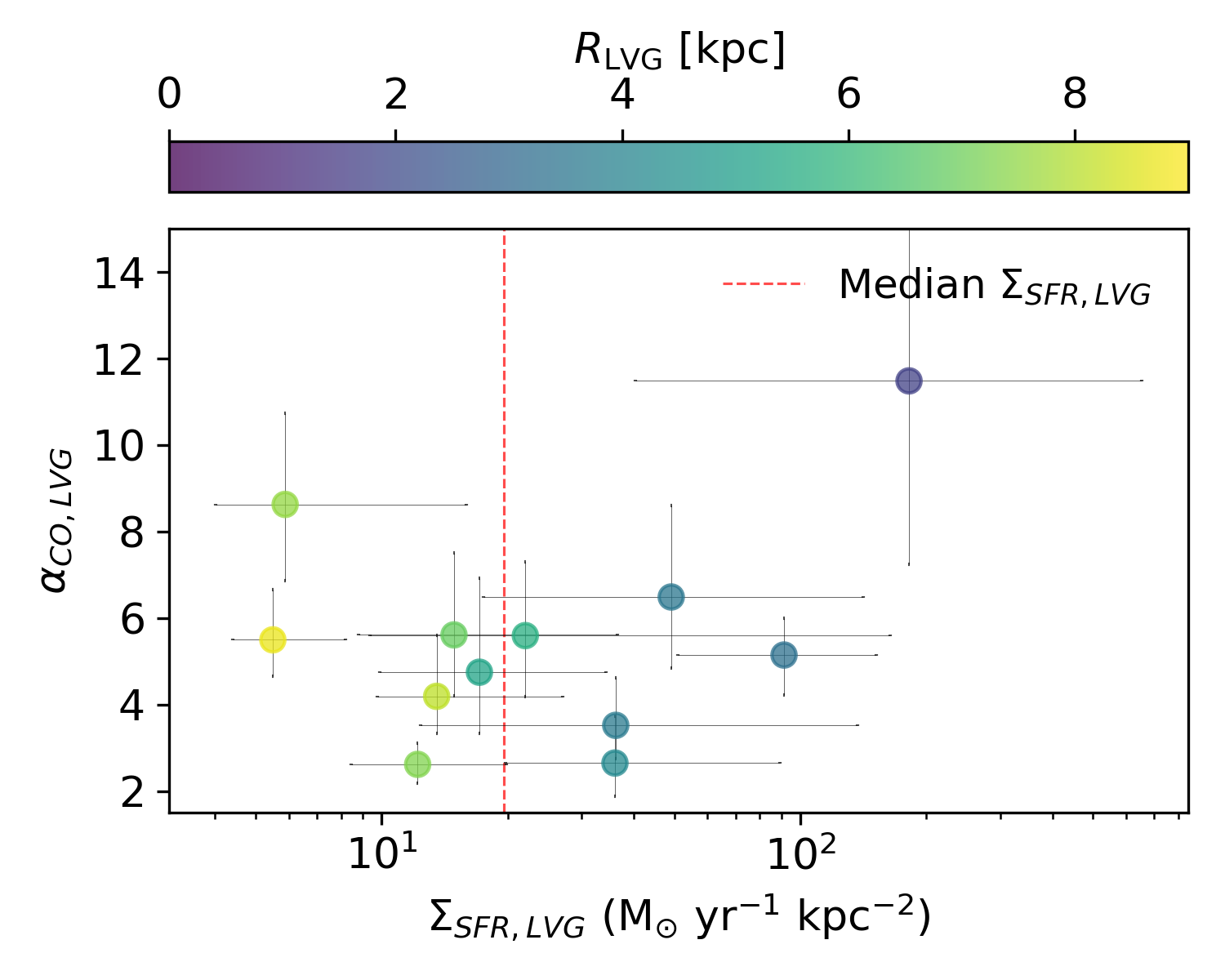}
\caption{Comparison of the TUNER LVG-derived CO excitation budget with the star formation rate surface density ($\Sigma_{\rm SFR}$). The CO excitation budget (y-axis) is defined as the ratio of the integrated CO line flux in the high-$J$ transitions ($J=7-15$) to that in the low-/mid-$J$ transitions ($J=1-6$). Left: $\Sigma_{\rm SFR}$ computed using the observed SFRs (8--1000~$\mu$m; \citealt{berta+2023}) and ALMA dust continuum sizes (Table~\ref{tab:dynamicalmasses}). Center: $\Sigma_{\rm SFR}$ computed using the TUNER-inferred total infrared luminosities (8--1000~$\mu$m) and LVG-derived source sizes. Right: TUNER LVG-derived \aco as a function of $\Sigma_{\rm SFR}$.}
\label{fig:surfacedensity}
\end{figure}

Due to their plausibly complex gas dynamics, DSFGs are expected to host high gas fractions, intense star formation, and elevated star-formation rate surface densities ($\Sigma_{\rm SFR}$). Figure~\ref{fig:surfacedensity} compares the observed and LVG-inferred $\Sigma_{\rm SFR}$ values and explores their relation to source size and gas excitation. 

The LVG-inferred results reveal a clear correlation between $\Sigma_{\rm SFR}$ and molecular gas excitation, with more compact galaxies exhibiting higher $\Sigma_{\rm SFR}$ and stronger high-$J$ ($J>6$) CO emission. While this trend is physically expected, it is much weaker when the observed dust continuum sizes are adopted, supporting the argument from $\S$~\ref{subsubsec:discussion_radius} that optical depth effects cause dust emission to underestimate the true extent of the ISM. We find a similar trend if we replace the $\Sigma_{\rm SFR}$ with the LVG-inferred molecular gas surface density, consistent with dense/warm ISM leading to higher-\textit{J} CO excitation. In contrast, we find no correlation between the \aco conversion factor and either $\Sigma_{\rm SFR}$ or source size (Figure~\ref{fig:surfacedensity}, right panel).

Using the observed dust sizes, $\Sigma_{\rm SFR}$ ranges from 20--400~$\rm{M}_{\odot}$~yr$^{-1}$~kpc$^{-2}$, with the most compact systems approaching the ``maximum starburst" regime. Adopting the larger ISM sizes inferred from the TUNER LVG analysis lowers this range to 0.5--200~$\rm{M}_{\odot}$~yr$^{-1}$~kpc$^{-2}$, although these galaxies remain capable of sustaining highly turbulent and intermittent ISM conditions, potentially shaped by processes such as cold gas inflows, merging gas streams, past/ongoing galaxy interactions, and feedback-driven strong winds.

Resolving the physical origin of the diversity between Group-1 and Group-2 systems, therefore, requires spatially resolved CO imaging. Such observations will constrain the spatial extent of the cold molecular gas and provide robust measurements of disk inclination, dynamical mass, and the relative contributions of rotational support, and turbulence in high-\textit{z} DSFG populations. At the same time, galaxy-integrated CO SLEDs anchored by at least \coonezero or CO(2--1), together with dust continuum and mid/high-\textit{J} CO observations, will remain essential for jointly constraining molecular gas masses and calibrating the characteristic spatial scales of these galaxies through semi-empirical approaches such as TUNER. Altogether, these complementary resolved and integrated observations will help reduce the current uncertainties in \aco and determine whether Group-1 and Group-2 sources in our sample reflect differences in their dynamical states, evolutionary stages, or both.

\subsection{$\alpha_{\rm CO}$ and distance from the star-forming main-sequence} \label{subsec:aCOdelMS}

Given the heavily dust-obscured nature of high-\textit{z} DSFGs, one of the main challenges is obtaining reliable stellar mass ($M_{\ast}$) estimates, which are required to determine their location relative to the SFR--$M_{\ast}$ main-sequence. Accurate stellar masses are also essential for estimating gas fractions and for deriving $\alpha_{\rm CO}$ upper bounds from dynamical masses after accounting for the stellar mass contribution (e.g., GN20 in Figure~\ref{fig:grid_mass3methods}). Recent rest-frame near-infrared observations of 13 such unlensed dusty galaxies using the JWST \citep{hodge2025} indicate high dust extinctions, which motivates expanding the observations up to mid-infrared to get robust stellar contributions. The observed high, total infrared luminosities in these sources are expected to obscure even strong nebular lines such as $\rm H{\alpha}$ \citep[see e.g., GN20;][]{ubler2024} and $\rm Pa{\alpha}$ \citep[e.g.,][]{bik2024}, making the stellar mass and metallicity estimates of such systems extremely challenging. Such a scenario also cautions about using alternative, independent \aco estimation techniques such as stellar \textit{mass-metallicity} relations \citep[e.g.,][]{genzel2015}. 

Even in local dusty merger systems, metallicity determinations remain subject to significant uncertainties arising from complex stellar populations, underlying Balmer absorption, merger-driven gas inflows, and spatial variations in chemical abundances. These challenges are likely to be even more severe at high redshift, where limited sensitivity and spatial resolution further complicate the interpretation of nebular emission-line diagnostics. Consequently, \aco estimates based on metallicity should be interpreted with caution and, whenever possible, cross-validated against independent constraints. In particular, at high redshift, such estimates should be compared with cold gas-based dynamical and semi-empirical (e.g., TUNER LVG) constraints.

Although the efforts to characterize the dust extinction and eventually stellar masses and metallicity, via multiwavelength SED fitting tools, rest-frame optical/near-infrared, and mid-infrared data for larger samples of high-\textit{z} DSFGs are currently in a nascent stage, facilities such as the JWST and ground-based $\gtrsim$10m-class telescopes, alongside future missions (e.g., PRIMA), hold great potential to explore this parameter space to reveal the nature of the high-\textit{z} dusty systems and their multi-phase ISM in the context of galaxy evolution over cosmic time.

\section{Summary} \label{sec:summary}

The CO--H$_2$ conversion factor ($\alpha_{\rm CO}$) remains the most discussed source of uncertainty in determining molecular gas masses of high-\textit{z} dusty star-forming galaxies (DSFGs). Using the largest to date homogeneous sample of 21 unlensed DSFGs at $z=1-4$ with integrated CO(1--0) measurements and resolved ALMA 1~mm continuum imaging, we revisit the $\alpha_{\rm CO}$ discrepancy. Restricting our analysis to the 12 galaxies with the tighter modeling constraints, we estimate molecular gas masses independently using dust SED modeling ($\S$~\ref{subsec:gasmass_gdmr}) and the TUNER multi-component LVG framework ($\S$~\ref{subsec:gasmass_tuner}). Although these approaches are not fully independent, as they share assumptions regarding the gas-to-dust ratio of 100 \citep[e.g.,][]{leroy2011}, dust opacity normalization \citep{draine2014}, and global dust modeling, they nevertheless yield consistent gas masses within their uncertainty. Together, they imply a wide range of $\alpha_{\rm CO} \sim 1.5-11.5$ (including Helium contribution) with median values of about 3.4 and 5.1, respectively, which are substantially larger than the commonly adopted $\alpha_{\rm CO}=0.8$ for high-\textit{z} DSFGs. 

In the context of \aco tension, we compare the resulting gas masses to dynamical constraints derived from both the isotropic ``virial" estimator ($\S$~\ref{subsec:mdynisovir}) and the proposed ``mixed" dynamical mass estimator ($\S$~\ref{subsec:mdynmixed}) that assumes DSFGs as rotating, pressure-supported thick disks. Our main findings are:

\begin{enumerate}[noitemsep]

    \item The isotropic virial estimator yields dynamical upper limits consistent with the independently derived gas masses once plausible cold-gas sizes are adopted (Figure~\ref{fig:virial_mdyn}), demonstrating that the available dynamical constraints do not strictly require \aco= 0.8.

    \item In contrast, the unresolved ``mixed'' dynamical estimator systematically underestimates dynamical masses, producing artificially low $\alpha_{\rm CO}$ upper limits (Table~\ref{tab:aco3methods_table}). Using the archetype GN20 as a case study ($\S$~\ref{subsec:gn20}), we show that incorporating resolved gas geometry, inclination, and the partition between ordered rotation and turbulent dispersion recovers the missing dynamical mass using the proposed ``mixed" estimator, bringing the dynamical constraints into agreement with the LVG-derived $\alpha_{\rm CO}$ \citep{boogard2026}.

    \item Under the adopted assumptions of a fixed gas-to-dust ratio (G/D = 100) and dust opacity normalization ($\kappa_{\nu}$; Table~\ref{tab:tunerlvg}), the TUNER model constrains the CO abundance required to reproduce the observed CO SLEDs. These results, therefore, suggest that uncertainties in ISM modeling and dynamical mass estimates may contribute substantially to the apparent global \aco discrepancy. Leaving the G/D ratio free in TUNER requires 9/12 DSFGs to have a G/D ratio between 200 and 380, which appears unrealistic for such dusty sources \citep[see][]{dunne+2022,Gururanjan2023}. Therefore, TUNER results presented here do not directly constrain the gas metallicity under the G/D = 100 assumption.

    \item TUNER modeling reveals a multi-phase molecular ISM in high-\textit{z} DSFGs: low-\textit{J} CO traces the most extended diffuse reservoir with LVG-inferred sizes $\sim 2\times$ larger than the observed 1~mm dust (Figure~\ref{fig:acoratio_rratio}), while higher-\textit{J} CO lines probe compact, dense star-forming gas that contains a substantial fraction of the total molecular mass. This naturally explains the elevated \aco $>$ 0.8, consistent with expectations for local ULIRGs \citep{papadopoulos2012b}.

\end{enumerate}

In summary, our analysis indicates that the available data do not require the commonly adopted \aco= 0.8 for high-\textit{z} DSFGs. Instead, intermediate ($\sim 2-3$) to near-Galactic values remain dynamically viable once plausible cold molecular gas extents and geometric uncertainties are taken into account. The conclusion is conditional on the adopted assumptions regarding the gas-to-dust ratio, dust opacity, gas geometry, and unresolved kinematics, but demonstrates that current dynamical constraints do not uniquely favor a low conversion factor such as \aco = 0.8 in these systems. 

Robust calibration of $\alpha_{\rm CO}$ will require physically motivated molecular gas mass estimates from multi-component ISM models such as TUNER, together with resolved measurements of cold-gas geometry and kinematics. Improved constraints on dust properties and gas-to-dust ratio in high-\textit{z} DSFGs will further complement these. While resolved CO(1--0) observations remain observationally demanding until the advent of the ngVLA, resolved studies of complementary tracers such as [CII], [CI], and $J=2-4$ CO lines can provide valuable intermediate constraints on the gas dynamics and molecular gas distribution. Establishing such observational benchmarks will be essential for robustly calibrating the global $\alpha_{\rm CO}$ of massive galaxies in the early universe.

\section*{Acknowledgments}
P.P., A.W., D.R., and E.R.D. acknowledge the Collaborative Research Center 1601 (SFB 1601 sub-projects C1, C2, C3, C5, and C6) funded by the Deutsche Forschungsgemeinschaft (DFG, German Research Foundation) -- 500700252. L.A.B. acknowledges support from the Dutch Research Council (NWO) under grant VI.Veni.242.055 (\url{https://doi.org/10.61686/LAJVP77714}). H.S.B.A. gratefully acknowledges support from Academia Sinica through grant AS-PD-1141-M01-2.

\bibliography{sample701}{}
\bibliographystyle{aasjournalv7}

\section{appendix} \label{appendix}

\subsection{Additional Unlensed Fields in the \vzgal} \label{app:sources_fig}

\begin{figure}[h]
    \centering
    \includegraphics[width=\textwidth]{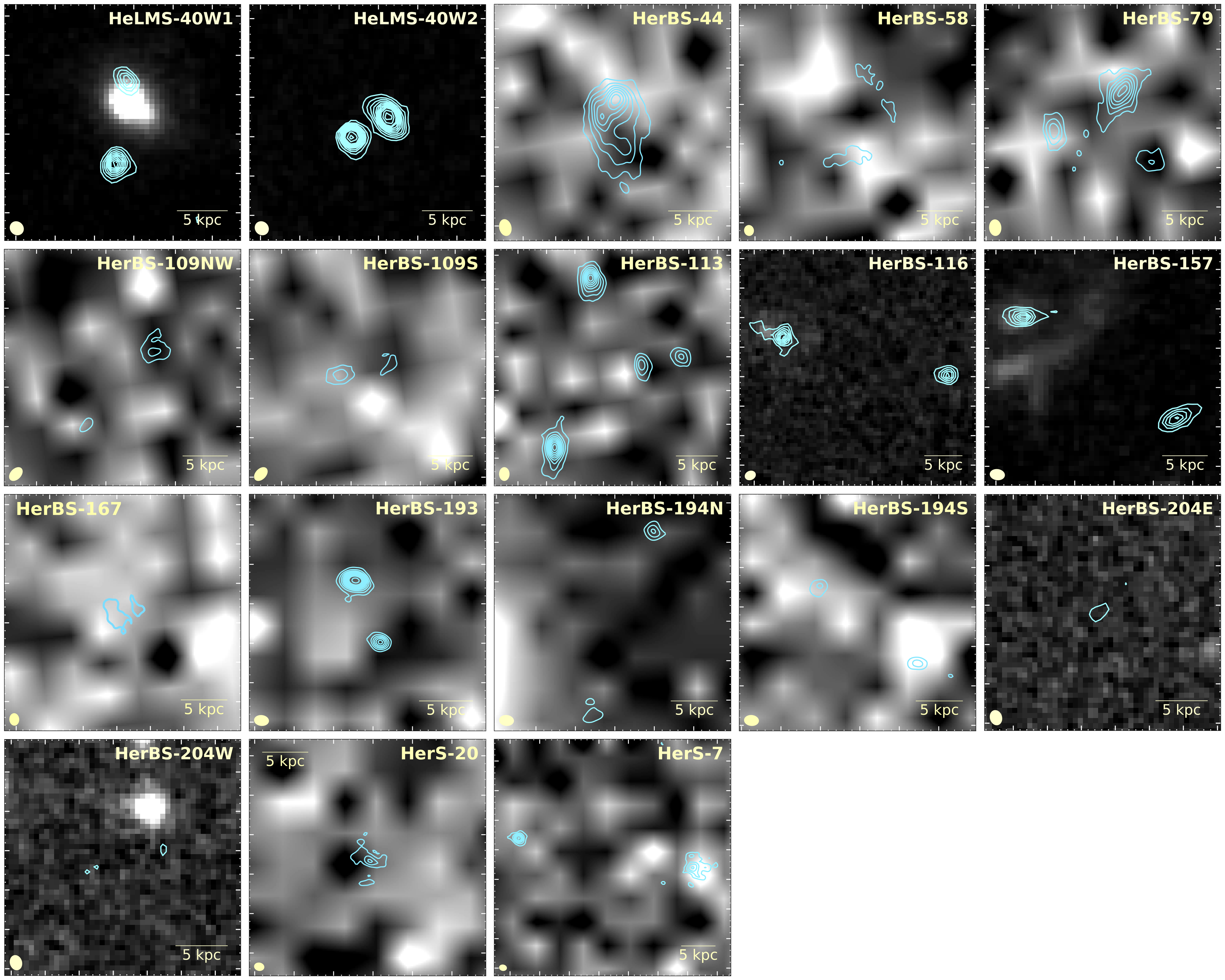}
\caption{Same as Figure~\ref{fig:source_grid}; however, with unlensed \vzgal fields with potentially multiple sources within the VLA beam.}
\label{fig:source_grid_unused}
\end{figure}

Figure~\ref{fig:source_grid_unused} shows the remaining unlensed \vzgal fields that exhibit multiple dust continuum components in the ALMA 1~mm imaging. Since the corresponding CO(1--0) emission remains unresolved within the VLA beam, the molecular gas cannot be reliably associated with individual dust components. These systems are therefore excluded from the present analysis. Details on the nature of these objects will be discussed by Bakx et al. (in prep.).

\clearpage

\subsection{\coonezero Spectra of DSFGs with Complex Line Profiles} \label{app:lineprofiles}

\begin{figure}[H]
\centering
\includegraphics[height=0.16\textwidth]{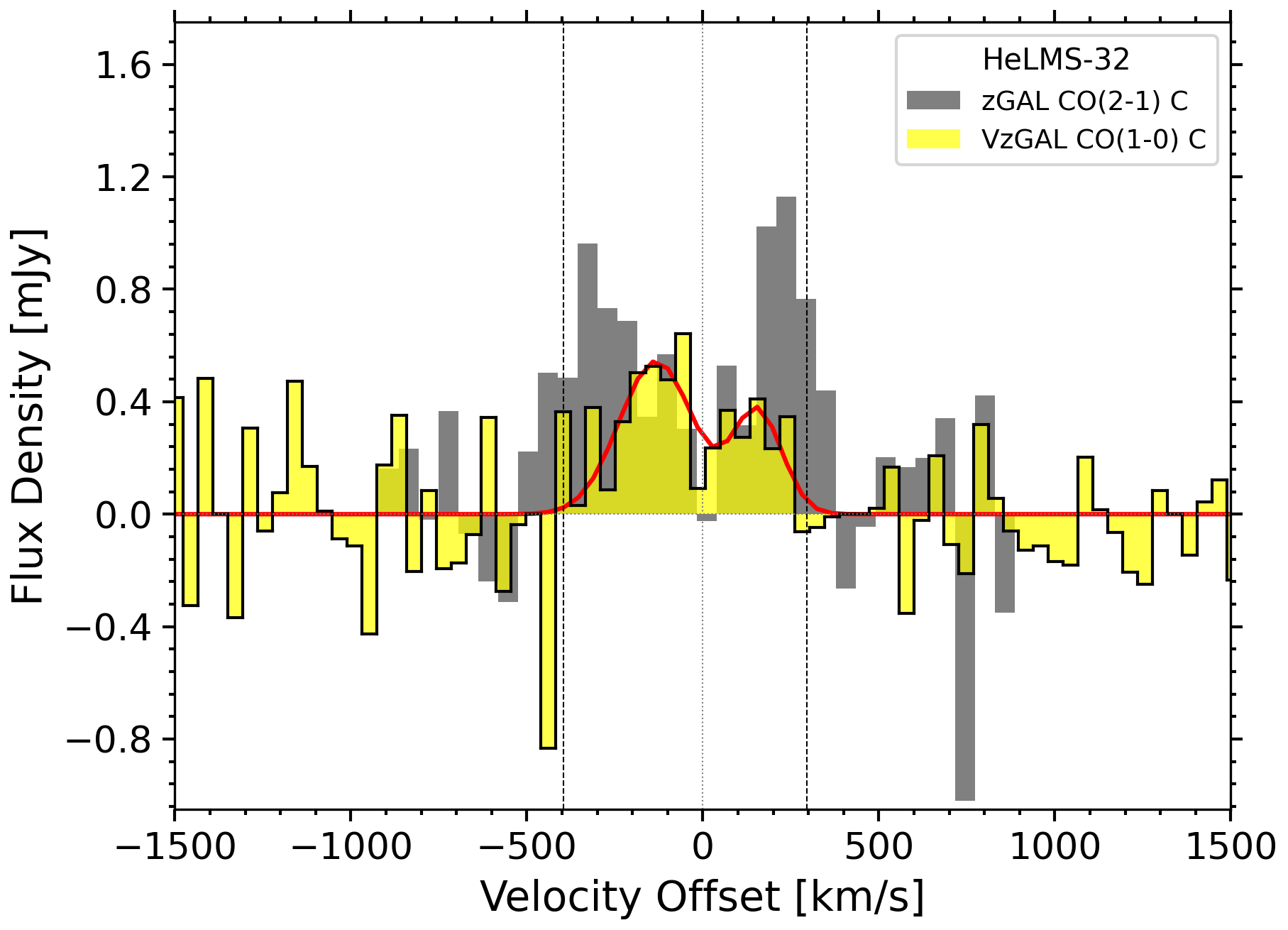}
\includegraphics[height=0.16\textwidth]{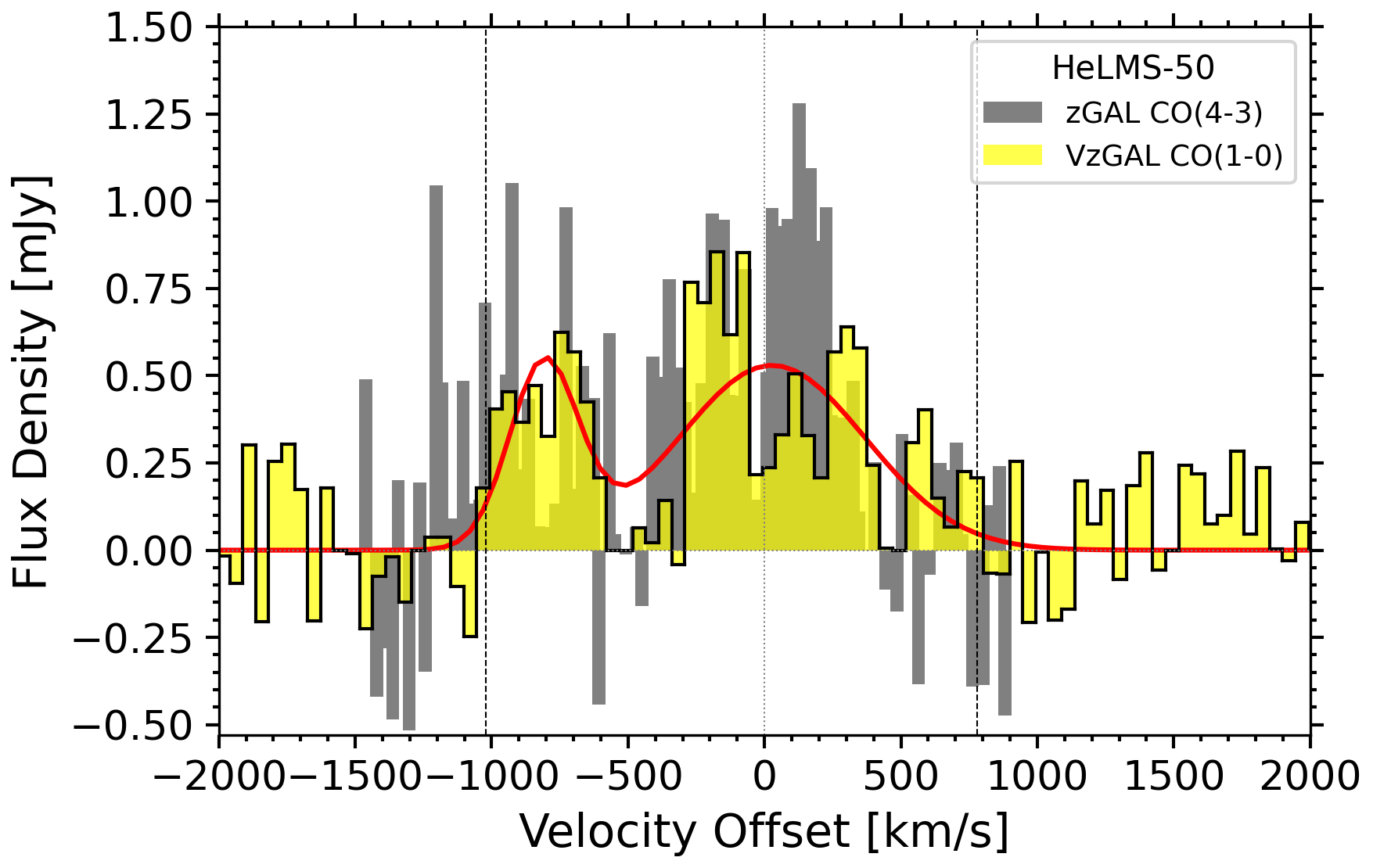}
\includegraphics[height=0.16\textwidth]{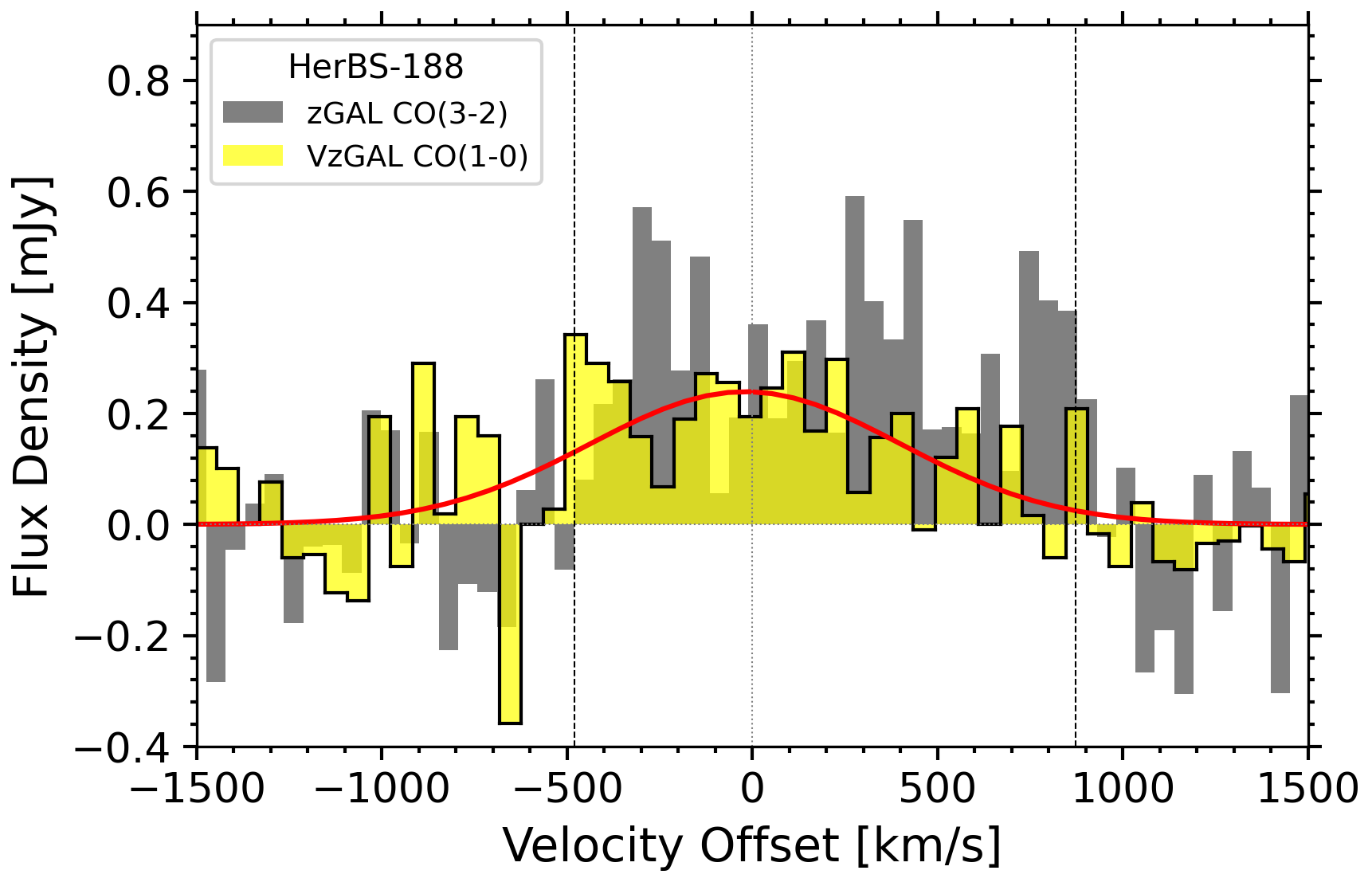}
\includegraphics[height=0.16\textwidth]{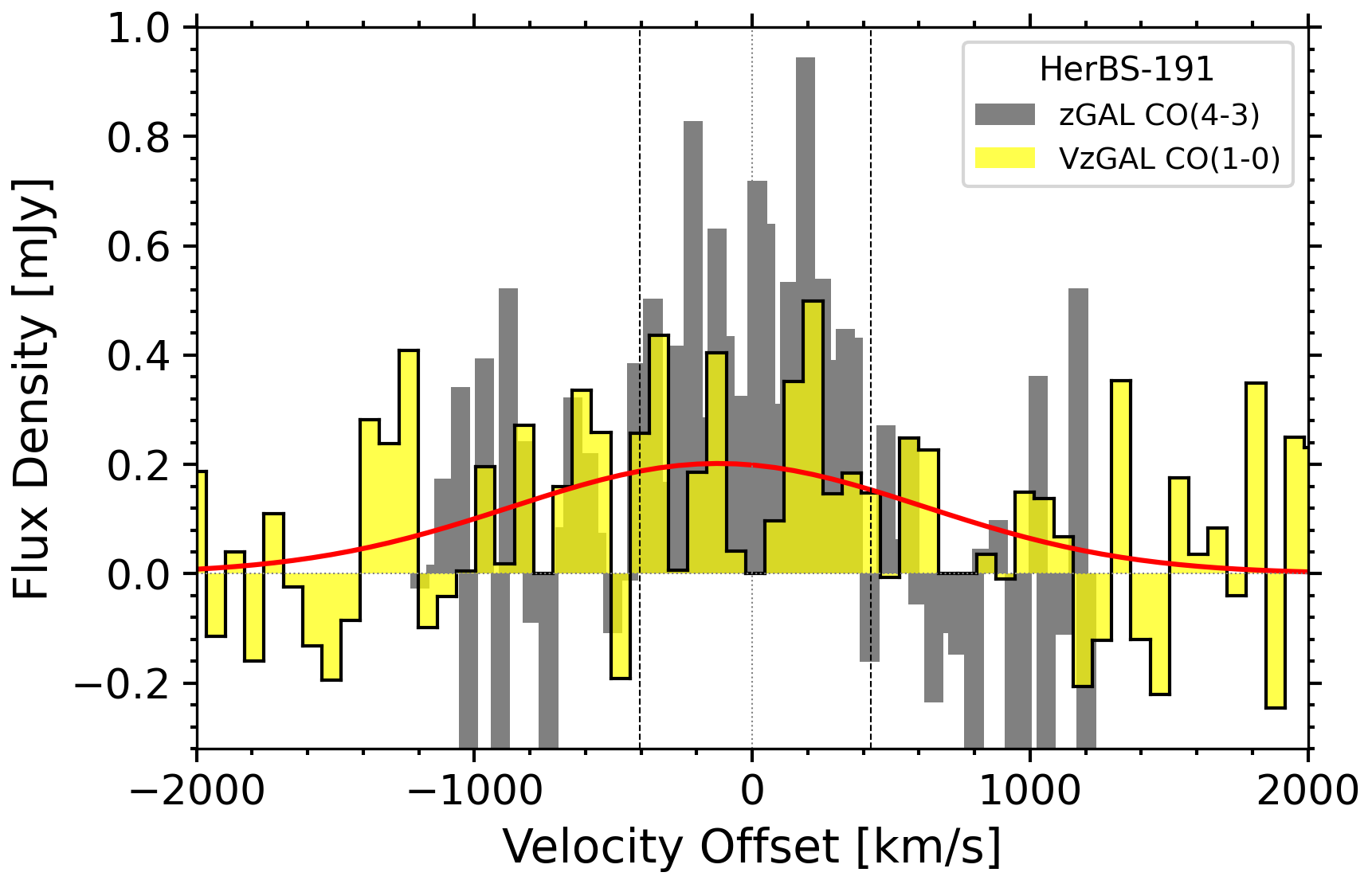}
\includegraphics[height=0.16\textwidth]{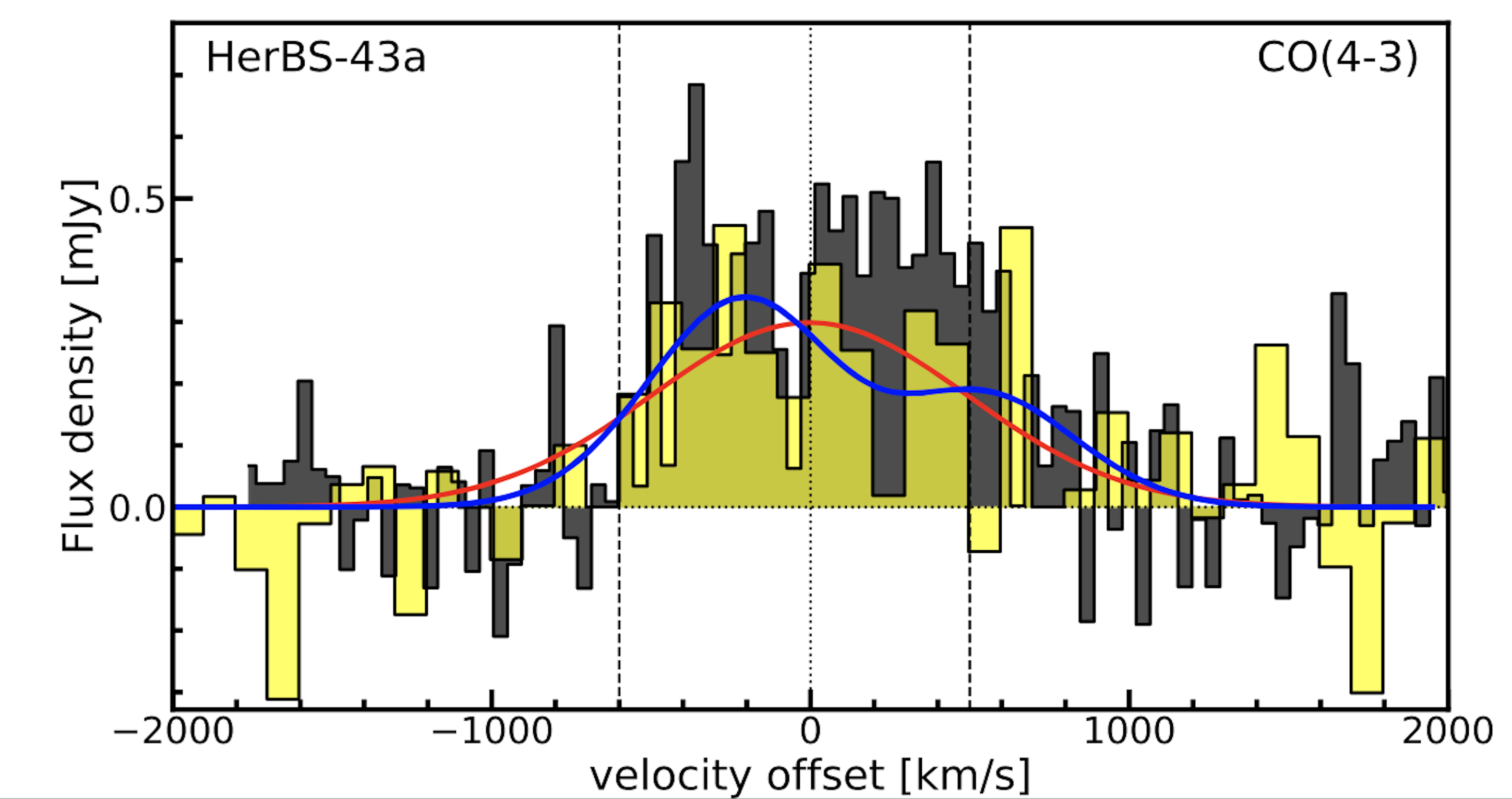}
\includegraphics[height=0.16\textwidth]{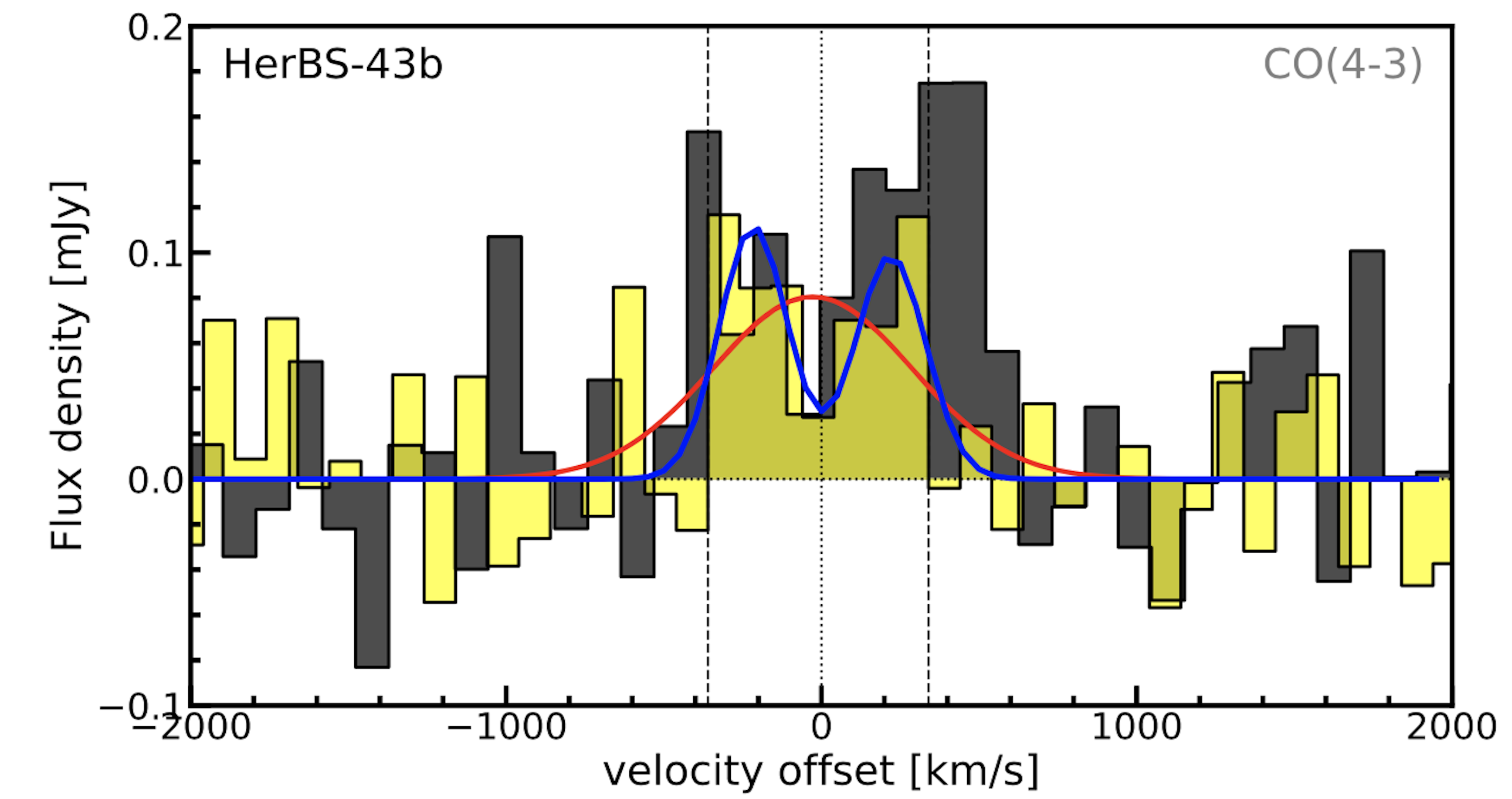}
\caption{\vzgal \coonezero spectra (yellow) of Group-2 targets that show complex \coonezero line profiles (see $\S$~\ref{subsec:co10obs} for more details). In gray are the higher-\textit{J} CO line profiles for each source to compare line shapes. The upper panel is adopted from \cite{prajapati2026}, and the pilot sources in the lower panel are from \cite{stanley+2023}.}
\label{fig:lineprofiles}
\end{figure}

CO spectra of sources with complex \coonezero line profiles, discussed in $\S$~\ref{subsec:co10obs}, are shown in Figure~\ref{fig:lineprofiles}. For HeLMS-32C and HerBS-191, we notice a difference in the shapes of \coonezero and higher-\textit{J} CO emission spectra, leading to approximately a factor-2 under- and over-estimation of their FWHM using CO(1--0). As we will discuss in the Appendix $\S$~\ref{subsec:allcontriaco}, Figure~\ref{fig:allparams_aco}, this could lead to $\sim 0.4$~dex over- and under-estimation of their \aco conversion factor, respectively.

\clearpage

\subsection{Radiative Transfer Comparison for Different Tracers of the Galaxy Extents} \label{app:sizeRT}

For optically thin dust emission, the specific intensity is given by:

\begin{equation}
    I_\nu^{\rm dust}(R_{\rm dust}) \simeq \kappa_\nu \times \Sigma_{\rm dust}(r) \times B_\nu[T_{\rm dust}(r)]
    \label{eqn:dustRT}
\end{equation}

which in the Rayleigh--Jeans limit reduces to \(I_\nu^{\rm dust} \propto \Sigma_{\rm dust}(r) \times T_{\rm dust}(r)\). Assuming a roughly constant gas-to-dust mass ratio (G/D), we would have \(\Sigma_{\rm dust}(r)\propto\Sigma_{\rm gas}(r)\). If both gas surface density and temperature decline with radius (e.g. \(\Sigma_{\rm gas} \propto e^{-r/R_{\rm gas}}\), \(T_{\rm dust} \propto e^{-r/R_{\rm Td}}\)), the observed continuum becomes more centrally concentrated with an effective scale length \(R_{\rm dust}^{-1} = R_{\rm gas}^{-1} + R_{\rm Td}^{-1}\), naturally yielding \(R_{\rm dust} < R_{\rm gas}\) even if dust traces the underlying molecular gas. In contrast, CO(1--0) emission is typically optically thick, with brightness temperature:

\begin{equation}
T_B \simeq \left[J_\nu(T_{\rm ex}) - J_\nu(T_{\rm CMB})\right](1 - e^{-\tau_{\rm CO(1-0)}})
\label{eqn:co10RT}
\end{equation}

such that for \(\tau_{\rm CO(1-0)} \gg 1\), \(T_B \approx J_\nu(T_{\rm ex}) - J_\nu(T_{\rm CMB})\) and becomes only weakly sensitive to molecular column density. Instead of tracing mass surface density directly, CO(1--0) primarily traces the projected emitting area, excitation conditions, velocity structure, and filling factor of the molecular gas. As a result, CO(1--0) can remain detectable in cooler and more diffuse gas where the dust continuum has faded substantially, leading to the commonly observed hierarchy \citep[see also][]{rivera2018,boogard2026}:

\begin{equation}
    R_{\rm dust}  \, <  \, R_{\rm CO(higher-J)} \, \lesssim \,  R_{\rm CO(1-0)}  \, \sim  \, {R_{\rm cold-gas}}
\label{eqn:radius_ineq}
\end{equation}

\clearpage

\subsection{Regarding randomly oriented disks} \label{app:s-rand}

For inclination defined in the usual sense ($0 \leq i \leq \pi/2$, with
$i = 0$ for a face-on disk and $i=\pi/2$ for an edge-on disk), the probability
that a randomly oriented disk will be observed to have a given inclination 
will be $p(i)\,di = {\rm sin}\,i\,di$.  Intuitively, if we station ourselves 
at the center of a disk galaxy, we find that all remote observers within a 
solid angle $2\pi\,{\rm sin}\,i\,di$ (out of a total $2\pi$) steradians on the 
sky will perceive {\it our} disk to have an inclination $i$.  The result for 
$p(i)$ then follows from the symmetry between one galaxy observed by an 
ensemble of astronomers and one astronomer observing an ensemble of galaxies.
$p(i)$ satisfies the normalization condition required of a probability
distribution, namely that when integrated over the full range of values taken
by $i$,
\begin{equation}
\int^{\pi/2}_0 p(i)\,di = \int^{\pi/2}_0 {\rm sin}\,i\,di = 1
\end{equation}
As a consequence, for any function $f(i)$ with the property that
$f(i)\,{\rm sin}\,i$ is nowhere infinite over the interval $[0,\pi/2]$, the
mean value of $f(i)$ for a random distribution of inclinations will be
\begin{equation}
\left<f(i)\right> = \int^{\pi/2}_0 f(i)\,{\rm sin}\,i\,di
\end{equation}
For the reader's convenience, we review a few idealized examples that occur in
the context of galaxy kinematics.  For an ensemble of randomly oriented disks,
the mean value of the ratio between the intrinsic velocity of an outflow
perpendicular to the disk and its observed line-of-sight component will be
\begin{equation}
\left<{\frac {v_{\rm out,los}}{v_{\rm out}}}\right> = \left<{\rm
cos}\,i\right> = {\frac 1 2}
\end{equation}

The mean ratio between the rotational velocity in the plane 
of the disk and its line-of-sight component will be
\begin{equation} \label{e-sin}
\left<{\frac {v_{\rm rot,los}}{v_{\rm rot}}}\right> = \left<{\rm
sin}\,i\right> = {\frac \pi 4}
\end{equation}
while the mean correction for inclination to a simple dynamical mass estimate
will be
\begin{equation} \label{e-sin2}
\left<{\frac {R v_{\rm rot,los}^2}{G M_{\rm dyn}}}\right> =
\left<{\rm sin}^2\,i\right> = {\frac 2 3}
\end{equation}
Noting that $\left<i\right> = 1$ (i.e., $57.3^\circ$), it is clear that in
general, $\left<f(i)\right> \neq f(\left<i\right>)$ and $\left<(f(i))^2\right>
\neq \left<f(i)\right>^2$.

In many situations, we are less interested in estimating $\left<f(i)\right>$
than in determining the form of the relationship $y = f(i,x_1,x_2,\dots)$ 
for an observed quantity $y$ that depends on inclination and (perhaps) a number 
of intrinsic galaxy parameters $\{x_1,x_2,\dots\}$.  In this context, it 
is convenient to exploit the availability of a statistical test --- the 
Kolmogorov-Smirnov (K-S) test --- for comparing theoretical and observed 
cumulative distribution functions (CDFs).  For measured $\{Y_1,Y_2,\dots\}$, 
the observed CDF($Y$) is defined as the fraction of the sample with $Y_i < Y$.
If $f(i,x_1,x_2,\dots)$ is a monotonically increasing function of $i$, the 
theoretical CDF($y$) takes a particularly simple form:
\begin{equation} \label{e-cdfup}
{\rm CDF}(y) = \int^{i_{\rm up}}_0 p(i)\,di  = \int^{i_{\rm up}}_0 {\rm 
sin}\,i\,di = 1 - {\rm cos}\,i_{\rm up}
\end{equation}
where the upper bound $i_{\rm up}$ is implicitly defined by
\begin{equation} \label{e-iup}
y = f(i_{\rm up},x_1,x_2,\dots)
\end{equation}
Likewise, if $f(i,x_1,x_2,\dots)$ is a monotonically decreasing function of 
$i$, we have 
\begin{equation} \label{e-cdflow}
{\rm CDF}(y) = \int^{\pi/2}_{i_{\rm low}} p(i)\,di = \int^{\pi/2}_{i_{\rm 
low}} {\rm sin}\,i\,di = {\rm cos}\,i_{\rm low}
\end{equation}
in terms of a lower bound $i_{\rm low}$ implicitly defined by
\begin{equation} \label{e-ilow}
y = f(i_{\rm low},x_1,x_2,\dots)
\end{equation}
Many problems of interest fall into one of these two simple categories; for
example, if a galaxy's observed line-of-sight velocity width $\Delta v$ is due
to a combination of a velocity dispersion term $v_{\rm disp}$ and a rotation term
$v_{\rm rot}\,{\rm sin}\,i$ added in quadrature
\begin{equation} \label{e-dv}
\Delta v\,(i,v_{\rm disp},v_{\rm rot}) \approx \Big(v_{\rm disp}^2 + v_{\rm 
rot}^2\,{\rm sin}^2 i\Big)^{\frac 1 2}
\end{equation}
then as a function of $i$, $\Delta v$ is monotonically increasing.  Solving
Equation \ref{e-iup} for $i_{\rm up}$ then allows us to substitute into Equation
\ref{e-cdfup} (as, in a different context, solving Equation \ref{e-ilow} for 
$i_{\rm low}$ would allow us to substitute into Equation \ref{e-cdflow}).  The
resulting CDF($y$) can then be compared to the observed CDF($Y$), for different
choices of $\{x_1,x_2\} = \{v_{\rm disp},v_{\rm rot}\}$, using the K-S test.  In
the present paper, we are merely interested in whether the inclinations we have
inferred for our DSFG sample are consistent with the expectations for a random
distribution (i.e., there are no other relevant parameters
$\{x_1,x_2,\dots\}$).  Thus, $f(i) = i$, which (trivially) increases
monotonically with $i$, and we can use Equation \ref{e-cdfup} to calculate a
theoretical CDF($y$). 

\begin{figure}[h]
\centering
\includegraphics[width=0.45\textwidth]{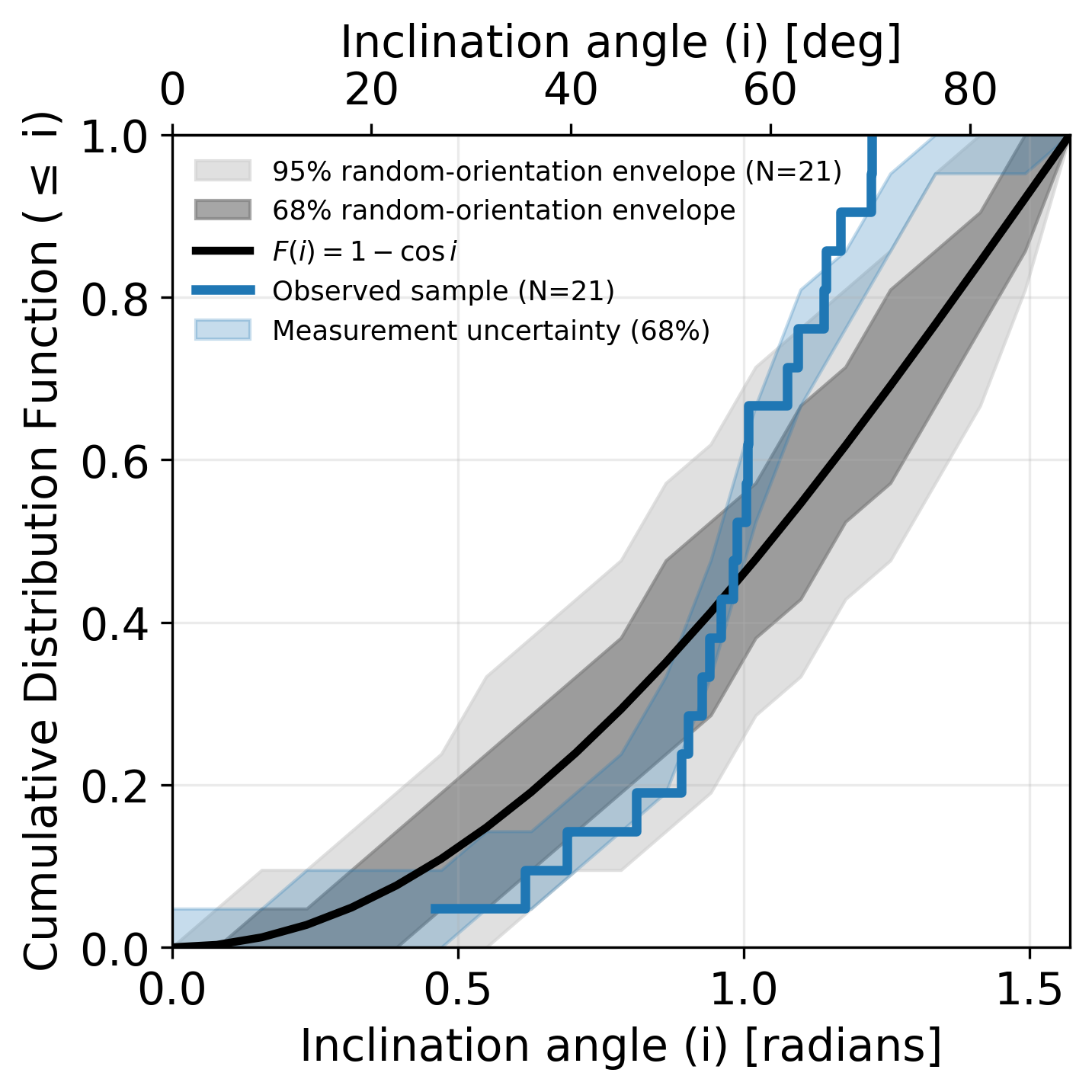}
\caption{Kolmogorov-Smirnov-Test or KS-test showing the cumulative distribution of dust inclination angles in blue. The shaded area in gray presents the expected spread for a sample of 21 randomly oriented inclinations, which ideally follows $1- {\rm cos}(i)$ for a sufficiently large sample size.}
\label{fig:incl_cdf_kstest}
\end{figure}

In Figure~\ref{fig:incl_cdf_kstest}, we show the observed CDF of our 21 \vzgal DSFGs considered for the analysis in this paper (Table~\ref{tab:dynamicalmasses}). The observed inclination distribution shows a moderate deviation from the expectation for randomly oriented disks (Kolmogorov-Smirnov or KS statistics; $D=0.34$ and $p=0.012$), which would formally reject the null hypothesis of random orientations at the ($\sim 2.5\sigma$) level. However, when measurement uncertainties in the axis ratios and derived inclinations are propagated through Monte Carlo realizations, the median KS test probability increases to ($p=0.061$) (16th-84th percentile range: 0.02--0.16). Therefore, the evidence against a random inclination distribution is weakened considerably, and the current sample of 21 galaxies remains at best only marginally inconsistent with random orientations. Furthermore, the inferred inclinations are not direct observables but are derived from dust-continuum axis ratios under assumptions regarding the intrinsic disk thickness ($\rm q_0$) and the correspondence between dust morphology and disk orientation. Given these additional systematic uncertainties, the current data do not provide compelling evidence against a random orientation distribution, and the hypothesis of randomly oriented disks cannot be robustly rejected.\footnote{Note also that dust geometry may not truly represent the cold gas disk and its inclination.}

To give proper historical credit, we note that Equations \ref{e-sin} and 
\ref{e-sin2} are specific examples of a more general result derived by 
\citet{chandrasekhar1950}: given an intrinsic quantity $x$ that is
related to the observed quantity $y \equiv x\,{\rm sin}\,i$ for inclination
$i$, the moments of the probability distribution $p(x)$ can be recovered from
the moments of the observed distribution $p(y)$ according to
\begin{eqnarray}
\overline{x} & = & {\frac 4 \pi}\,\overline{y} \\
\overline{x^2} & = & {\frac 3 2}\,\overline{y^2} \\
\overline{x^3} & = & {\frac {16}{3\pi}}\,\overline{y^3}
\end{eqnarray}
As the first three moments do not capture all information about a probability
distribution, \cite{chandrasekhar1950} go on to derive exact relations
between intrinsic $p(x)$ and observed $p(y)$ for various specific forms of
$p(x)$, including a delta function $p(x) = \delta(x-x_1)$.  The approach we
have outlined above is complementary: by effectively fixing $p(x) \equiv
\delta(x-x_1)$ (and thereby gaining the ability to generalize to multiple
parameters), it becomes possible to evaluate many possible forms of $y=f(i,x)$
that differ from the simple $y = x\,{\rm sin}\,i$.

\clearpage

\subsection{SED fits and Corner Plots from the TUNER LVG modeling} \label{herbs165_corner}

\begin{figure}[h]
\centering
\includegraphics[width=\textwidth]{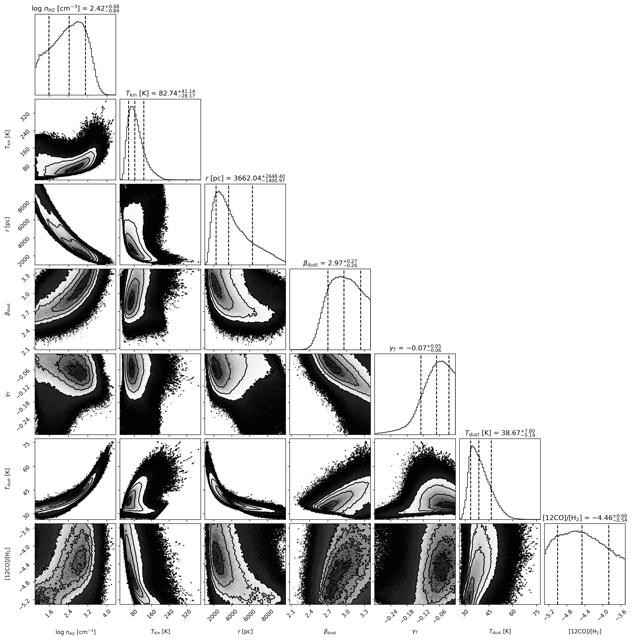}
\caption{HerBS-51 as an example to show TUNER-derived corner plots \citep{ForemanMackey2016} of the posterior for the model (Figure~\ref{fig:tuner_herbs51}). The top panels show the 1D marginalized posteriors with the ${50}^{\rm th}$ (median), ${16}^{\rm th}$, and ${84}^{\rm th}$ percentiles, while the lower panels show the 2D marginalized covariances.}
\label{fig:herbs51corner}
\end{figure}

Corner plots of the posteriors for the TUNER model consist of the explored free parameters, namely ISM density, kinetic temperature, LVG-inferred radius of the ISM ($R_{\rm LVG}$), $\beta_{\rm dust}$, $\gamma_{\rm T}$, dust temperature, and $^{12}$CO-to-H$_2$ abundance, for each source. In Figure~\ref{fig:herbs51corner}, we show an example of HerBS-51. Refer to $\S$~\ref{subsec:gasmass_tuner} and Table~\ref{tab:tunerlvg} for exploring the results for the 12/21 sources explored with TUNER. 

Further, Figure~\ref{fig:sedfits} shows TUNER-fitted dust and CO SEDs for all the remaining 11 sources, similar to those shown for HerBS-51 in Figure~\ref{fig:tuner_herbs51}.

\begin{figure}[H]
\centering
\includegraphics[width=0.6\textwidth]{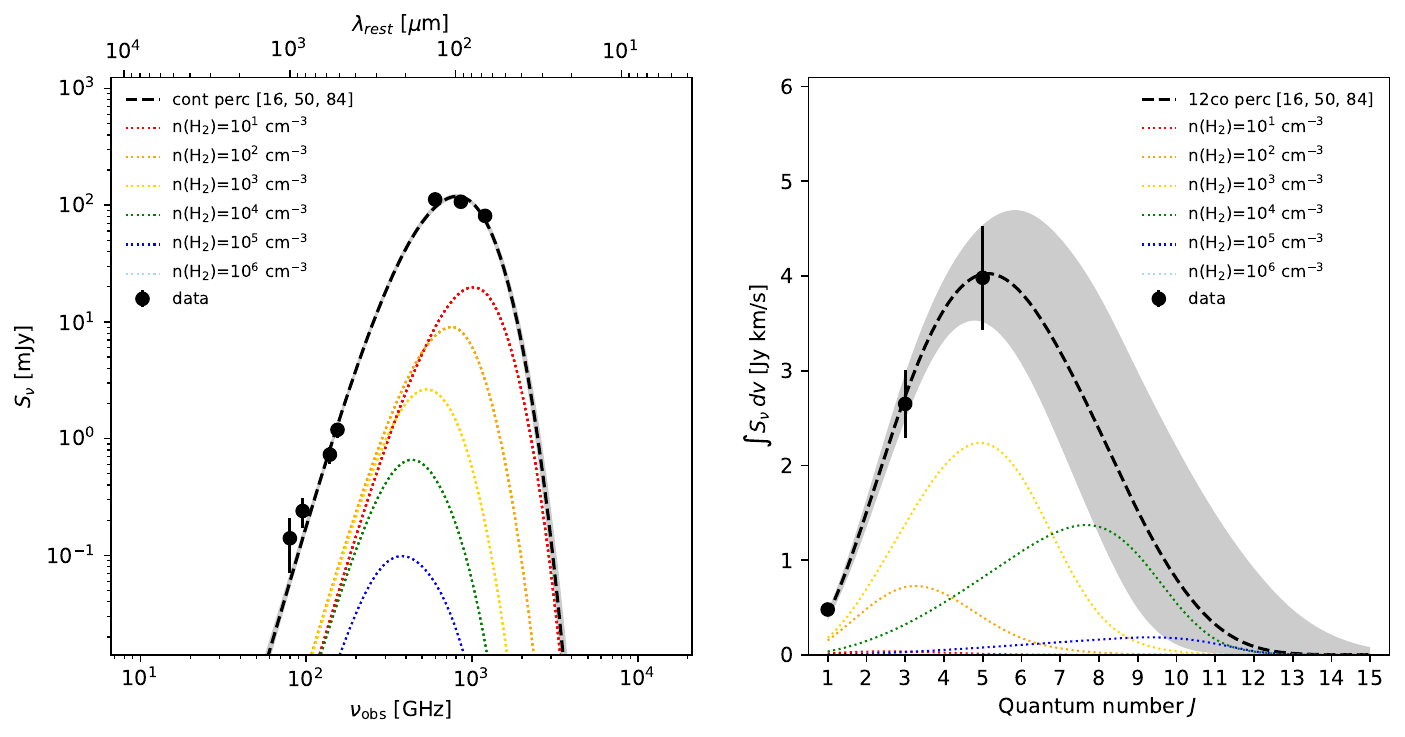}
\includegraphics[width=0.6\textwidth]{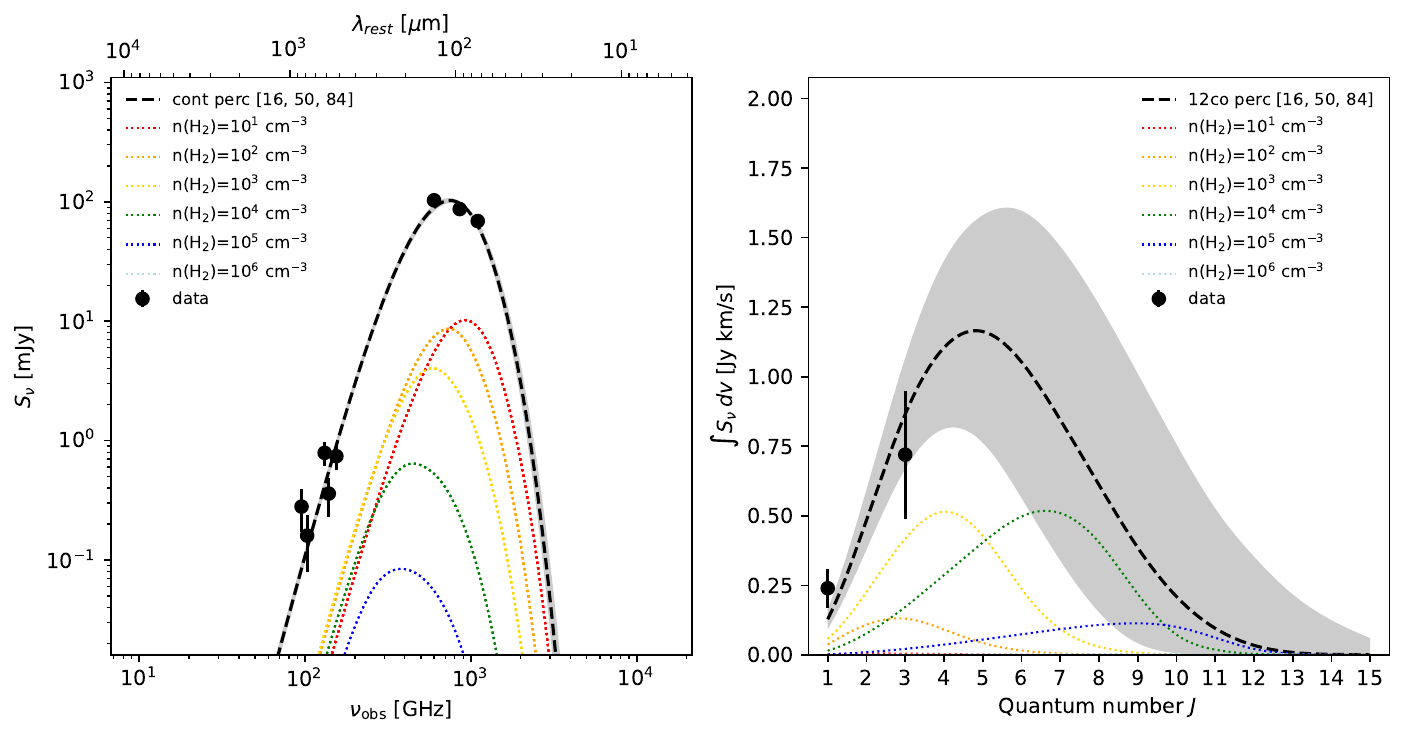}
\includegraphics[width=0.6\textwidth]{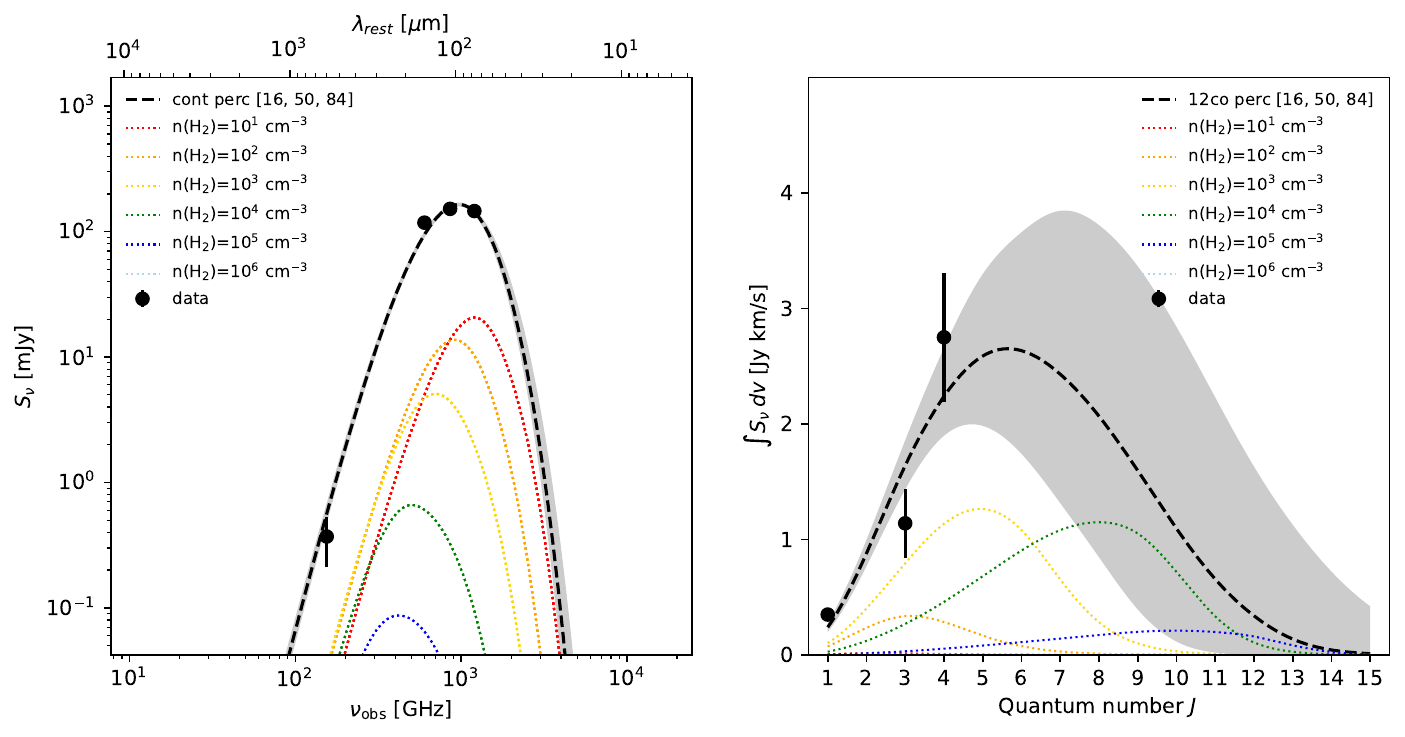}
\includegraphics[width=0.6\textwidth]{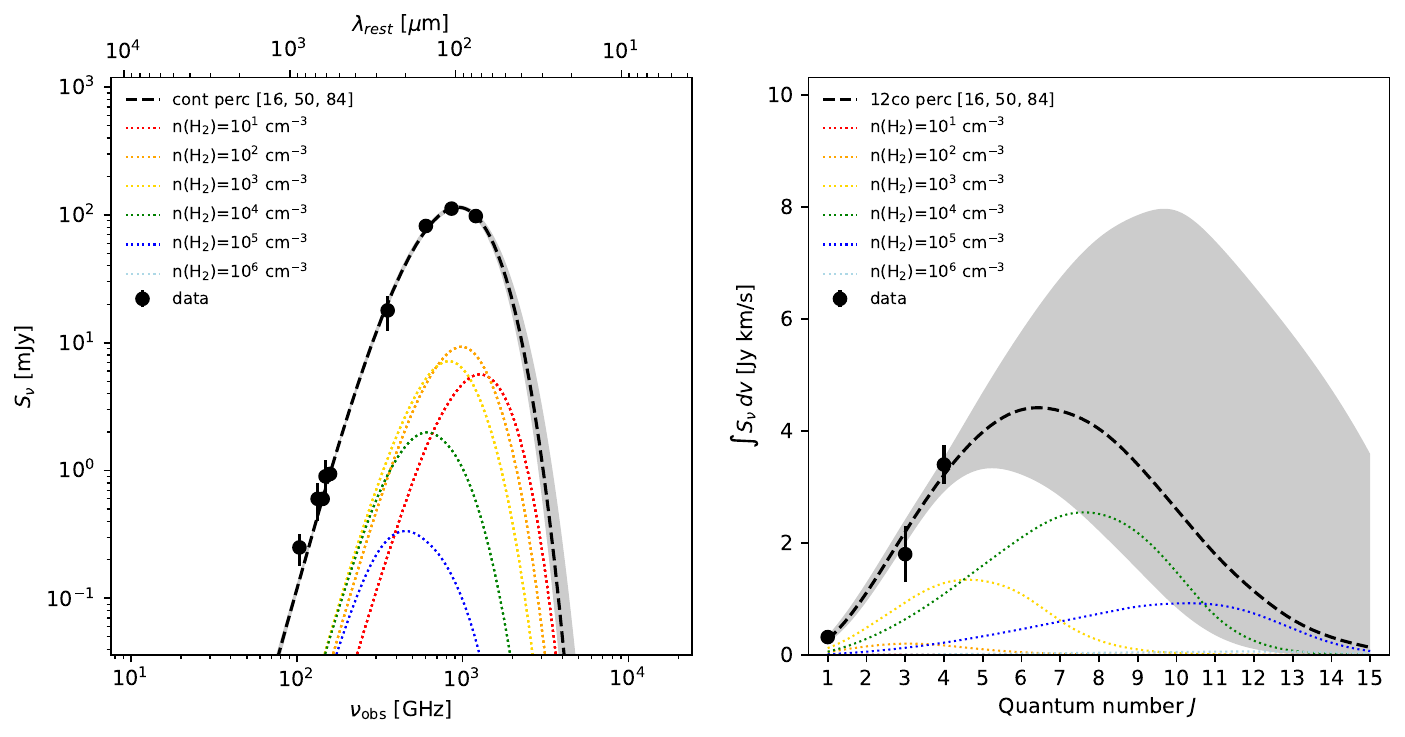}

\caption{Similar to Figure~\ref{fig:tuner_herbs51} for HerBS-51, here shown are the dust and CO SED fits from TUNER for Group-1 sources, namely HeLMS-39, HeLMS-54, HerS-8, and HerBS-70E.}
\label{fig:sedfits}
\end{figure}

\begin{figure}[H]
    \centering
\includegraphics[width=0.6\textwidth]{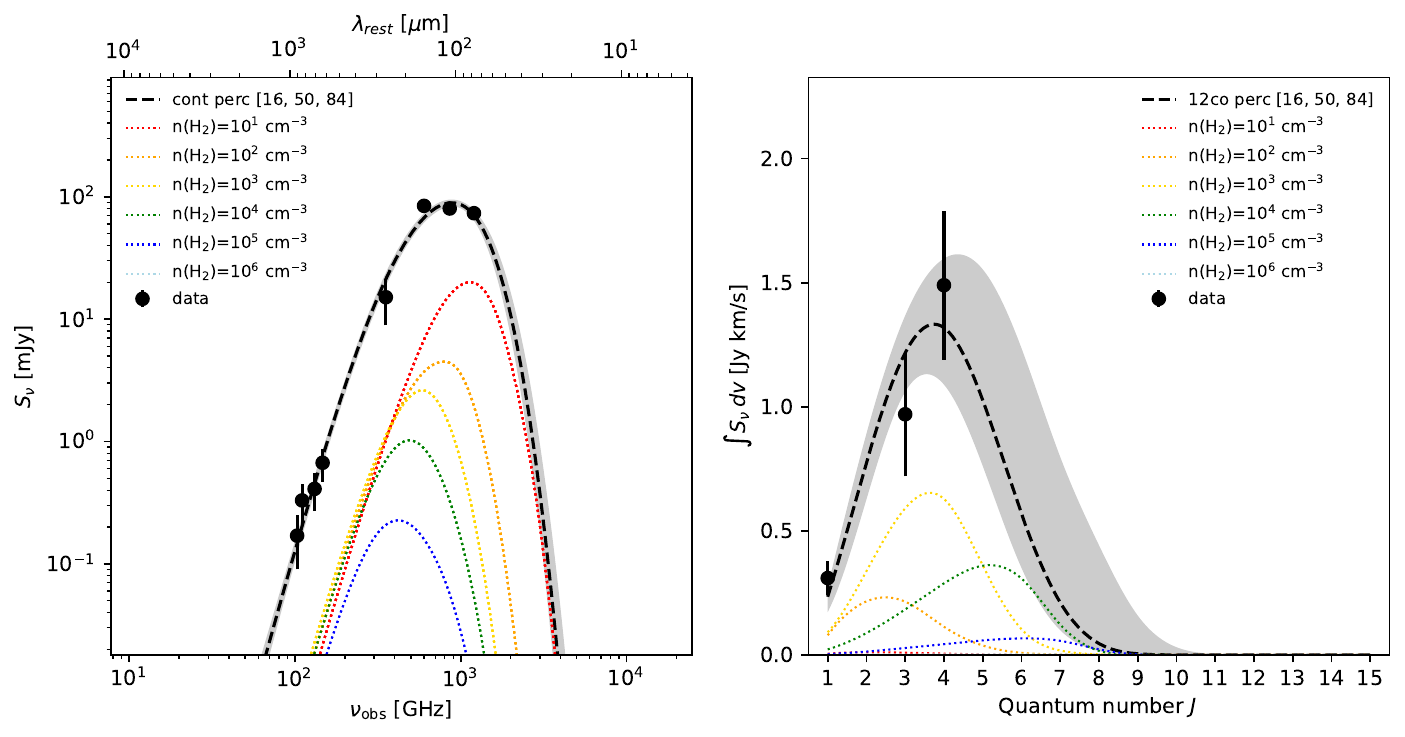}
\includegraphics[width=0.6\textwidth]{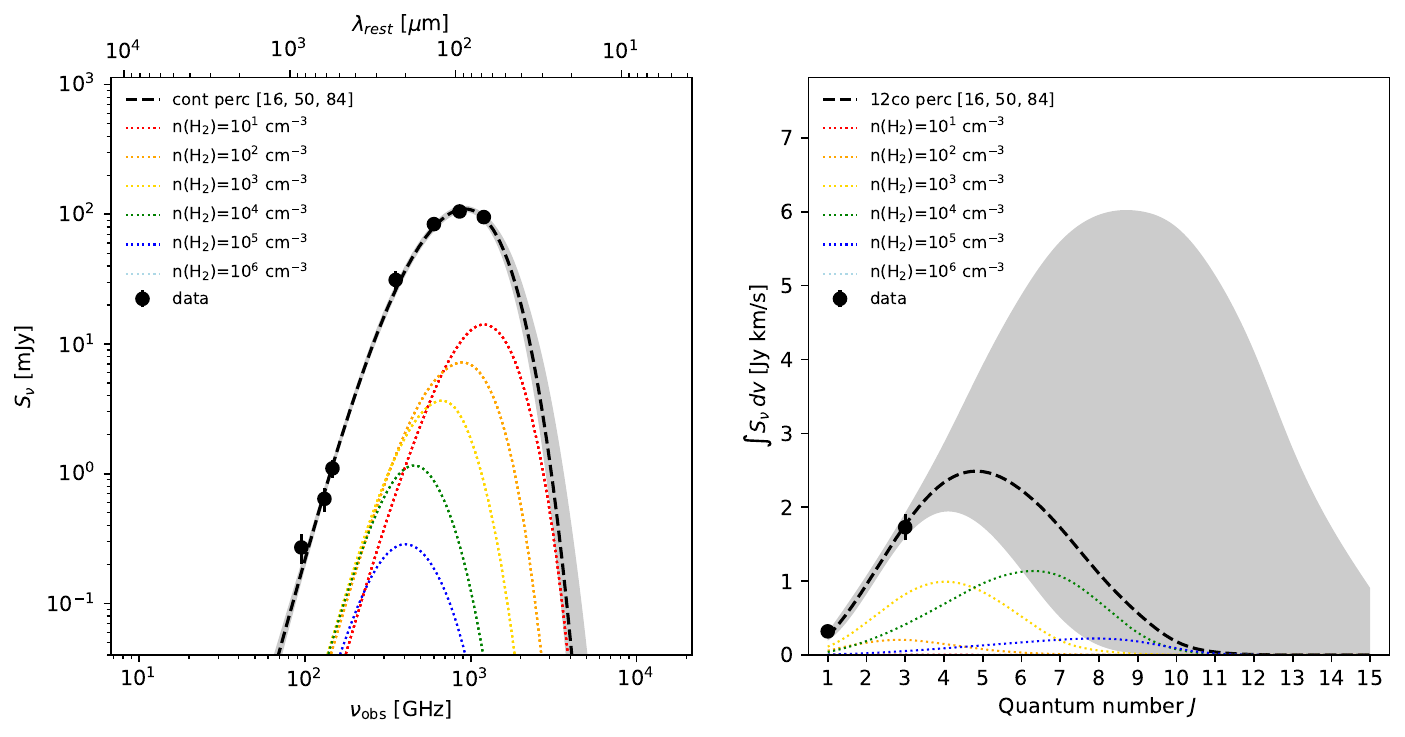}
\includegraphics[width=0.6\textwidth]{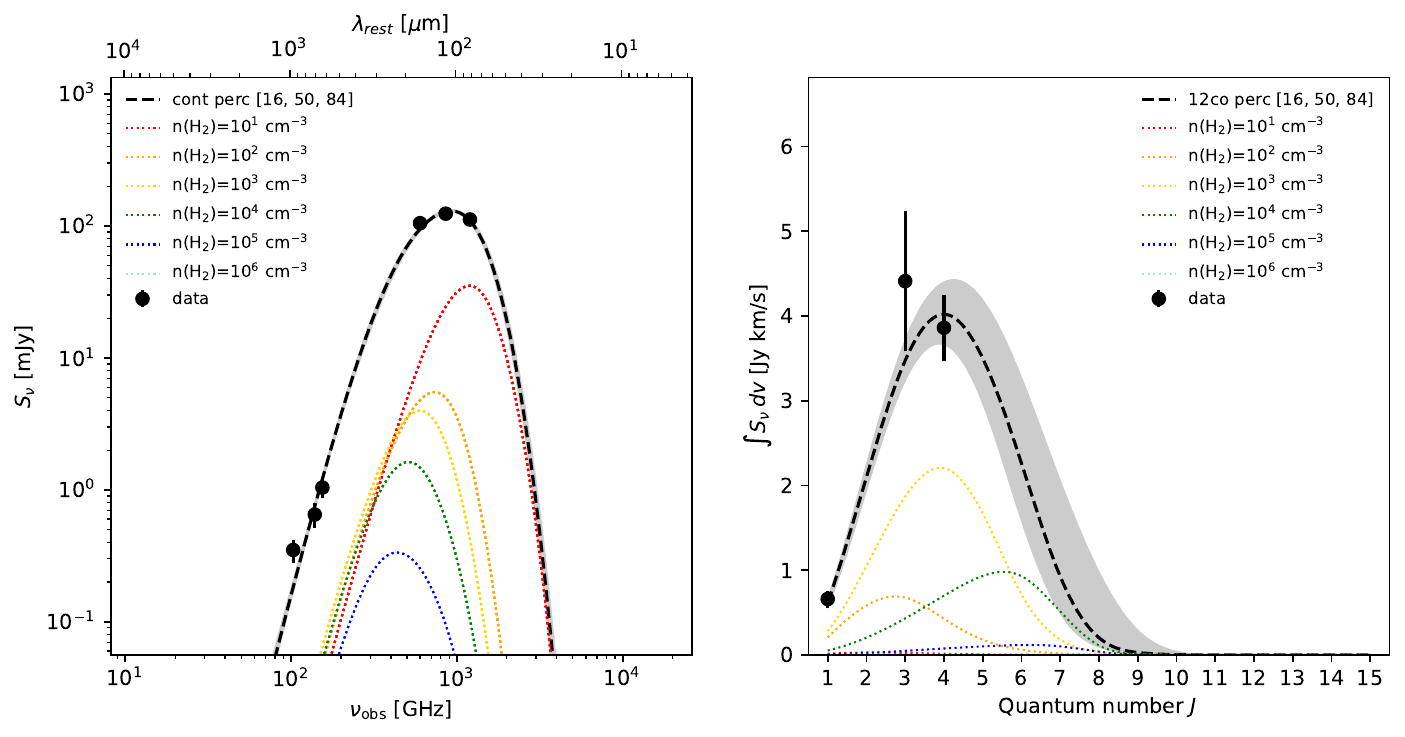}
\includegraphics[width=0.6\textwidth]{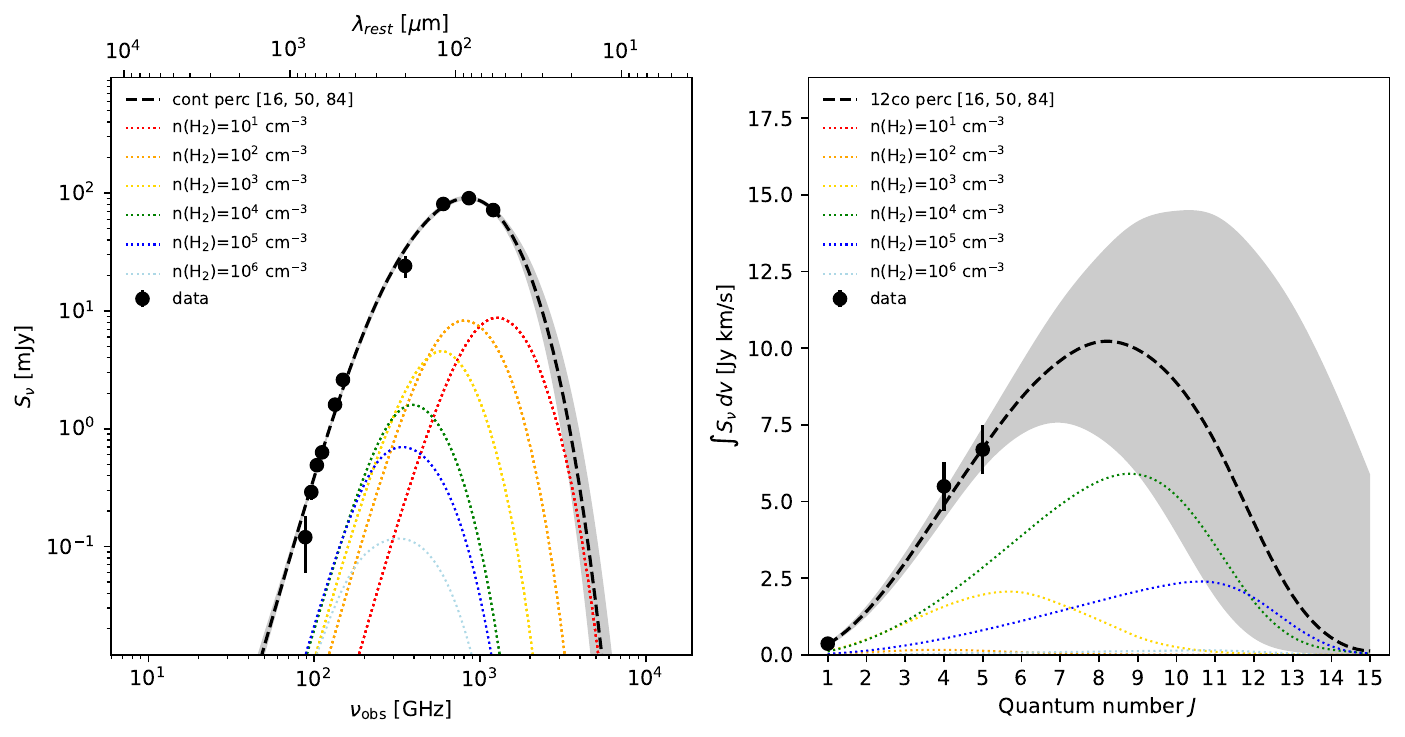}

    \addtocounter{figure}{-1}
\caption{{(continued) TUNER fits for HerBS-165 and HerBS-169 from Group-1, and HeLMS-50 and HerBS-43a from Group-2.}}
\end{figure}

\begin{figure}[H]
    \centering
\includegraphics[width=0.6\textwidth]{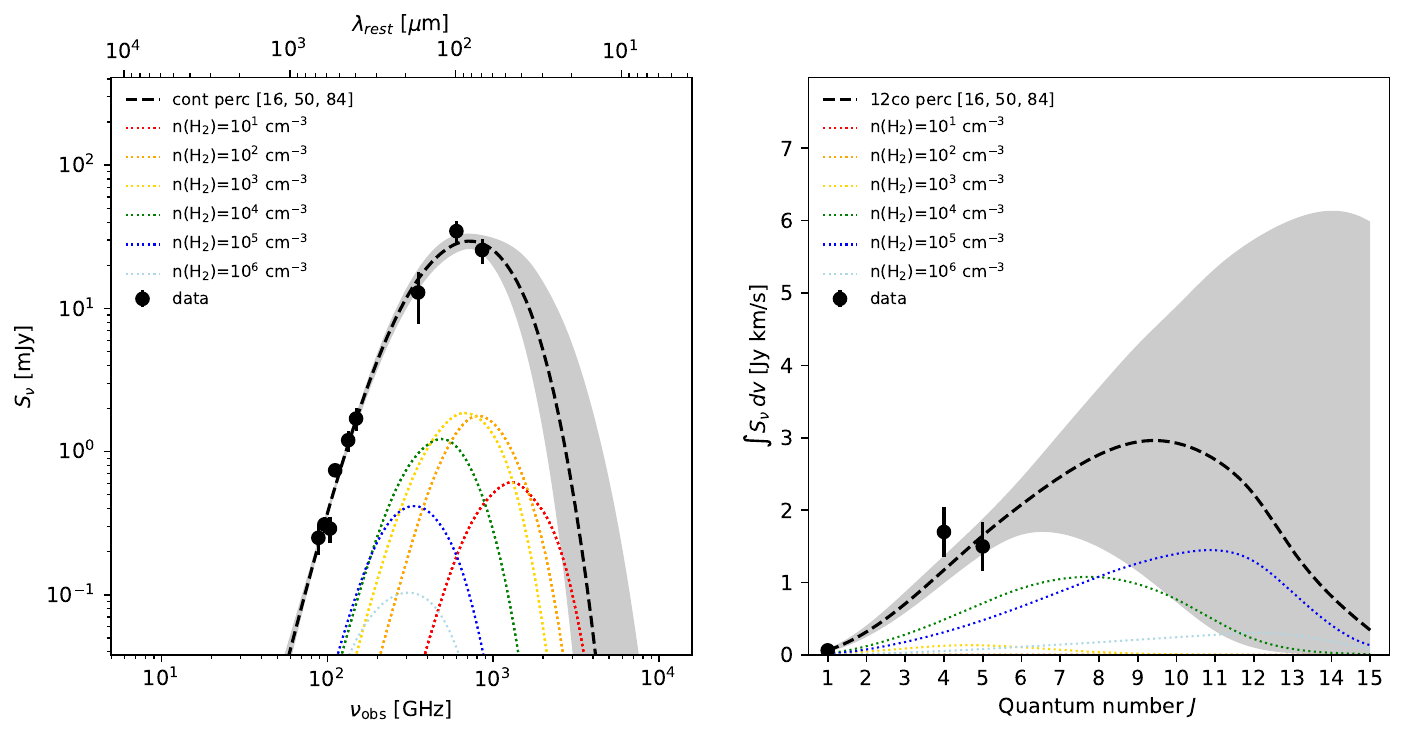}
\includegraphics[width=0.6\textwidth]{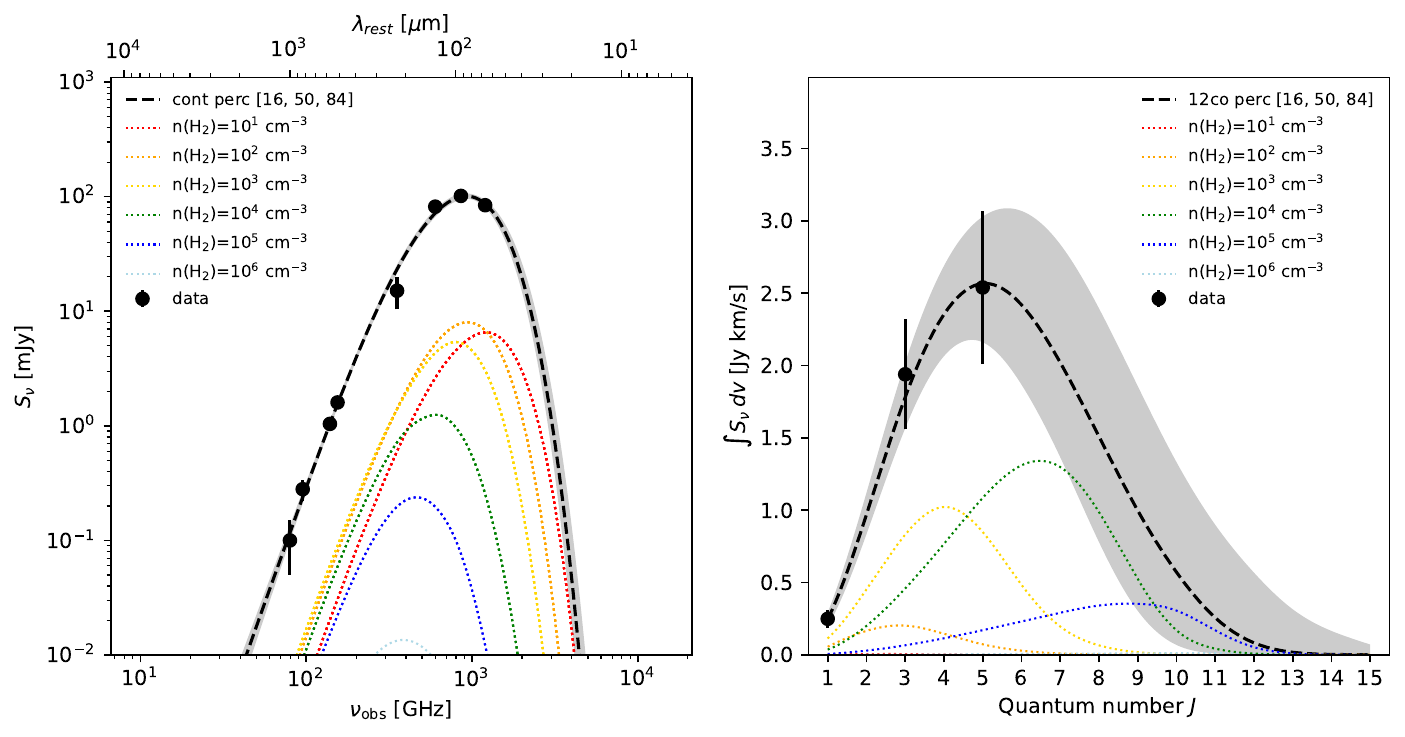}
\includegraphics[width=0.6\textwidth]{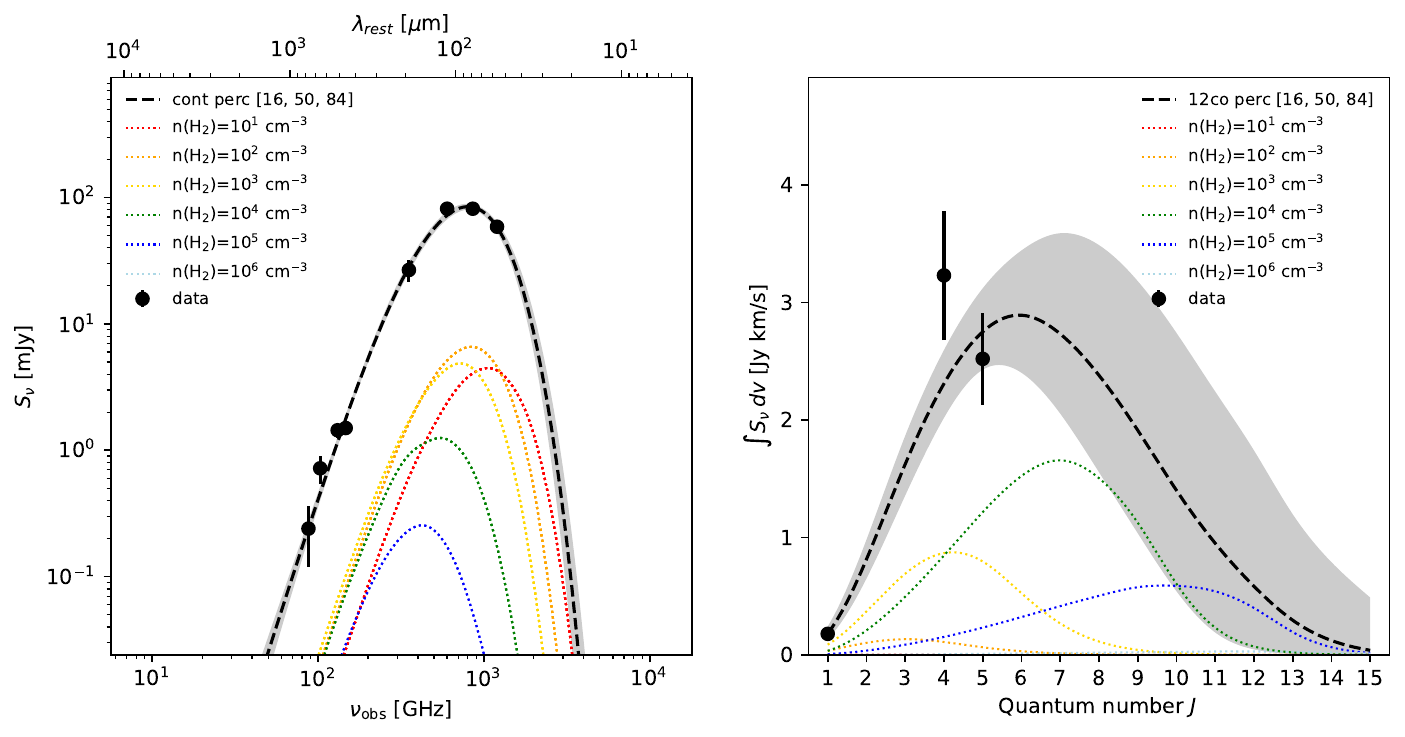}
    \addtocounter{figure}{-1}
\caption{{(continued) TUNER fits for Group-2 sources, namely HerBS-43b, HerBS-188, and HerBS-191.}}
\end{figure}

\begin{figure}[H]
\centering
\includegraphics[width=0.63\textwidth]{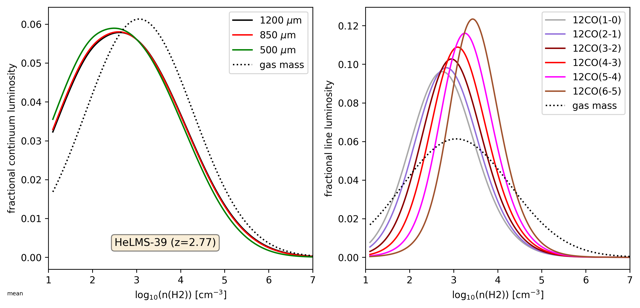}
\includegraphics[width=0.3\textwidth]{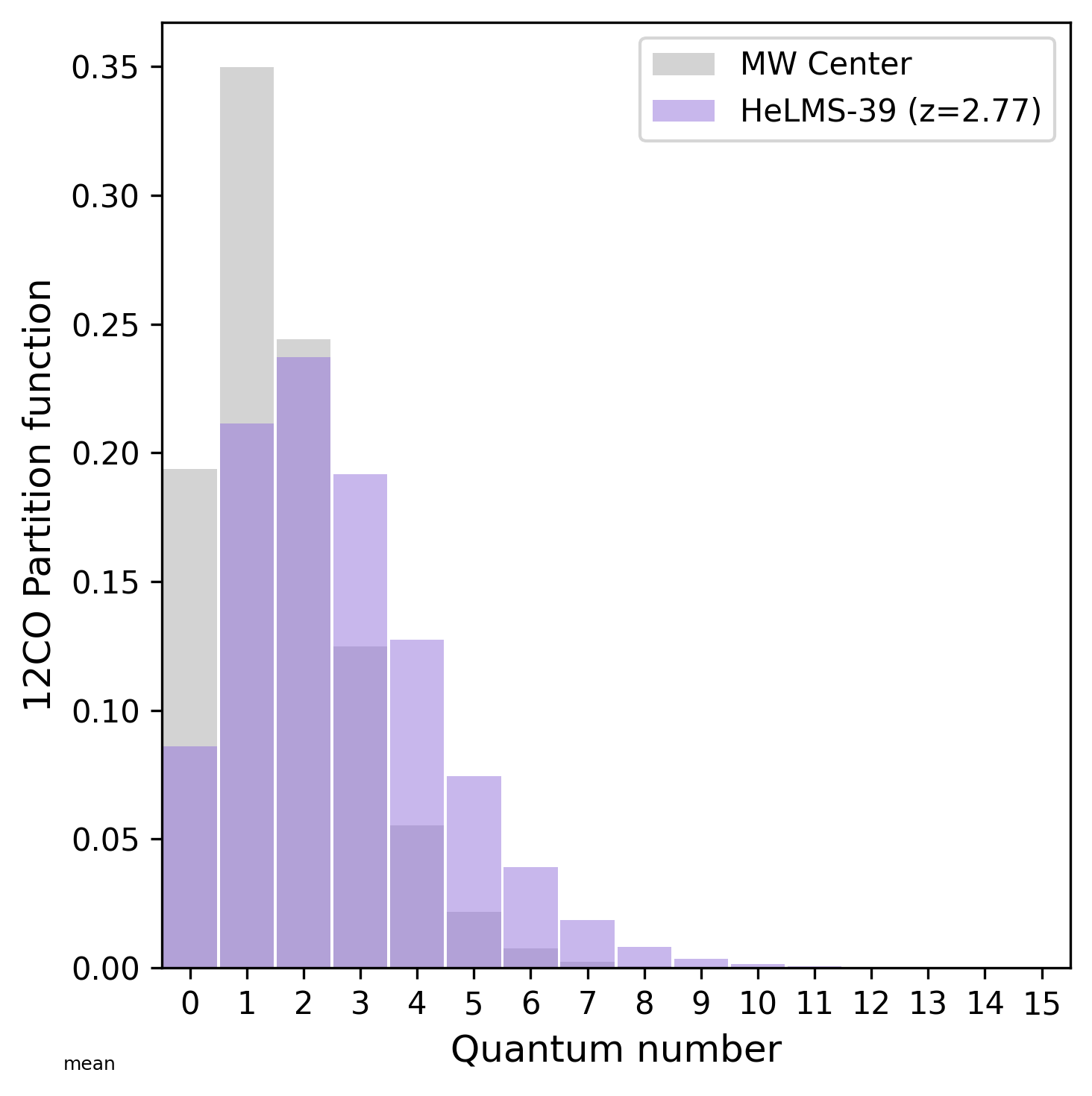}
\includegraphics[width=0.63\textwidth]{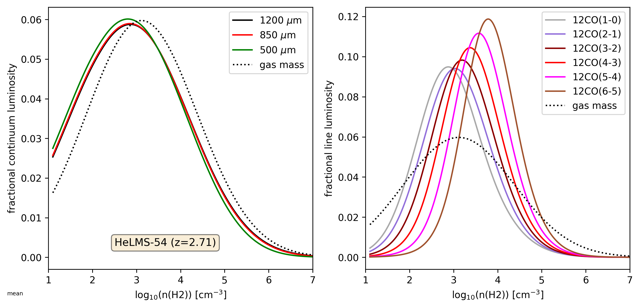}
\includegraphics[width=0.33\textwidth]{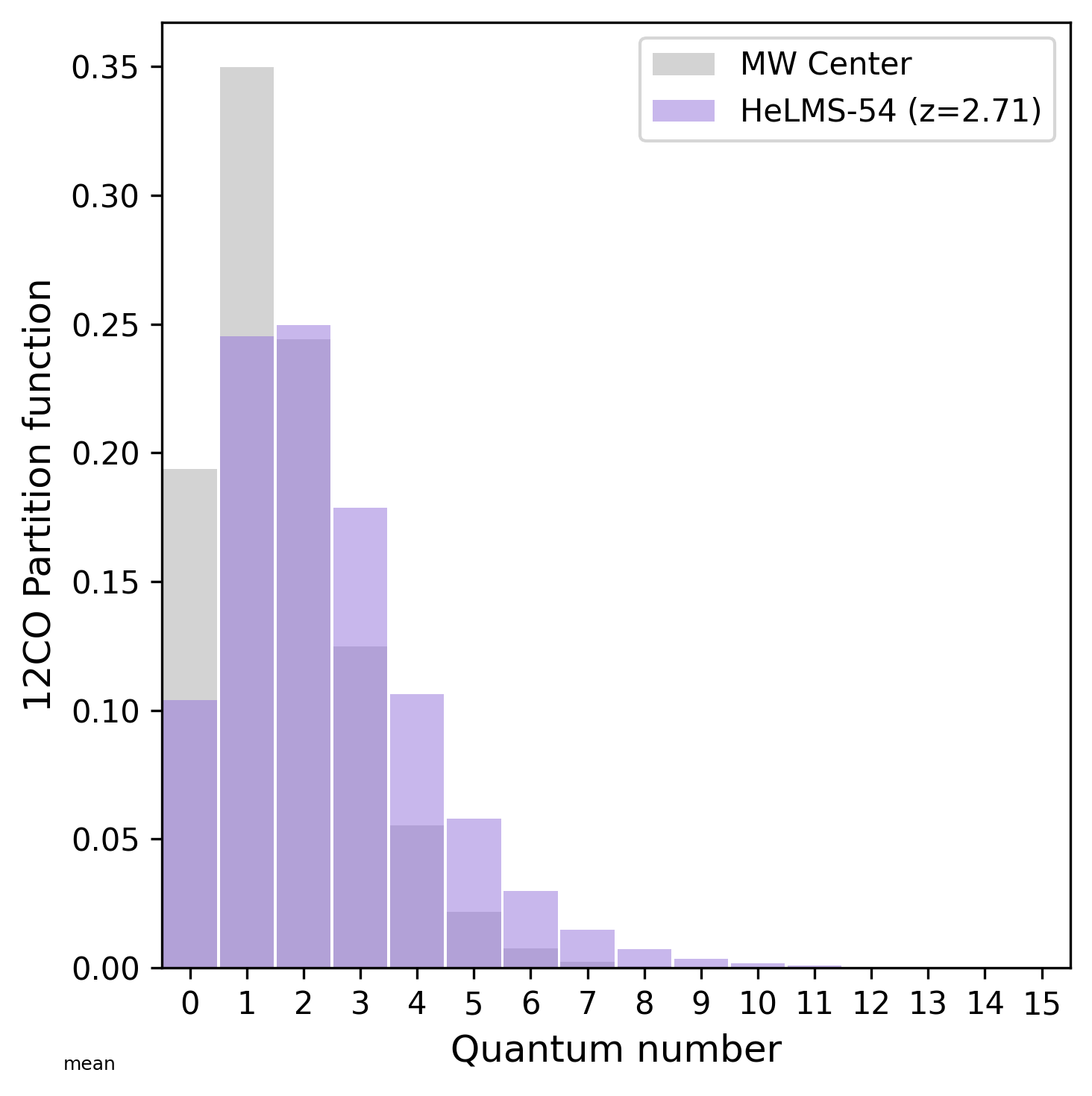}
\includegraphics[width=0.63\textwidth]{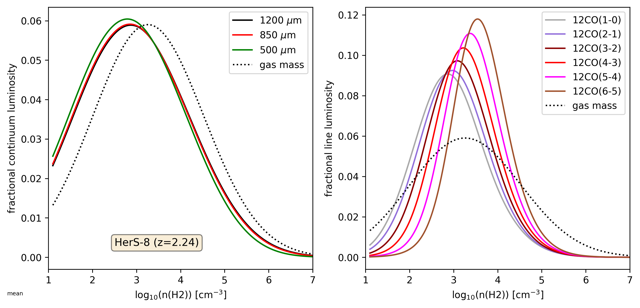}
\includegraphics[width=0.3\textwidth]{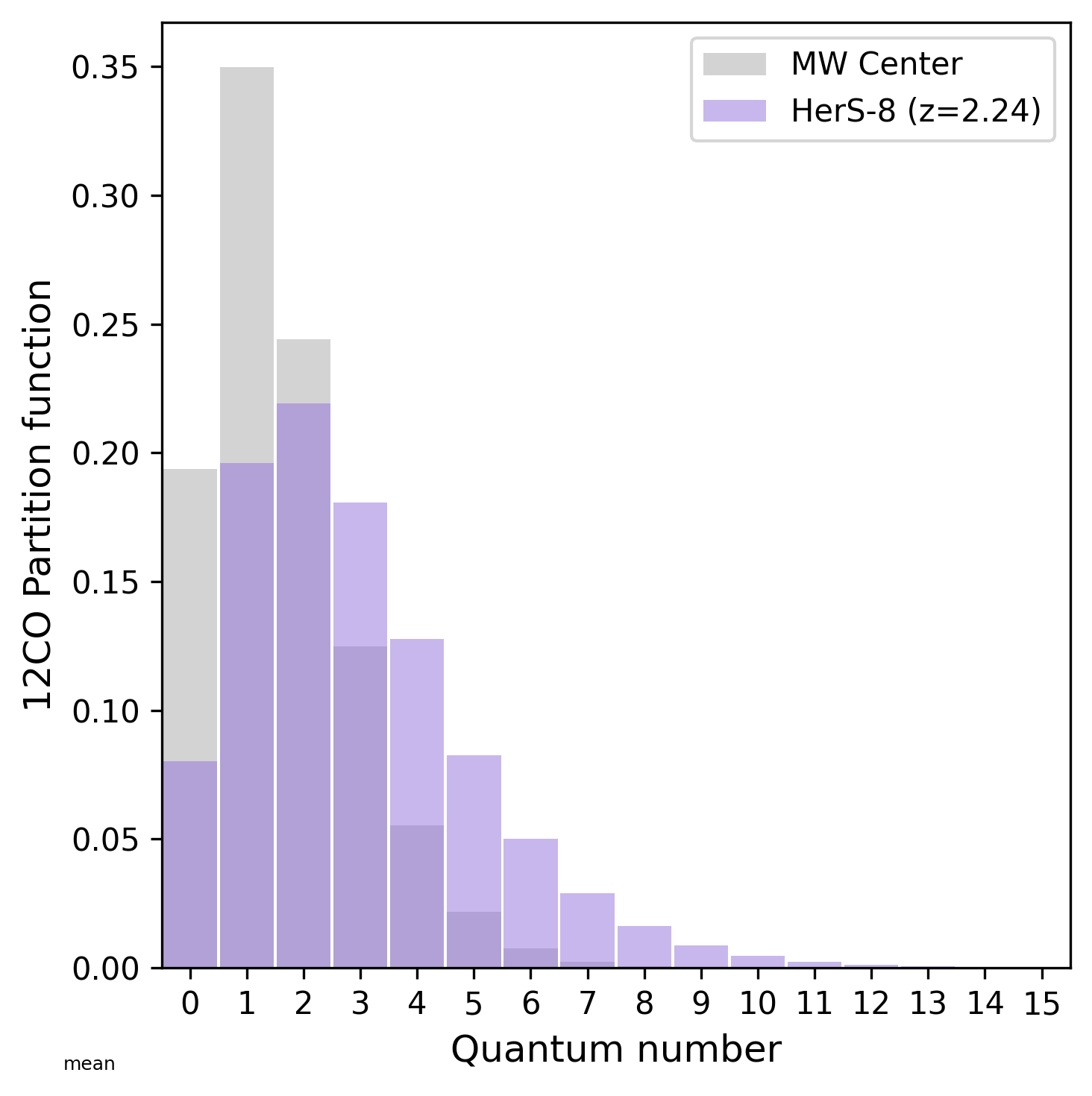}
\includegraphics[width=0.63\textwidth]{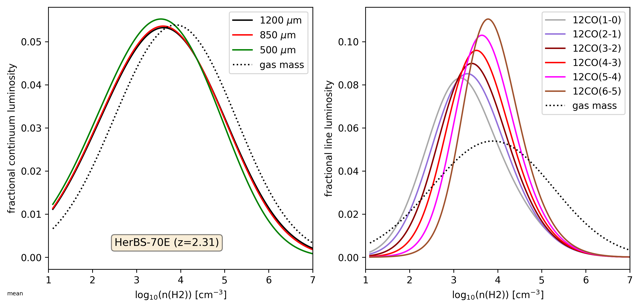}
\includegraphics[width=0.3\textwidth]{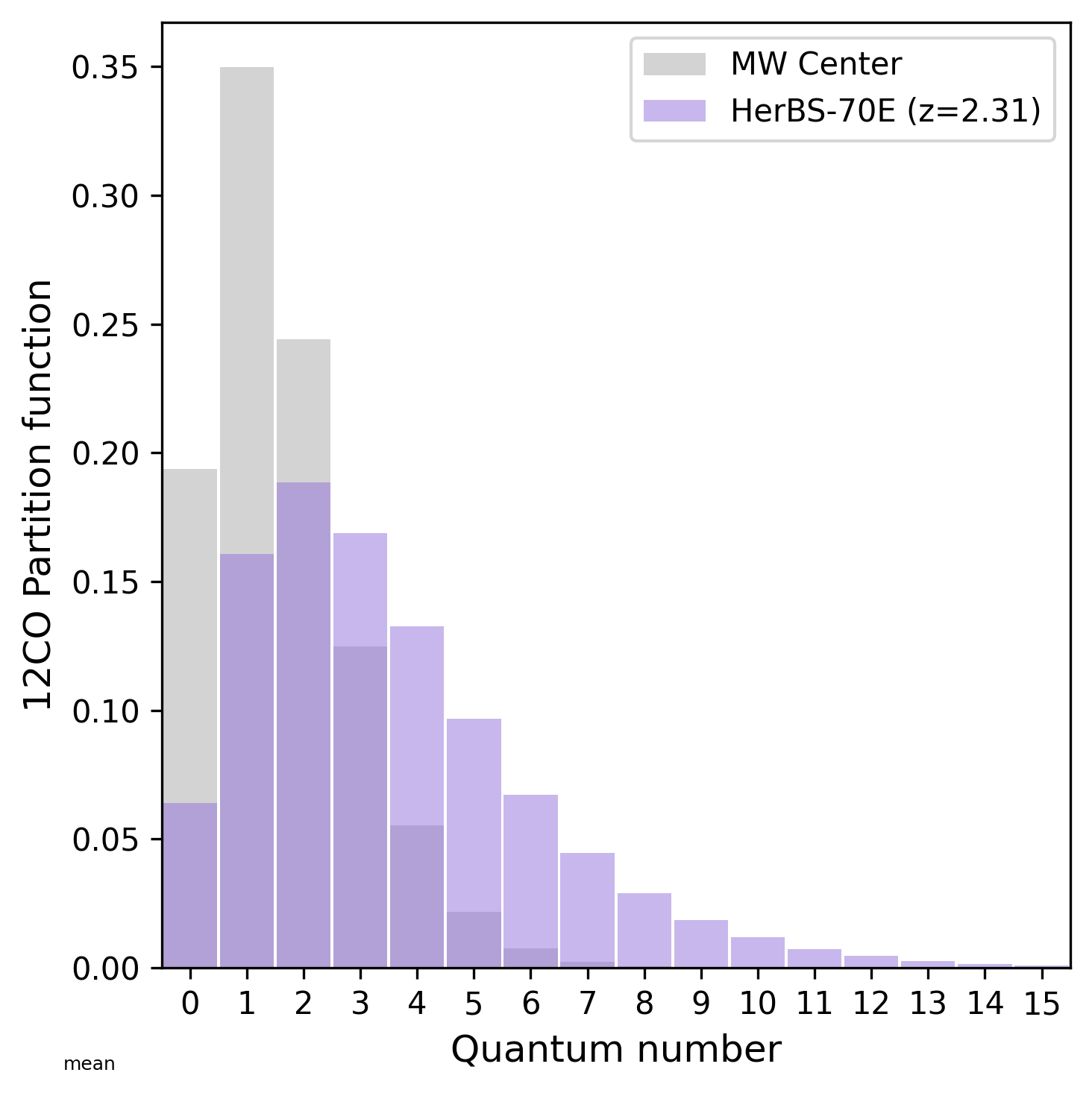}

\caption{Similar to Figure~\ref{fig:tuner_herbs51} for HerBS-51, here shown are the continuum/line luminosity versus density plots, with the overplotted gas mass distribution from TUNER. Group-1 sources, namely HeLMS-39, HeLMS-54, HerS-8, and HerBS-70E are covered here.}
\label{fig:levelpop}
\end{figure}

\begin{figure}[H]
    \centering
\includegraphics[width=0.63\textwidth]{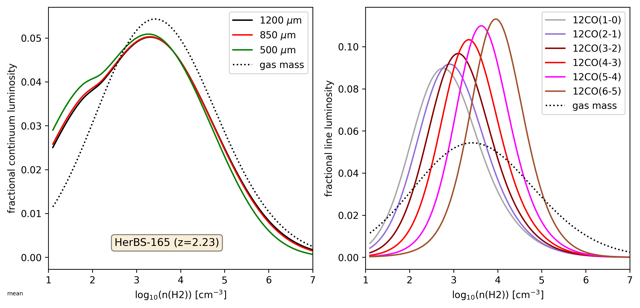}
\includegraphics[width=0.3\textwidth]{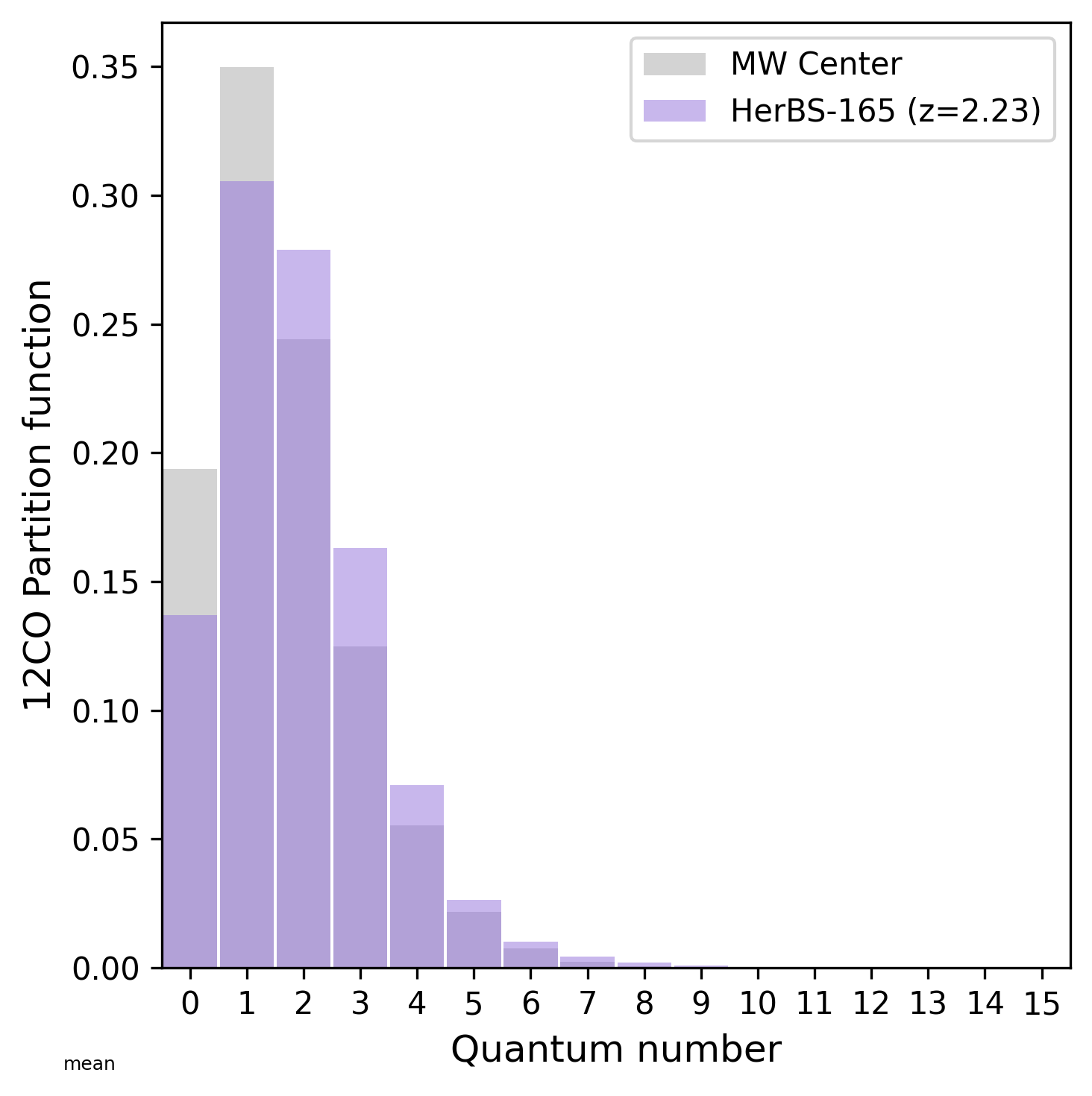}
\includegraphics[width=0.63\textwidth]{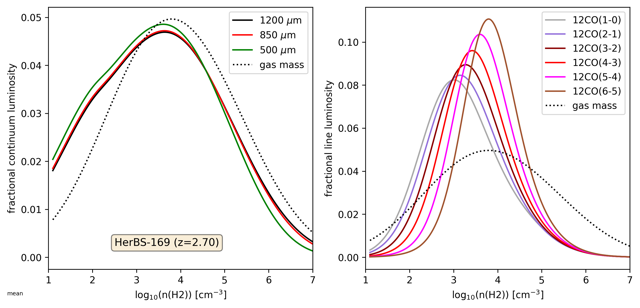}
\includegraphics[width=0.33\textwidth]{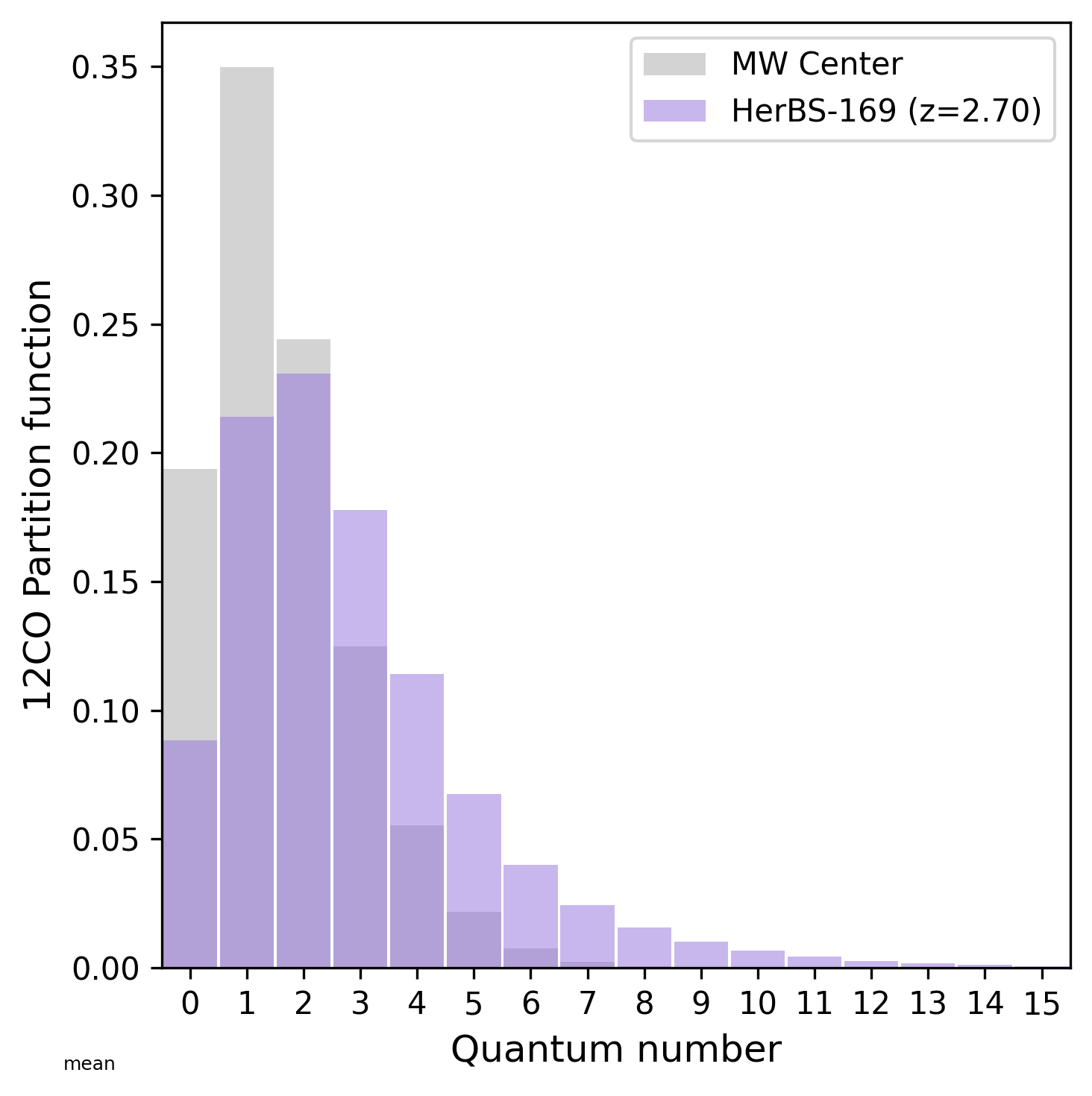}
\includegraphics[width=0.63\textwidth]{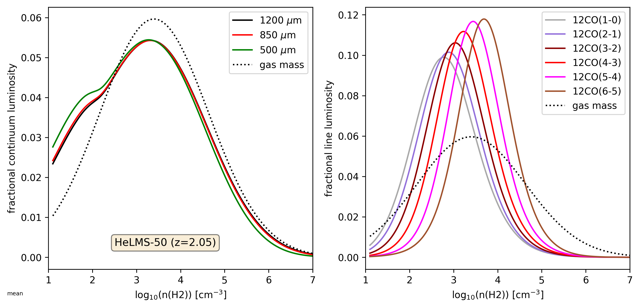}
\includegraphics[width=0.3\textwidth]{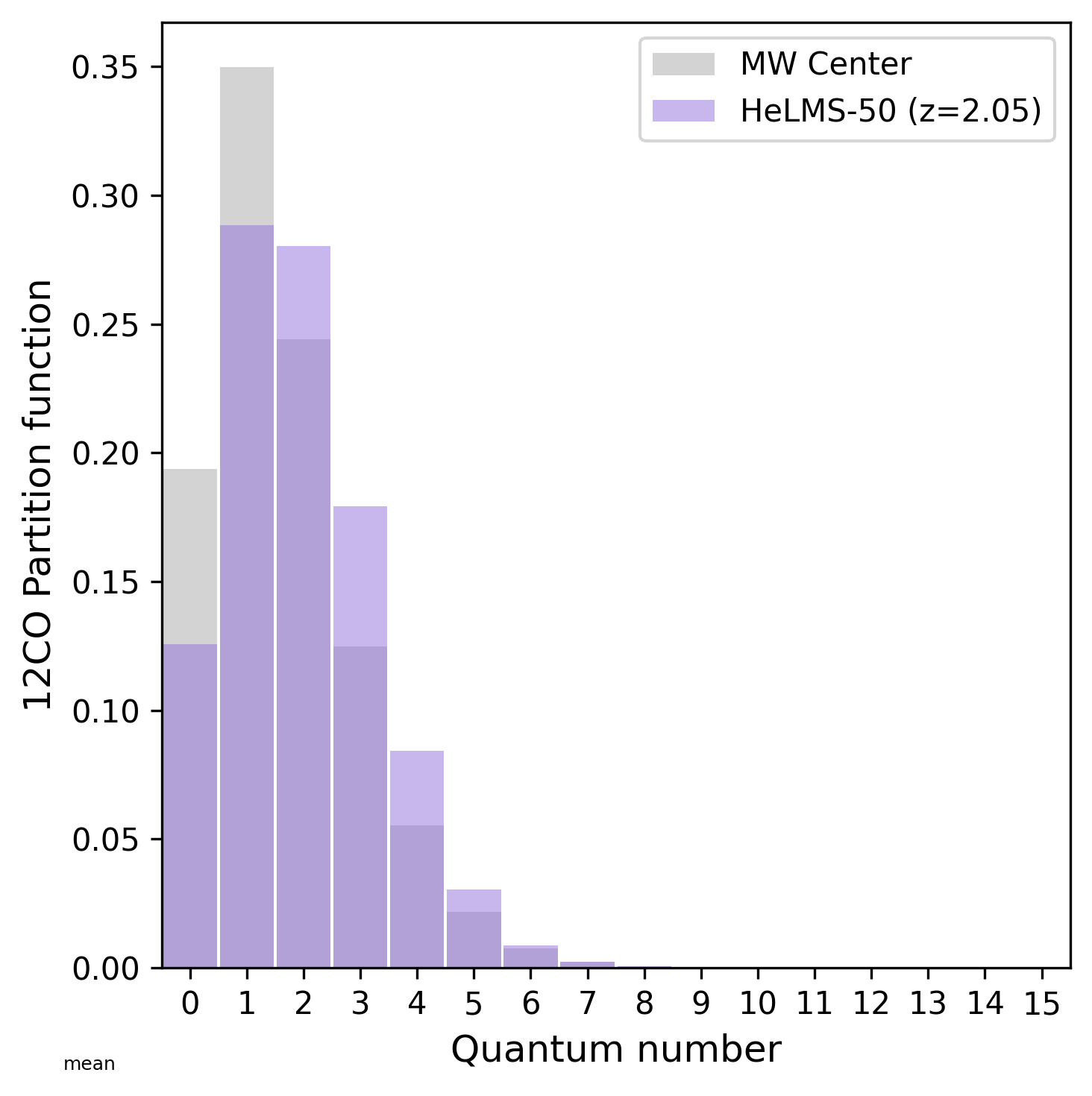}
\includegraphics[width=0.63\textwidth]{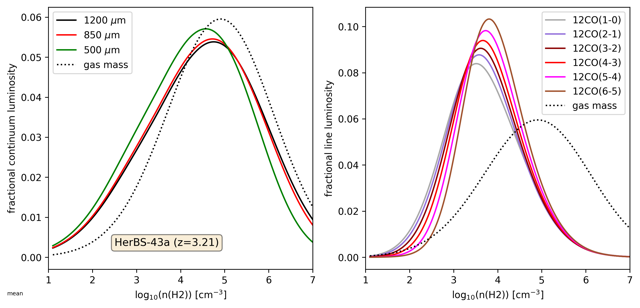}
\includegraphics[width=0.3\textwidth]{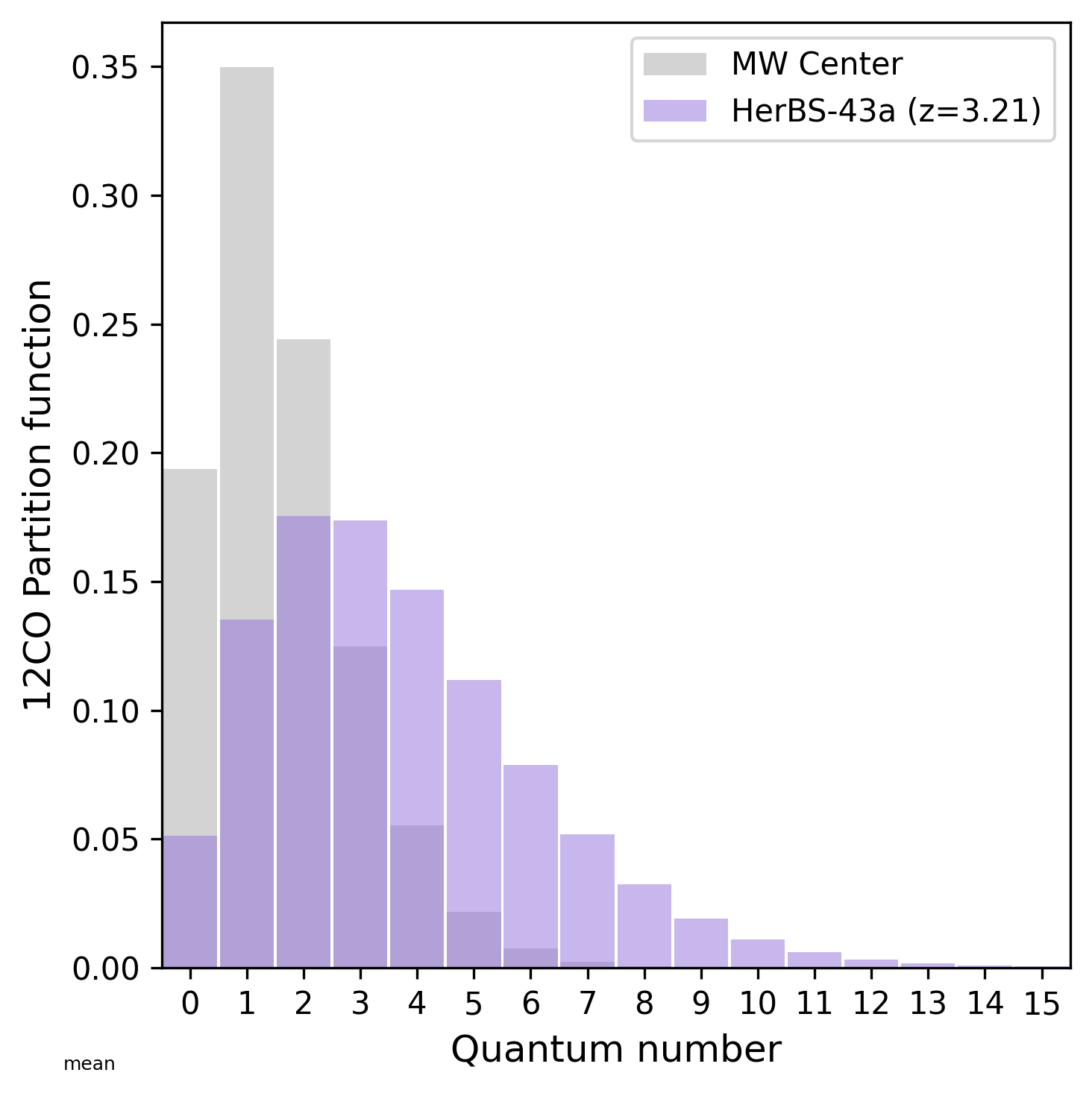}

    \addtocounter{figure}{-1}
\caption{{(continued) TUNER fits for HerBS-165 and HerBS-169 from Group-1, and HeLMS-50 and HerBS-43a from Group-2.}}
\end{figure}

\begin{figure}[H]
    \centering
\includegraphics[width=0.63\textwidth]{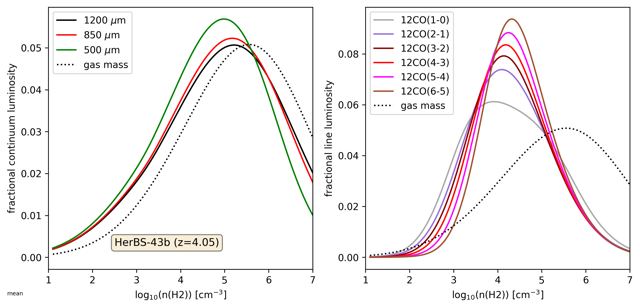}
\includegraphics[width=0.3\textwidth]{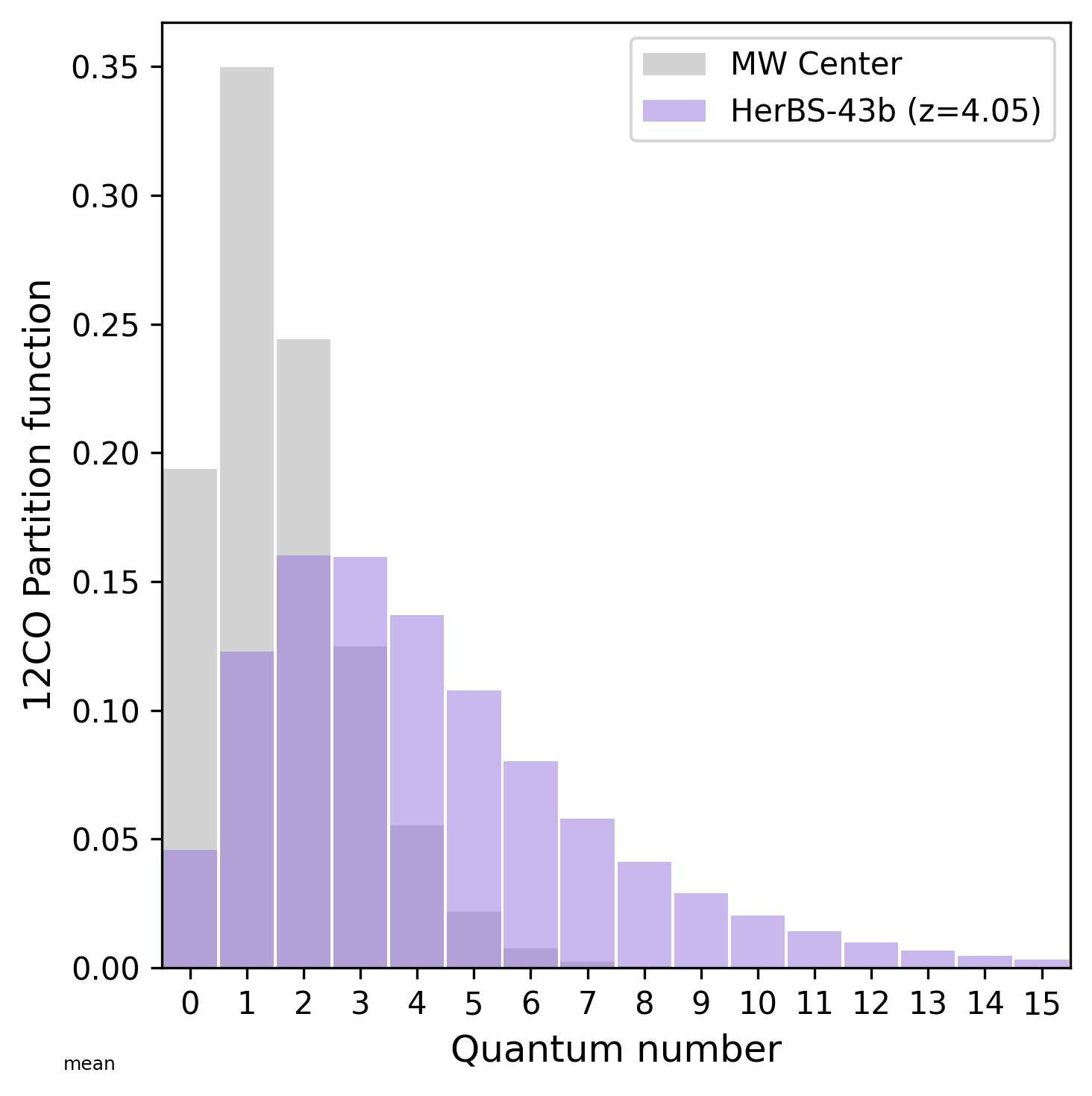}
\includegraphics[width=0.63\textwidth]{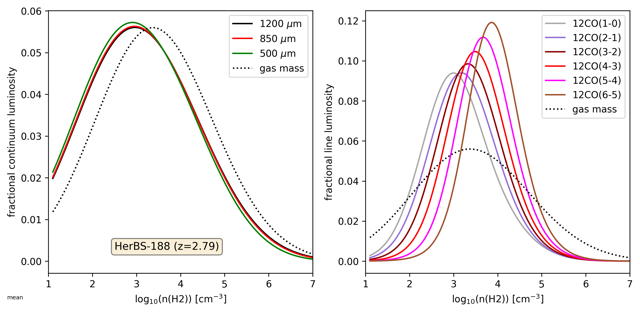}
\includegraphics[width=0.33\textwidth]{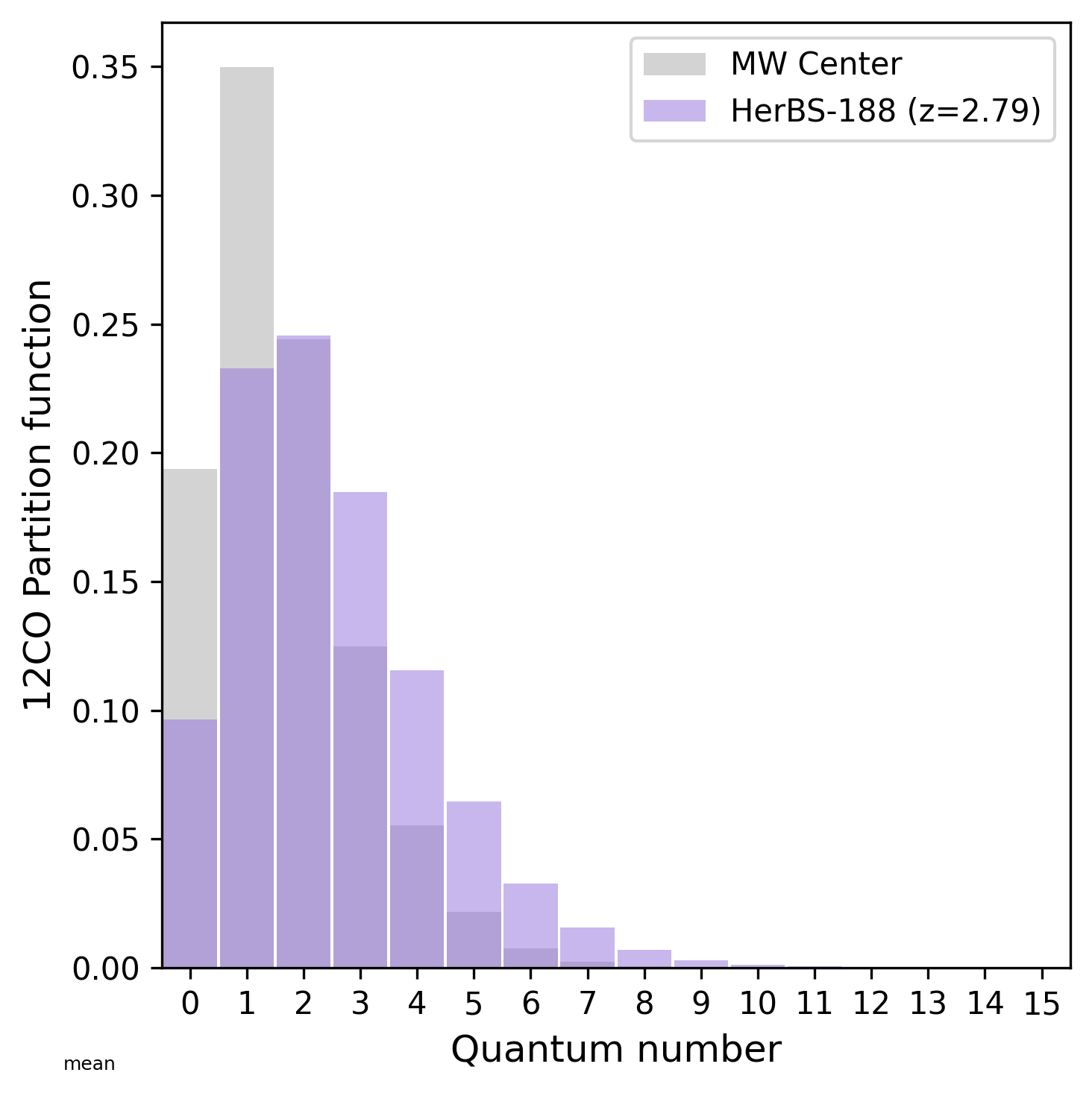}
\includegraphics[width=0.63\textwidth]{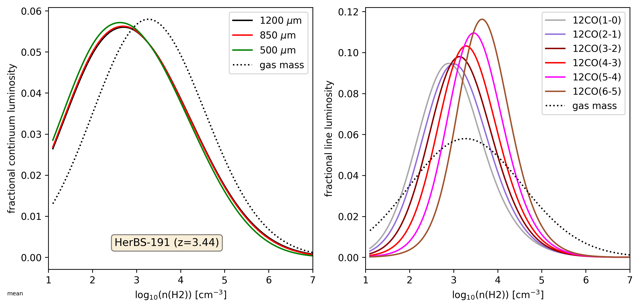}
\includegraphics[width=0.3\textwidth]{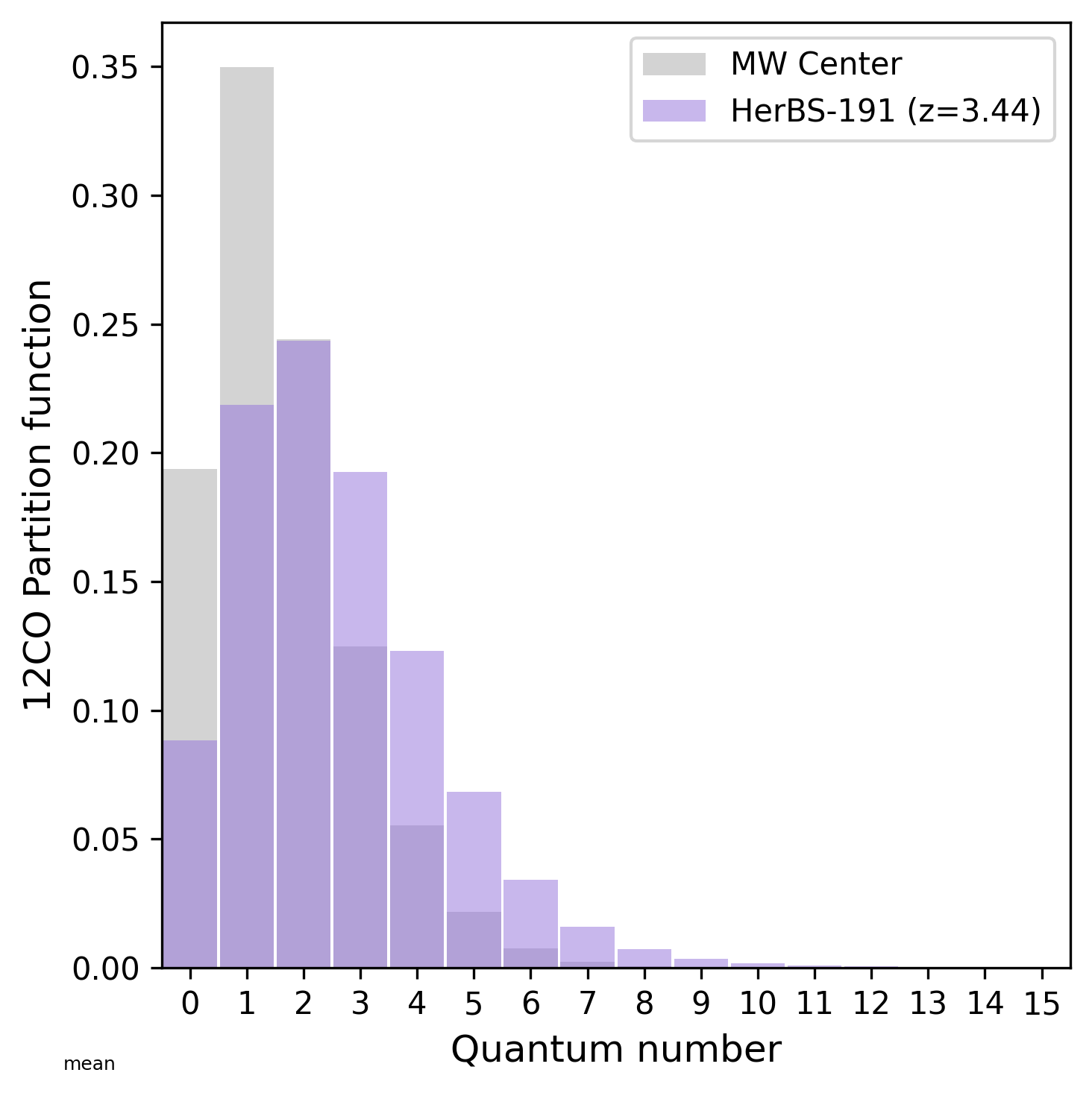}
    \addtocounter{figure}{-1}
\caption{{(continued) TUNER fits for Group-2 sources, namely HerBS-43b, HerBS-188, and HerBS-191.}}
\end{figure}

\clearpage

\subsection{Contributors to the Dynamical Mass Uncertainties and ``\aco Tension"} \label{subsec:allcontriaco}

\begin{figure}[h]
\centering
\includegraphics[width=0.45\textwidth]{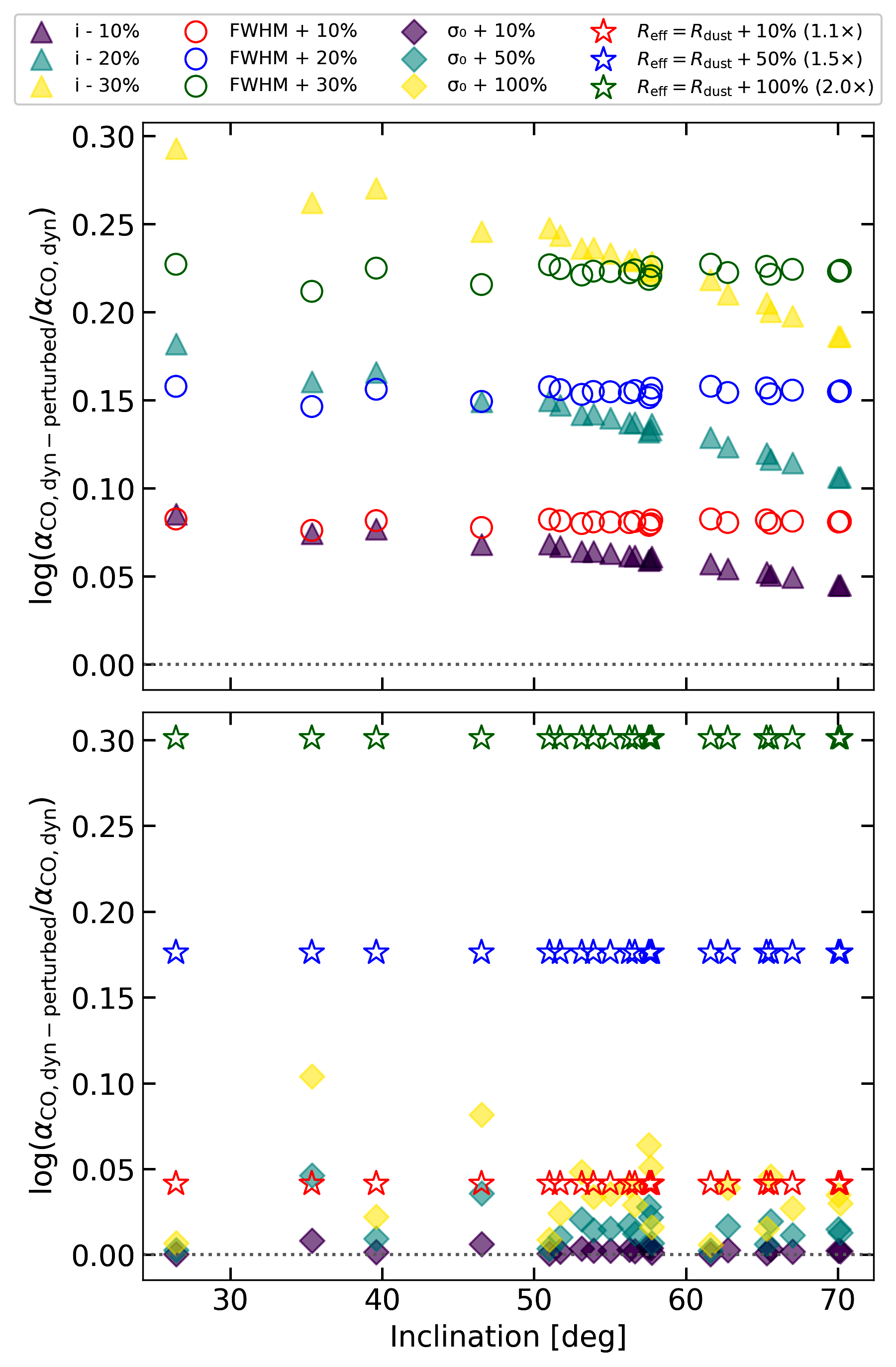}
\includegraphics[width=0.45\textwidth]{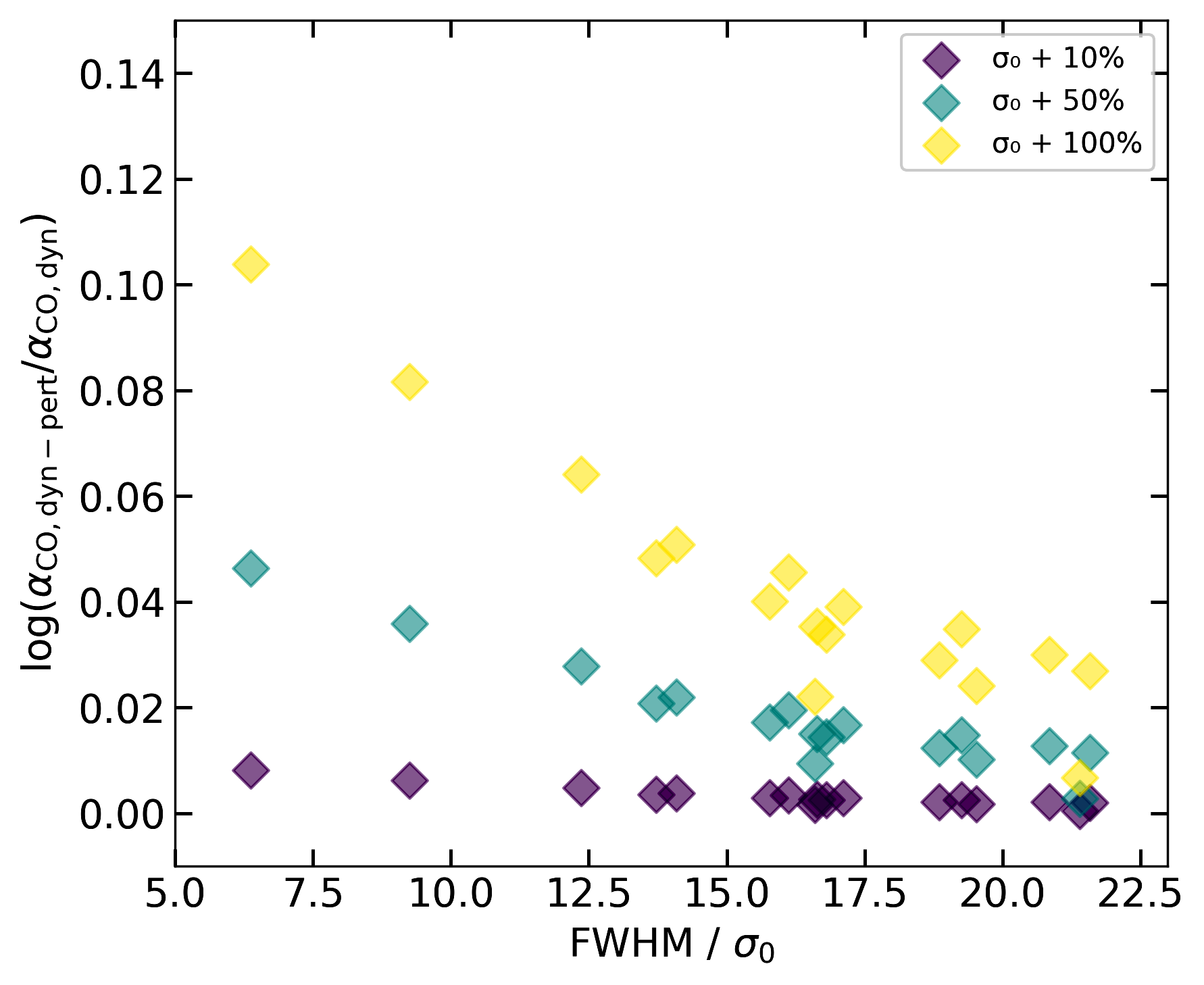}
\caption{Similar to Figure~\ref{fig:gn20} for GN20, for our 21 targeted DSFGs (Table~\ref{tab:dynamicalmasses}), we here quantify the contributions from perturbations in various parameters that go into deriving \aco from the dynamical mass estimates (Eqn.~\ref{eqn:Mdyn}). The ratios of the perturbed to baseline \aco values are presented on a logarithmic scale to emphasize order-of-magnitude variations.}
\label{fig:allparams_aco}
\end{figure}

In the absence of resolved cold gas data, the dynamical mass estimation using the proposed ``mixed" (i.e., rotating, turbulent, thick-disk) estimator in Eqn.~\ref{eqn:Mdyn} involves four key observables, namely inclination and \coonezero FWHM for $v_{\rm rot}$, velocity dispersion, and radius. In Figure~\ref{fig:allparams_aco}, we quantitatively show the contributions of uncertainty in each of these parameters that can propagate to the inferred dynamical upper limits on the $\alpha_{\rm CO}$. As expected, \aco is more sensitive to the \coonezero line FWHM and inclination angle --- both contributing through the rotational velocity assumption --- followed by radius being the next large uncertainty.

\end{document}